%% file: main.tex
\documentclass[11pt]{article}

\usepackage[moderate]{savetrees}

\usepackage{graphicx}
\usepackage{amsmath}
\usepackage{setspace}

\usepackage{multirow}

\usepackage{threeparttable}
\usepackage{booktabs, tabularx}
\usepackage{color}

\usepackage{chngcntr}

\usepackage[T1]{fontenc}

\usepackage[left=1in,top=1in,bottom=1in,right=1in]{geometry}

\usepackage[style=authoryear,citetracker=true,maxcitenames=1,natbib=true]{biblatex}
\AtEveryCitekey{\ifciteseen{}{\defcounter{maxnames}{2}}}
\newcommand{\num}[1]{#1}

\usepackage{hyperref}

\newcommand{\note}[1]{}

\newcommand{\appref}[1]{\ref{#1}}
\newcommand{\refFromApp}[1]{\ref{#1}}

\begin{document}

\title{Choosing and Using Information in Evaluation Decisions\thanks{
\noindent{\footnotesize {Pre-registered: AEARCTR-0010969. We received funding from Harvard Business School. We would like to thank May Hong, Yoona Kim, Emma Ronzetti, and Douglas Turner for excellent research assistance, and Lucas Coffman, Markus Eyting, Alex Imas, Nagore Iriberri, James Peck, Matthew Rabin, David Sappington, Joshua Schwartzstein, Andrei Shleifer, and Sevgi Yuksel for their helpful feedback and suggestions. We benefited from comments and questions from seminar audiences at Wharton, Ohio State, Cornell, NYU Abu Dhabi, Florida State, and Booth as well as conference participants at the Motivation and Incentives Conference at Harvard Business School, the 2023 Spring Behavioral and Experimental Economics Research Conference, the 2023 BRIQ Beliefs Workshop, the 2023 ESA Europe Meeting, and the 2023 SITE Gender Conference.}}}\\
}
\bigskip

\author{Katherine B. Coffman\\
Scott Kostyshak\\
Perihan O. Saygin\footnote{Coffman: \href{mailto:kcoffman@hbs.edu}{kcoffman@hbs.edu}, Harvard Business School; Kostyshak: \href{mailto:scott.kostyshak@upf.edu}{scott.kostyshak@upf.edu}, Visiting Professor at the Department of Economics and Business, Universitat Pompeu Fabra; Saygin: \href{mailto:peri.saygin@uab.cat}{peri.saygin@uab.cat}, Department of Applied Economics, Universitat Aut\`onoma de Barcelona.} \\
}

\date{\today}
\maketitle

\begin{abstract}

We use a controlled experiment to study how information acquisition impacts candidate evaluations. We provide evaluators with group-level information on performance and the opportunity to acquire additional, individual-level performance information before making a final evaluation. We find that, on average, evaluators under-acquire individual-level information, leading to more stereotypical evaluations of candidates. Consistent with stereotyping, we find that (irrelevant) group-level comparisons have a significant impact on how candidates are evaluated; group-level comparisons bias initial assessments, responses to information, and final evaluations. This leads to under-recognition of talented candidates from comparatively weaker groups and over-selection of untalented candidates from comparatively stronger groups.

\end{abstract}

\spacing{1.306}

	\newpage

\section{Introduction}

Evaluations play an important role in determining economic outcomes. In academic contexts, teachers provide grades to students, admissions officers select new cohorts of university students, advisors identify promising graduate students, and editors and referees decide which papers will be published. In professional contexts, hiring managers choose candidates and supervisors review performance. Even outside of school and work, life trajectories are shaped by the discretionary judgements of evaluators such as mortgage lenders, criminal justice officers, and doctors.

Across these evaluation contexts, social scientists have explored the prevalence, sources, and consequences of discrimination. Perhaps the most common approach to documenting discrimination is to compare the treatment of two otherwise identical candidates, isolating a role for gender (or race, age, etc.) in driving disparate outcomes. For instance, audit studies such as the seminal work by \citet{bertrand2004emily} send identical resumes to different job openings, varying only the name attached to the resume. They find that resumes receive significantly fewer callbacks when they are labeled with a Black-sounding name rather than a white-sounding name. Studies of this type provide clean, compelling evidence of discrimination: \textit{given} identical information about two candidates, those candidates are treated differently based upon their race or gender.

Our project builds on this important body of work by exploring what happens when information is not \textit{given} but instead must be \textit{acquired}.\footnote{Our work is connected to recent projects that have explored discrimination via endogenous allocation of attention, see \citet{bartovs2016attention} and \citet{10.1093/ej/uead079}, discussed in detail in Section \ref{sec.lit}.} We consider a decision-maker who must choose how much to learn about a candidate before making their evaluation. For instance, a hiring manager must choose how many references to call before deciding whether to interview the candidate, a loan officer must decide how many years of employment history to review carefully before deciding whether to extend a mortgage, or a doctor must decide how many tests to order before making a diagnosis. Our hypothesis is that these endogenous decisions about how much information to acquire have the potential to fuel discrimination. 

We study this question within the context of a simulated hiring experiment. Evaluators are presented with limited information about a candidate. They make an initial assessment of that candidate's likely quality. Then, they decide whether to acquire additional information about the candidate or instead make a final evaluation. Acquiring additional information is costly, but arguably only modestly so. We use this environment to study (i) evaluators' initial beliefs about candidate quality, (ii) when evaluators choose to acquire more information, (iii) how evaluators update their beliefs in response to that information, and (iv) evaluators' final assessments of candidate quality. We ask how each of these decisions depends upon candidate gender and the characteristics of the pool of candidates. In particular, we ask how the extent of gender discrimination depends upon the underlying gender gaps in performance in the candidate pool.

In a preliminary study, we collect performance data from a group of workers via the online platform Prolific. These workers serve as the candidates in our main evaluation study. For each candidate, we measure performance on a brief multiple-choice math and science test. Each candidate also answers some brief questions about themselves, including their gender. Based upon their performance, we classify each candidate as either Above-the-Bar or Below-the-Bar.

In our main study, we recruit new participants from Prolific to serve as evaluators. The task of evaluators is to accurately classify candidates as Above or Below-the-Bar. For each candidate, an evaluator sees a brief profile that includes the gender of the candidate. In our endogenous treatment, evaluators choose whether to receive additional information about the candidate prior to making their classification. This additional information takes the form of noisy binary signals about the candidate's performance. Signals are imperfect but informative, with Above-the-Bar candidates generating a positive signal more often than Below-the-Bar candidates. Evaluators can choose to acquire up to 5 signals about a given candidate before making their classification. In our exogenous treatment, evaluators must view all 5 signals, sequentially, before making their classification. Comparisons across these two treatments isolate the role of endogenizing information acquisition. This design also allows us to separately consider both evaluator selection into receiving more information and biased responses to information absent these selection effects. 

We incentivize evaluators to make accurate judgements by offering a bonus for a correct final classification of a candidate. We take steps to encourage information acquisition by assigning only a small cost to each signal (5 cents) and by minimizing the time cost of viewing additional signals. We also ask for the evaluator's initial belief of the candidate being Above-the-Bar, based purely on the candidate profile, and then elicit an updated belief after each additional signal acquired. 

This design yields a wealth of interesting data to explore. From an evaluation standpoint, we can analyze initial assessments, how beliefs are updated, and the accuracy of final classifications. From an information standpoint, we can unpack the endogeneity of information acquisition, asking what predicts the decision to seek more information about a candidate and how this impacts final classifications. Our goal is to understand how biases emerge from the choice and use of information. 

A critical piece of our design is the construction of four different candidate pools for evaluators to consider. All pools have 50 candidates, 25 male and 25 female. But, the pools vary in the share of male and female candidates within the pool that are Above-the-Bar: 60\% of men and 40\% of women; 60\% of both men and women; 40\% of both men and women; and, finally, 40\% of men and 60\% of women. We will refer to these pools as $60-40$, $60-60$, $40-40$, and $40-60$, respectively.

We provide evaluators with the relevant shares for their pool, and this information remains available to them throughout their participation. We choose to provide these shares instead of relying on the (potentially biased) prior beliefs of evaluators so that we can induce exogenous variation in beliefs about gender differences. This is critical for understanding what role beliefs about population-level gender differences play in driving discriminatory behavior. By exploring how gender discrimination varies across pools, we can back out the extent to which relative comparisons between men and women are biasing evaluator decision-making. %

Discrimination on the basis of relative comparisons is deeply tied to a recent literature on the importance of contrast across groups in shaping beliefs. The \citet{Bordalo_etal_2016} stereotypes model predicts that beliefs about a group are biased toward types that are diagnostic or representative relative to a reference group---those types that are \textit{relatively} more frequent in the target group compared to the reference group.\footnote{Evidence consistent with this model of stereotyping has been found in a variety of contexts. See \citet{alesina2023immigration}, \citet{Bordalo_etal_2019}, \citet{carlana2019implicit}, \citet{hebert2020gender}, and \citet{coffman2014evidence}, among others.} This framework makes a clear prediction for our setting. Consider the case in which 60\% of men are Above-the-Bar. Stereotyping predicts that beliefs about men will be more favorable when they are in a pool where only 40\% of women are Above-the-Bar than when they are in a pool where 60\% of women are also Above-the-Bar. Evaluator beliefs for a given group would thus depend not only on the true proportion of that group that is Above-the-Bar, but also on the relative advantage that group has compared to the other. We test this prediction in our setting by considering how the assessment of a candidate of a given gender depends upon their group's relative advantage (or disadvantage) within the pool.

In this same spirit, \citet{esponda2023seeing} document that contrast between groups influences how new information is perceived. Relying on an abstract setting with green and orange groups, they show that evaluator perceptions of a noisy visual signal depend on differences in priors across groups: the same visual signal is perceived as more favorable when it is associated with the advantaged group. Our project advances this important line of work by considering how stereotyping shapes not only how evaluators use information, but also how they choose how much of it to acquire. Relative to \citet{esponda2023seeing}, we study a setting where there is arguably less wiggle room in how a signal could be perceived. And, by using a less abstract setting, we explore the explanatory power of this important channel for gender discrimination. 

Collapsing across the four candidate pools, we find no overall evidence of discrimination against female or male candidates. On average across the pools, evaluators assess female candidates no less favorably than male candidates, both in terms of initial assessments and final classifications.

However, we observe significant and substantial evidence of contrast-driven stereotyping: evaluator assessments are biased by comparisons across groups. Holding fixed a candidate's prior likelihood of being Above-the-Bar, a candidate is treated significantly more favorably when their gender has a relative advantage. For example, men are significantly more likely to be classified as Above-the-Bar in the $60-40$ pool, where a greater share of men are top performers, than in the $60-60$ pool, where both men and women have the same chance of being a top performer. Importantly, female candidates also benefit from relative advantage when it is their group that is advantaged: conditional on having a 60\% chance of being Above-the-Bar, they are evaluated significantly more favorably when male candidates have only a 40\% chance of being Above-the-Bar relative to when both groups have a 60\% chance of being Above-the-Bar. Just as candidates with a 60\% chance of being Above-the-Bar are viewed more positively when their group has the relative advantage, candidates with a 40\% chance of being Above-the-Bar are viewed more negatively when their group has the relative disadvantage: evaluators provide more negative assessments to low-prior candidates from the unequal pools than to those same candidates in the $40-40$ pools. 

Relative advantage has a sizable impact on initial assessments, prior to the receipt of any signals: the gap in initial assessments between high and low-prior candidates is 75\% larger when evaluators face candidate pools with unequal groups.\footnote{We compute this by contrasting the comparisons of the average assessment of low and high-prior candidates from the $40-40$ and $60-60$ pools to the assessments of low and high-prior candidates from the $60-40$ and $40-60$ pools.} Relative advantage continues to matter in final assessments, even after evaluators have had the chance to acquire more information. The gap in final classifications between high and low-prior candidates is more than 50\% larger when evaluators face candidate pools with unequal groups. 

Final evaluations are biased, in part, because of under-acquisition of information. In nearly 25\% of cases, evaluators choose to receive no signals about the candidate and the median number of signals acquired is 2.\footnote{By contrast, a Bayesian decision-maker would always acquire at least 3 signals per candidate. We show this formally in Section \ref{sec.infoacq}.} There is significant selection into information. The initial assessments of evaluators who choose to receive no signals about the candidate are significantly more biased by relative advantage on average. That is, individuals who are more biased \textit{initially} are less likely to then acquire the information that could eventually help to de-bias them.

In the endogenous treatment, high-prior candidates from the advantaged group are 6pp more likely to be classified as Above-the-Bar compared to high-prior candidates in the $60-60$ pool. And, evaluators are quicker to classify them as such, requiring significantly fewer signals on average before deciding to make their final evaluation. Similarly, low-prior candidates from the disadvantaged group are 6pp less likely to be classified as Above-the-Bar compared to low-prior candidates in the $40-40$ pool, and evaluators require significantly more signals on average before making this positive determination when the candidate is disadvantaged. Note that these findings rely on comparisons of candidates \textit{with the same prior likelihood of being Above-the-Bar}, who produce the same signals in expectation. These otherwise identical candidates are treated differently depending upon whether their group happens to be advantaged or disadvantaged relative to the other. 

Relative advantage shapes not only how much information evaluators choose to acquire, but also how they use the information they receive. By shifting focus to the exogenous treatment, we can study biases in belief updating absent selection into information. Consider a high-prior candidate. When an evaluator views a negative signal about that candidate, they remain significantly more optimistic when the candidate is from a relatively advantaged group ($60-40$ relative to $60-60$). Casually speaking, relative advantage seems to provide high-prior candidates more benefit of the doubt. On the flip side, when an evaluator views a positive signal for a low-prior candidate, they update their beliefs significantly less favorably when that candidate is relatively disadvantaged ($40-60$ relative to $40-40$). Informally, low-prior candidates seem to have a harder time convincing evaluators of their high quality when they are at a relative disadvantage.

While overall accuracy of final classifications is high, endogenizing information acquisition significantly increases the rate of mistakes, from 16\% in the exogenous treatment to 25\% in the endogenous treatment. Not surprisingly, mistakes are more common for atypical candidates: Above-the-Bar candidates from low-prior groups and Below-the-Bar candidates from high-prior groups are significantly less likely to be classified correctly. These mistakes are more common (i)~in the endogenous treatment, and (ii) in pools where the groups are unequal. That is, a talented candidate from a low-prior group is least likely to be identified as such when their group is disadvantaged relative to the other, and when their evaluator gets to choose whether to seek out more information. Similarly, evaluators are less likely to recognize that a candidate from a high-prior group is actually Below-the-Bar when that candidate is from a relatively advantaged group and when they can choose not to receive additional information.

Together, these results illustrate the reinforcing role of relative advantage. Contrast between groups makes it more likely that high-prior candidates are classified as Above-the-Bar and that low-prior candidates are classified as Below-the-Bar. Allowing for endogenous information acquisition exacerbates the impact of this form of stereotyping, with group advantage reducing the number of signals evaluators acquire before reaching a stereotypical judgment. While these judgements are correct on average, stereotyping overlooks the significant heterogeneity within groups, at the expense of talented candidates from disadvantaged groups (and to the benefit of less talented candidates from advantaged groups). 

\section{Related Literature}\label{sec.lit}

Our work contributes to a growing literature on belief-based discrimination, under which potentially biased beliefs drive discriminatory behavior (\cite{Bordalo_etal_2016}; \cite{Bohren_etal_2019}; \cite{bohren2019inaccurate}; \cite{Bordalo_etal_2019}; \cite{Coffman_etal_2021}). Much of this work considers fixed information sets: the researcher asks, holding fixed a given set of information about a candidate, are male and female candidates evaluated differently?\footnote{In their recent working paper, \citet{bohren2022systemic} refer to this as ``direct'' discrimination, distinguishing it from systemic forms of discrimination that may give rise to differences in candidate qualifications prior to the evaluation stage.} We build on this important body of work by expanding the decision problem to consider endogenous acquisition of information, asking how evaluators choose whether to get more information and how evaluators use that information to update their beliefs. 

Our study of endogenous information acquisition is related to the idea of attention discrimination introduced by \citet{bartovs2016attention}. They build a model of rational attention allocation, showing that both in theory and in their experiments evaluators spend more effort scrutinizing minorities in a lemon-dropping market, and less effort considering minorities in a cherry-picking market, leading to discrimination. Building on this idea, \citet{10.1093/ej/uead079} explore the idea of selective monitoring, where an employer is more likely to monitor and subsequently punish Black employees relative to white. A related literature explores how lack of information may perpetuate discrimination; for instance, past discrimination against a stigmatized group may lead to under-sampling, depriving decision-makers of the information needed to de-bias their beliefs.\footnote{See the theoretical approaches of \citet{komiyama2020statistical} and \citet{lepage2021endogenous} as well as experimental evidence from \citet{gupta2023essays} and \citet{mccrea2023shortlist}.} Even closer to our work, \citet{eyting2023we} uses a series of online experiments to study how motivated reasoning shapes information acquisition and discriminatory behavior. He finds that individuals are more likely to seek out information that matches their motivation, using this to justify discriminatory behavior. 

Relative to this past work, our experiment highlights that under-acquisition of information can amplify discrimination, even when groups receive the same amount of attention. In our setting, biases arise not from over or under-scrutinizing one group but rather through an over-reliance on group-level information at the expense of easily accessible, highly informative individual-level information for all candidates. In addition, we introduce an important comparison across endogenous and exogenous information treatments, allowing us to show that the patterns we observe are hard to square with rational models of information acquisition. Finally, our design and analysis are informed by cognitive, rather than motivated, explanations for behavior; indeed, our results suggest that cognitive biases alone are enough to generate discrimination through biased acquisition and biased processing of information.

Our work is also connected to a wealth of recent literature on biases in belief updating, particularly as they relate to gender and discrimination. One strand of this work documents that stereotypes shape how men and women update their beliefs about themselves in response to noisy signals (\cite{mobius2022managing}, \cite{coffman2023sbu}, \cite{coffman2023zafar}).\footnote{Our design allows us to study a similar concept in the context of evaluating others, identifying the impact of gender stereotyping on belief updating.} The second relevant strand of this work looks at biased processing of information about others.\footnote{A flood of recent work has begun to unpack the various mechanisms that may contribute to these belief biases. For instance, \citet{RUZZIER2023379} explore confirmation bias, a la \citet{rabin1999first}, \citet{campos2023non} consider conservatism, \citet{schwartzstein2021using} consider endogenously chosen models for organizing data, and \citet{bordalo2023memory} and \citet{bordalo2023people} consider selective attention and memory. Our results will be most consistent with stereotyping as a driver (\citet{Bordalo_etal_2016}).} In a stylized experiment on racial bias and hiring, \citet{rackstraw2022bias} shows that decision-makers give more weight to stereotype-consistent signals when updating their beliefs. Using evidence from both abstract and gender-based experiments, \citet{campos2023non} show that employer conservatism (under-reaction to information) leads to more discrimination against members of disadvantaged groups relative to a Bayesian benchmark. Moving to the field, \citet{egan2022harry} show asymmetry in how misconduct is punished among financial advisors, with women facing greater punishment than men, inconsistent with a Bayesian response. Similarly, \citet{sarsons2017interpreting} finds that referrals to female surgeons, relative to male surgeons, fall more sharply after a negative patient outcome.  

We build on this excellent recent work by studying endogenous information acquisition and biased information processing in a unified framework, with treatments that allow us to tease apart potential mechanisms. Our findings are both remarkably consistent with and shed more light on many of these past findings. We find support for both biased acquisition and processing of information, driven not by bias against women or motivated reasoning, but rather by a contrast-based form of stereotyping that exaggerates differences across groups.

\section{Experimental Design}

We study evaluator decision-making in a controlled environment that allows for causal identification of key factors that predict behavior. To do so, we construct a simulated labor market using candidates and evaluators recruited from Prolific. Below, we describe the design. The full experimental instructions are available in the Appendix; the experiment received IRB approval from Harvard Business School and was pre-registered (AEARCTR-0010969).

\subsection{Candidate Sourcing}
We field candidates by running a preliminary study with 200 participants.\footnote{The study was advertised as 10 minutes, with \$2 in guaranteed pay and the possibility of incentive pay. It was conducted on Prolific in June 2022.} Our goal was to collect performance data and individual characteristics that could be used to build candidate profiles. Participants begin the study by taking a 10-question math and science skills test. Each multiple-choice question appears on a separate page of the study, in a random order. They have up to 20 seconds to answer each question; the time limit is designed to discourage looking up answers on the internet. We incentivize participants by offering 20 cents in additional compensation for each question answered correctly.

Following completion of the test, participants answer 25 simple questions about their preferences. Each question has only two possible answers. For example, participants are asked whether they prefer beach vacations or mountain vacations, whether they prefer spring or fall, etc. The goal is to have information about each potential candidate that can serve as profile ``filler,'' additional information about the person that increases evaluator engagement and differentiates individuals beyond gender. We intended for these questions to be unrelated (both in perception and in reality) to performance on the test and to gender, simplifying our analysis of evaluator decisions. 

After collecting the candidate data, we used the distribution of performance on the test to classify each participant as having demonstrated Above-the-Bar performance (top-half of performers) or Below-the-Bar performance (bottom-half of performers). Using a binary measure of candidate performance simplifies the design and empirical analysis. We use the Above-the-Bar terminology because we felt it would be intuitive in a hiring context and that it would be perceived (accurately) as binary.

\subsection{Constructing Candidate Pools}

We use the data from the candidate survey to construct pools of candidates for evaluators to assess. We build four different pools. Each pool has exactly 50 members: 25 women and 25 men. Across pools, we vary the frequency of men and women within the pool who are Above-the-Bar. Table~\ref{t:pools} displays the design. In the $60-60$ pool, both 60\% of men and 60\% of women within the pool are Above-the-Bar. In the $60-40$ pool, 60\% of men but only 40\% of women are Above-the-Bar, while in the $40-60$ pool, 40\% of men and 60\% of women are Above-the-Bar. Finally, in the $40-40$ pool, only 40\% of both men and women are Above-the-Bar.  

\begin{table}
\hfill
\begin{center}
     \caption{Candidate Pools}
      \label{t:pools} \medskip

\begin{tabular}{ |p{2cm}||p{2cm}|p{2cm}|p{2cm}|  }
 \hline
 \multicolumn{4}{|c|}{Probability that Candidate from Group is Above-the-Bar} \\
 \hline
& &  \multicolumn{2}{c|}{Female Candidate}\\
 \hline
& &40\% &60\%\\
 \hline
 \multirow{2}{4em}{Male Candidate} & 40\%  & 40-40 & 40-60 \\ 
& 60\% & 60-40 & 60-60 \\ 
 \hline
\end{tabular}
\end{center}
  \hfill
	\medskip %
\begin{minipage}{0.99\textwidth} %
{\footnotesize Note: This table presents the construction of four different pools with the corresponding share of candidates who are Above-the-Bar for female and male candidates.\par}
\end{minipage}
\end{table}

We build a candidate profile for each candidate within the pool. This profile consists of their gender and their answers to exactly four preference questions from the candidate study: whether they prefer a beach vacation or a mountain vacation, whether they prefer lions or tigers, whether they prefer light or dark colors, and whether they prefer spring or fall.

\subsection{Evaluator Study}

In our main evaluation study, evaluators are tasked with assessing whether candidates were Above-the-Bar based upon their profile and additional signals of performance. We provide evaluators with information about the candidate study. We explain that the candidates they will evaluate are real Prolific participants who completed a past study for us. We explain that these participants each completed a test that contained math and science questions, and that, based upon their score on this test, we classified each candidate as either Above-the-Bar or Below-the-Bar. We explain that, as evaluators, their job is to use information about a candidate to guess whether that candidate was Above-the-Bar.

We randomly assign evaluators to one of the four pools. In addition, across subject, we vary whether the evaluator can choose whether to receive additional signals of performance (endogenous) or if instead the signals are provided with certainty (exogenous). Evaluators assess five candidates from their assigned pool, drawn at random with replacement. 

\subsubsection{Pool Information}

We provide each evaluator with specific information about their  candidate pool. We inform them of the overall share of candidates who were Above-the-Bar within their pool (60\% for the $60-60$ pool, 50\% for the $60-40$ and $40-60$ pools, and 40\% for the $40-40$ pool). We tell them that there are exactly 25 men and 25 women in the candidate pool. And, most critically, we provide them with the share of men and women within their pool that were Above-the-Bar. We communicate this information with a pie chart. They see two pie charts side-by-side, one for men and one for women, with green wedges corresponding to the share of men (women) who are Above-the-Bar. Following the pie charts, we summarize the information in a table. Throughout their evaluations, evaluators can see this table of information about the candidate pool by reading the Instructions Summary at the bottom of each page. Figures \appref{fig:piechart} and \appref{fig:summarytable} in the Appendix present examples.\footnote{We also provide them with a breakdown according to each of the other profile characteristics. For example, we inform them what share of candidates who preferred mountain vacations were Above-the-Bar versus what share of candidates who preferred beach vacations were Above-the-Bar. We do the same for spring versus fall, lions versus tigers, and light colors versus dark colors. Note that by design, within each pool, the share of Above-the-Bar candidates of each of these preferences types is very similar and close to the pool average. For example, within the $40-40$ pool, we aimed to have approximately 40\% of both beach vacation types and mountain vacation types be Above-the-Bar.}

The goal is to provide clear, accurate, accessible information about the relevant candidate pool. In doing so, we aim to dictate evaluator beliefs of gender differences in performance.\footnote{We do not observe participants' ex-ante beliefs about gender differences in this task. Past work suggests that for a math and science test of this type, which carries a more masculine stereotype, participants are likely to believe that men outperform women on average, see \cite{Bordalo_etal_2019} and \cite{Coffman_etal_2021}. While we do not observe ex-ante beliefs of gender differences, our approach (if we take our estimates at face-value) allows us to extrapolate and predict behavior of an evaluator with any given prior belief of the gender difference in the likelihood of being Above-the-Bar.} The pools allow us to induce exogenous variation in these beliefs. 

For ease of exposition, we use the term \textbf{candidate prior}, or sometimes just \textbf{prior}, to refer to the share of candidates of a given gender who are Above-the-Bar within a pool. Because there are just two possible candidate priors (40 or 60), we write \textbf{low-prior} to refer to a candidate prior of 40, and \textbf{high-prior} to refer to a candidate prior of 60.\footnote{Note that these semantic choices abstract away from a candidate's membership in groups defined by characteristics other than their gender (such as beach vacation lovers, or dark color fans), though those characteristics will be accounted for in our empirical approach. Throughout, again for ease of exposition, the reader can interpret \textbf{group} as defined by gender.} 

\subsubsection{Initial Candidate Assessment}

For each evaluation, the computer chooses one candidate at random from the pool of 50 candidates and shows the candidate profile to the evaluator. Figure \appref{fig:profile} provides a screenshot of an example candidate profile. The evaluator then guesses the likelihood that the candidate is Above-the-Bar. To indicate their answer, they use a slider that ranges from 0 percent likely to 100 percent likely; we use color-coding (moving from red to green) and intuitive labels on top of precise likelihoods (moving from not at all likely to extremely likely) to improve comprehension. 

\subsubsection{Information Acquisition}

We explain to evaluators that, in order to help them make a better guess, they can pay to acquire (\textbf{endogenous} treatment) or will be given (\textbf{exogenous} treatment) additional information about the candidate. Additional information about candidates takes the form of noisy signals of performance. The signals are simple and binary, either \textbf{positive} or \textbf{negative}. These signals are randomly generated by the computer, with the probability of a positive signal dependent upon whether the candidate is Above-the-Bar. For Above-the-Bar candidates, the probability of a positive signal is 0.75; for Below-the-Bar candidates, the probability of a positive signal is 0.25. We explain these probabilities to evaluators using both percentages and expected frequencies. We use color-coding (green and red) and icons (plus sign or minus sign) to increase comprehension.  

In the endogenous treatment, after completing the initial likelihood slider for a candidate, evaluators must decide whether to ``Make Their Evaluation'' or ``Receive More Information.'' If they choose to receive more information, they pay a small cost (5 cents, explained in more detail below) and receive one signal. Then, they guess again the likelihood that the candidate is Above-the-Bar using another slider. This process repeats, with evaluators being asked again whether they want to ``Make Their Evaluation'' or ``Receive More Information.'' Evaluators can choose to receive up to five signals about each candidate.

The exogenous treatment operates identically, except instead of choosing whether to receive each signal, the evaluator must view all 5 signals for each candidate. Just as in the endogenous treatment, signals are shown sequentially and the evaluator must complete a new likelihood slider after each signal.

\subsubsection{Final Evaluations}

When an evaluator in the endogenous treatment chooses to forgo additional information and make their evaluation, they are asked to make a final binary classification of the candidate. They indicate whether they believe the candidate is Above-the-Bar, or not. Evaluators in the exogenous treatment make this evaluation after 5 signals for the candidate. Following their final evaluation, evaluators receive no feedback before moving to the next candidate.

In addition, to discourage evaluators in the endogenous treatment from skipping signals to save time in the study, we introduced a time penalty for each non-acquired signal.\footnote{For example, an evaluator who chose to view two signals for a candidate would be asked to complete three~($5-2$) time penalty tasks following their final evaluation of that candidate. We mention these filler tasks in the instructions.} The time penalty was designed to be similar to viewing and responding to a signal, both in terms of format and in terms of length of time. For each non-acquired signal for a candidate, the evaluator was shown a random number between 0 and 100, had to indicate whether that number was above or below 50, and then had to drag a slider to the chosen number.\footnote{Our data suggests that this filler task eliminates, if not reverses, the time-cost of acquiring signals: participants randomly-assigned to our exogenous treatment, who must view and respond to all 5 signals, spend significantly \textit{less} time in the study on average compared to those in the exogenous treatment who can choose to view fewer signals.}  
 
\subsubsection{Payments}

Evaluator decisions have no impact on the payment of participants from the candidate study. All evaluators receive \$3.50 in guaranteed pay for completing the 20-minute study. In addition, we pay 1 out of every 10 participants a bonus based on their evaluation of the candidates. At the conclusion of the study, the computer randomly chooses 1 of the 5 candidates that the participant evaluated to determine the bonus payment. If the evaluator correctly classified the chosen candidate, they earn \$7 in additional bonus payment. If they incorrectly classified that candidate, they earn 50 cents in additional bonus payment. In the endogenous treatment, for each signal acquired for the chosen candidate, we deduct 5 cents from the bonus payment. 

In addition, evaluators chosen to earn incentive pay also earn money depending on their answer to one randomly-chosen likelihood slider question. One of their slider answers is chosen as the ``slider-that-counts.'' If their slider-that-counts is within 10 percentage points of the true probability, they receive an additional bonus payment of \$1.\footnote{We calculate the ``true probability'' by first calculating the prior as the fraction of candidates in the pool whose observed characteristics match the candidate under evaluation. We then apply Bayes' rule to calculate the posterior based on the realized signals observed by the participant at that point in time.}

\section{Hypotheses and Empirical Approach} \label{Hyp}

We study how evaluators use candidate gender in their assessments and how endogenizing the information acquisition process amplifies (or mitigates) any bias we observe. In this section, we lay out an empirical approach for analyzing these research questions.

Previous literature documents discrimination against women in many field contexts of interest (see \cite{bertrand2017field} for an overview). However, in many cases, it is difficult to determine what drives this discrimination. One particular challenge is that, in part due to past discrimination, many of these environments feature significant gender gaps; as a result, we disproportionately observe evaluators making decisions in contexts where they believe (either accurately or inaccurately) that men have a greater likelihood of being qualified on average. This can make it hard to identify whether discrimination against women is driven by their status as a disadvantaged group---a group with a \textit{relatively} lower likelihood of being qualified. 

Our controlled experiment can tease out this channel. Our approach separates bias based purely upon a candidate's gender from a more general form of bias that stems from (potentially implicit) comparisons across groups. In our setting, bias based purely upon a candidate's gender would take the form of male candidates receiving more favorable assessments than female candidates, holding fixed other candidate-level characteristics. If, on the other hand, it is less about gender per se and more about relative comparisons, we would expect that candidate assessments depend not on gender but rather on how the candidate's group compares to the other group. 

We refer to this key measure as a candidate's \textbf{relative advantage}, defined as the difference in priors across groups within the pool, signed so that positive values favor the candidate's own group. Thus, a male candidate has a relative advantage of $20$ within the $60-40$ pool, of $-20$ within the $40-60$ pool, and of $0$ within the $40-40$ and $60-60$ pools; a female candidate's relative advantage is $-20$, $20$, and $0$ within those same pools, respectively. 

We rely on a workhorse specification that predicts an evaluator's assessment of a candidate from the candidate's gender and relative advantage. We do so controlling for the candidate prior ($40$ or $60$) and an indicator for whether the evaluator was assigned to the exogenous treatment. We cluster standard errors at the evaluator level. Formally, we predict an evaluation of a candidate $i$ by evaluator $j$:
\begin{equation}
	Evaluation_{i,j}= \beta_0 + \beta_1 \textit{Male}_{i} + \beta_2 \textit{RelAdv}_{i} + \beta_3 \textit{Prior}_{i} + \beta_4 \textit{Exog}_{j} + \textbf{X}_i + \epsilon_{i,j}
\end{equation}
\noindent where $\textit{Male}_{i}$ is an indicator for candidate $i$ being male, $\textit{RelAdv}_{i} \in \{20,0,-20\}$ is the candidate's relative advantage within the pool, $\textit{Prior}_{i} \in \{40,60\}$ is the share of candidates of $i$'s gender who are Above-the-Bar, and $\textit{Exog}_{j}$ is an indicator for the evaluator's treatment assignment. A positive $\beta_1$ represents gender bias, while a positive $\beta_2$ represents a bias that stems from contrast between groups. We can add interactions of the exogenous treatment indicator with $\textit{Male}_{i}$ and $\textit{RelAdv}_{i}$ to study how the magnitudes of these biases depend upon whether information acquisition is endogenous. To the extent that we observe gender bias and/or relative advantage bias, we can also ask how these two biases interact: do men benefit more from being relatively advantaged (or suffer less from being relatively disadvantaged) compared to women?  

Recall that evaluators see four characteristics of each candidate in addition to gender: preference between fall/spring, preference between beach/mountain vacation, preference between light/dark colors, and preference between lions/tigers. Our control vector, $\textbf{X}_i$, contains candidate $i$'s relative advantage on each of these four dimensions.\footnote{For instance, suppose within a pool that $42$\% of candidates who liked lions but only $38$\% of candidates who liked tigers were Above-the-Bar, and candidate $i$ liked lions. Then, we would define this candidate's relative advantage on the lion/tiger dimension as $4$. A candidate within this same pool who instead preferred tigers would have a relative advantage of $-4$ on this dimension. Relative advantage on the other dimensions is determined analogously. Each relative advantage enters the specification independently, so that each candidate has a relative advantage defined in terms of gender, plus four additional relative advantage terms corresponding to the four preference characteristics. Note that because we designed the pools to minimize differences in performance across these other dimensions, we have only minimal variation in relative advantage on these dimensions. This makes it difficult for us to benchmark the impact of relative advantage as defined by gender against relative advantage defined by these other dimensions.} We control for these other relative advantages to rule out that differences among men and women along these other dimensions could drive our results.    

An $Evaluation_{i,j}$ can correspond to different types of assessments. First, we observe the initial likelihood estimate the evaluator provides for the candidate, after seeing the candidate's profile but before receiving any signals. We refer to this as the evaluator's \textbf{initial assessment}. We also observe the evaluator's updated likelihood assessment after each additional signal they view. Finally, we observe the evaluator's final (binary) classification of the candidate as either Above- or Below-the-Bar. For evaluators in the exogenous treatment, this final classification is made after all five signals are viewed. In the endogenous treatment, evaluators make their final classification whenever they choose to stop receiving signals for the candidate and move to this stage. We refer to this binary classification as the evaluator's \textbf{final evaluation}. 

We then study information acquisition choices for evaluators in the endogenous treatment. We explore when evaluators choose to acquire another signal, and how these decisions depend upon a candidate's gender and/or relative advantage. To explore this, we predict the decision to acquire signal $t$, where $t \in \{1,2,3,4,5\}$. We estimate the following specification:
\begin{equation}
	Acquire_{i,j,t} = \beta_0 + \beta_1 \textit{Male}_{i} + \beta_2 \textit{RelAdv}_{i} + \beta_3 \textit{Prior}_{i} + \beta_4 \textit{Likelihood}_{i,j,0} + \textbf{X}_i + \epsilon_{i,j,t}
\end{equation}
\noindent where $Acquire_{i,j,t}$ is an indicator that takes 1 if evaluator $j$ choses to receive signal $t$ for candidate $i$, conditional on having viewed signal $t-1$. We control for an evaluator's initial assessment of a candidate, $\textit{Likelihood}_{i,j,0}$, accounting for any impact of relative advantage or candidate gender that influences initial beliefs. We can run this specification separately by signal history. We will define \textbf{confirming signals}: confirming signals are consistent with the candidate prior (negative signals for low-prior candidates and positive signals for high-prior candidates). With this framework, we can explore whether candidate gender or relative advantage plays a larger role in decisions after confirming signals, when the accumulated evidence about the candidate tells a consistent story, or whether instead these biases are larger when the evidence is more mixed.

Finally, we study biases in belief updating. We restrict attention to only evaluators in the exogenous treatment. This cleanly isolates responses to information, shutting down any selection into receiving more information. For these exogenous treatment evaluators, we predict the updated likelihood assessment of evaluator $j$ for candidate $i$ after observing $t$ signals, where $t \in \{0,1,2,3,4,5\}$. To focus on the role of how beliefs are updated, we control for an evaluator's initial assessment of a candidate, $\textit{Likelihood}_{i,j,0}$. Formally,
\begin{equation}
	Likelihood_{i,j,t} = \beta_0 + \beta_1 \textit{Male}_{i} + \beta_2 \textit{RelAdv}_{i} + \beta_3 \textit{Prior}_{i} + \beta_4 \textit{Likelihood}_{i,j,0} + \textbf{X}_i + \epsilon_{i,j,t}
\end{equation}
This allows us to ask whether, conditional on initial assessment, evaluators assess male candidates or candidates with a greater relative advantage more favorably after observing new information. Again, we can run this specification separately by candidate prior and signal history. This allows us to explore whether candidate gender or relative advantage plays a smaller role in belief updating after confirming signals, when there is less uncertainty about candidate quality.   

We close by asking how much evidence evaluators require before making a final evaluation, and whether these requirements seem to be a function of a candidate's gender and/or relative advantage. To explore this, we predict the total number of signals acquired by an evaluator before they decide to make their final evaluation. To allow for simple comparisons, we start by considering low and high-prior candidates separately. 

We estimate the following specification, restricting attention to only low (or high) prior candidates:
\begin{equation}
	TotalSignals_{i,j} = \beta_0 + \beta_1 \textit{Male}_{i} + \beta_2 \textit{RelAdv}_{i} + \textbf{X}_i + \epsilon_{i,j}
\end{equation}
\noindent where $TotalSignals_{i,j}$ is the total number of signals acquired by evaluator $j$ before making a final evaluation for candidate $i$. We make the important distinction of splitting the sample based upon whether candidate $i$ was classified as Above-the-Bar or Below-the-Bar by evaluator $j$. This allows us to ask how quickly evaluators classify different candidates as Above or Below-the-Bar.

Consider the hypotheses for low-prior candidates. First, we focus on candidates who were classified as Below-the-Bar. In this model, $\beta_1 > 0$ would indicate that evaluators required more signals before classifying low-prior male candidates as Below-the-Bar relative to low-prior female candidates. Similarly, $\beta_2 > 0$ would indicate that evaluators required more signals before classifying low-prior candidates from the $40-40$ pool as Below-the-Bar relative to low-prior candidates from the $40-60$ pool. Put differently, we can ask whether low-prior female candidates and/or relatively disadvantaged candidates are quicker to be classified as Below-the-Bar; do evaluators require less evidence before dismissing these candidates as Below-the-Bar? Analogously, by estimating this same specification for the set of low-prior candidates who were classified as Above-the-Bar, we can ask whether evaluators are \textit{slower} to classify low-prior female candidates and/or relatively disadvantaged candidates as Above-the-Bar; do evaluators demand more evidence before being willing to classify these candidates as Above-the-Bar? Note that restricting to candidates from low-prior groups rules out that any differences are driven by different signal realizations. All candidates in these specifications have the same true likelihood of being Above-the-Bar and therefore produce the same signals in expectation. 

A similar set of hypotheses can be tested for high-prior candidates. First, we can focus on high-prior candidates who were classified as Above-the-Bar, asking whether evaluators require fewer signals before classifying male candidates (relative to female candidates) or advantaged candidates (relative to candidates in the $60-60$ pool) as Above-the-Bar. And, on the flip side, we can restrict attention to high-prior candidates who were classified as Below-the-Bar, hypothesizing that evaluators may be slower to classify male candidates and advantaged candidates as Below-the-Bar.

\section{Data Overview and Quality}

We ran the evaluator study in March 2023 with 3,200 Prolific participants, 400 in each treatment, for a total of 16,000 unique candidate evaluation observations. The median age of our evaluators is 37, approximately 50\% of our evaluators are female, and just under 75\% of our sample is white.  

Prolific has been found to have higher data quality compared to other popular online data collection sources \citep{douglas2023data}. \citet{Rigotti_etal} find that the proportion of participants making dominated decisions on Prolific is no higher than in a high quality physical laboratory sample, at roughly 11--12\%. To further ensure attentiveness and understanding, we include seven understanding questions during the instructions that participants must answer correctly before they can proceed. These questions include asking participants what it means for a candidate to be Above-the-Bar and how their payments are determined. We also offer interested participants the option to watch a closed-captioned demo video explaining how the study works; approximately 1/4 of participants choose to view this video. We include an instructions summary on each page of the study. The median duration of study participation is just under 18 minutes, with more than 90\% of evaluators taking at least 10 minutes to complete the study. In the Online Appendix, we show that our results replicate with a restricted sample of evaluators that drops participants in the top and bottom tails of study duration, see Tables \appref{t:first_final_slider_and_eval_rob_dur},   \appref{t:info_acquisition_dur_rob}, \appref{t:whentoacquire_dur_rob},  \appref{t:sliders_two_signals_by_prior_exog_dur_rob}, and \appref{t:sliders_beyond_two_signals_dur_rob}. 

Our data is suggestive of relatively high levels of comprehension and attention. Evaluators classify approximately 80\% of candidates correctly in their final evaluations, well-above the random benchmark. We observe more than 58,000 cases of evaluators updating their beliefs in response to a binary signal. In 92\% of cases, evaluators update their beliefs in the correct direction after observing a signal. More than half of our evaluators update their beliefs in the correct direction every time, and 90\% of evaluators update their beliefs in the correct direction at least 75\% of the time. They are also largely internally consistent. Define an internally consistent final evaluation as either (i) classifying the candidate as Above-the-Bar and having a final slider likelihood greater than or equal to 50\%, or (ii) classifying the candidate as Below-the-Bar and having a final slider likelihood less than or equal to 50\%. By this definition, 97\% of final evaluations are internally consistent.

\section{Results}

We start by providing an overview of our evidence on assessments. On average, evaluators believe that a candidate has approximately a 54\% chance of being Above-the-Bar initially; they are, as expected, more optimistic about candidates from high prior groups (believing they have a 60\% chance of being Above-the-Bar initially) than candidates from low prior groups (who they estimate to have a 47\% chance of being Above-the-Bar initially). The modal belief for each candidate-type is the group prior, with approximately 17\% of evaluators providing the group prior as their initial assessment for the candidate; however, there is significant dispersion. This dispersion could reflect noise, but may also indicate inference from other, non-gender characteristics, homophily, or discrimination based on gender or relative advantage. We unpack these forces below.

\begin{figure}[!ht] %
  \hfill
    \begin{center}  
     \caption{Initial Beliefs and Final Evaluations by Gender}
      \label{fig:initial_final_by_gender}
      \medskip
      
       \centering	
       Panel A: Initial Assessments\\
       \includegraphics[scale=0.3]{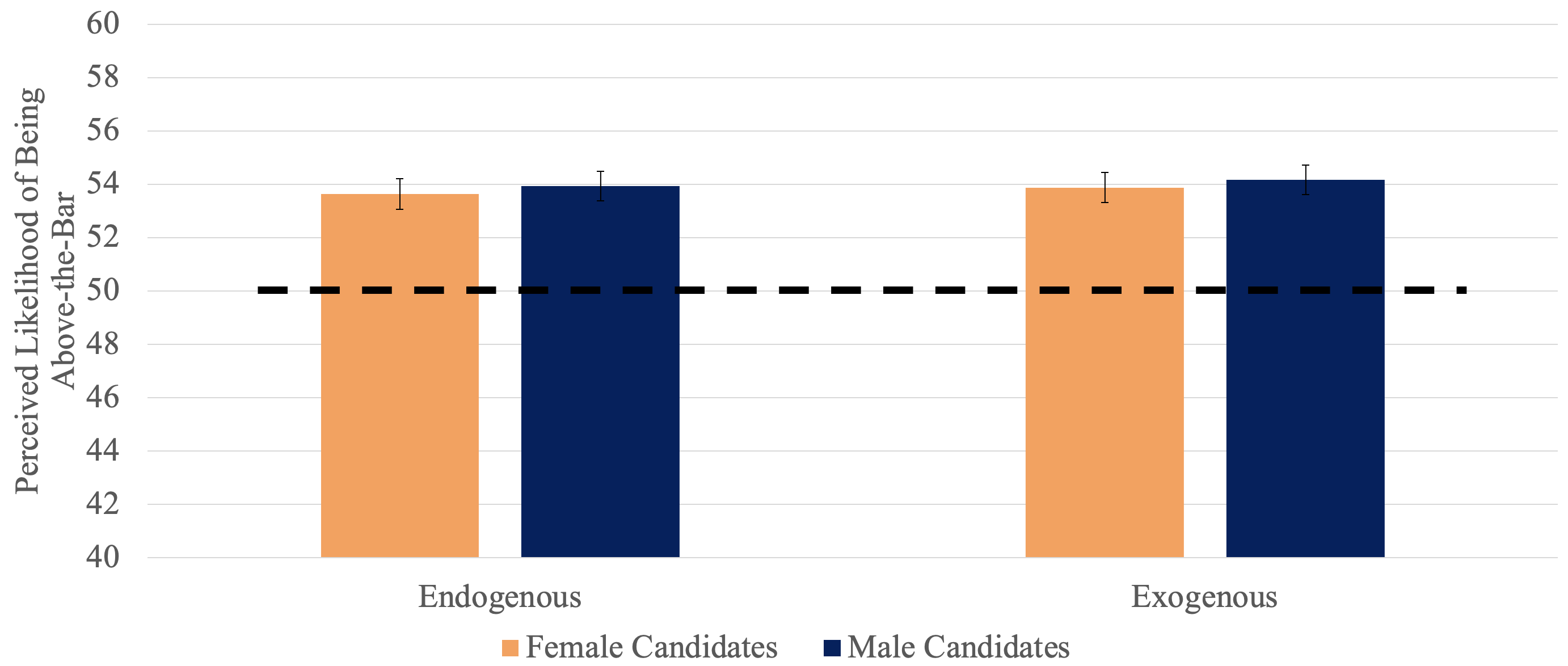} 
       	\medskip
       Panel B: Final Evaluations\\
       \includegraphics[scale=0.3]{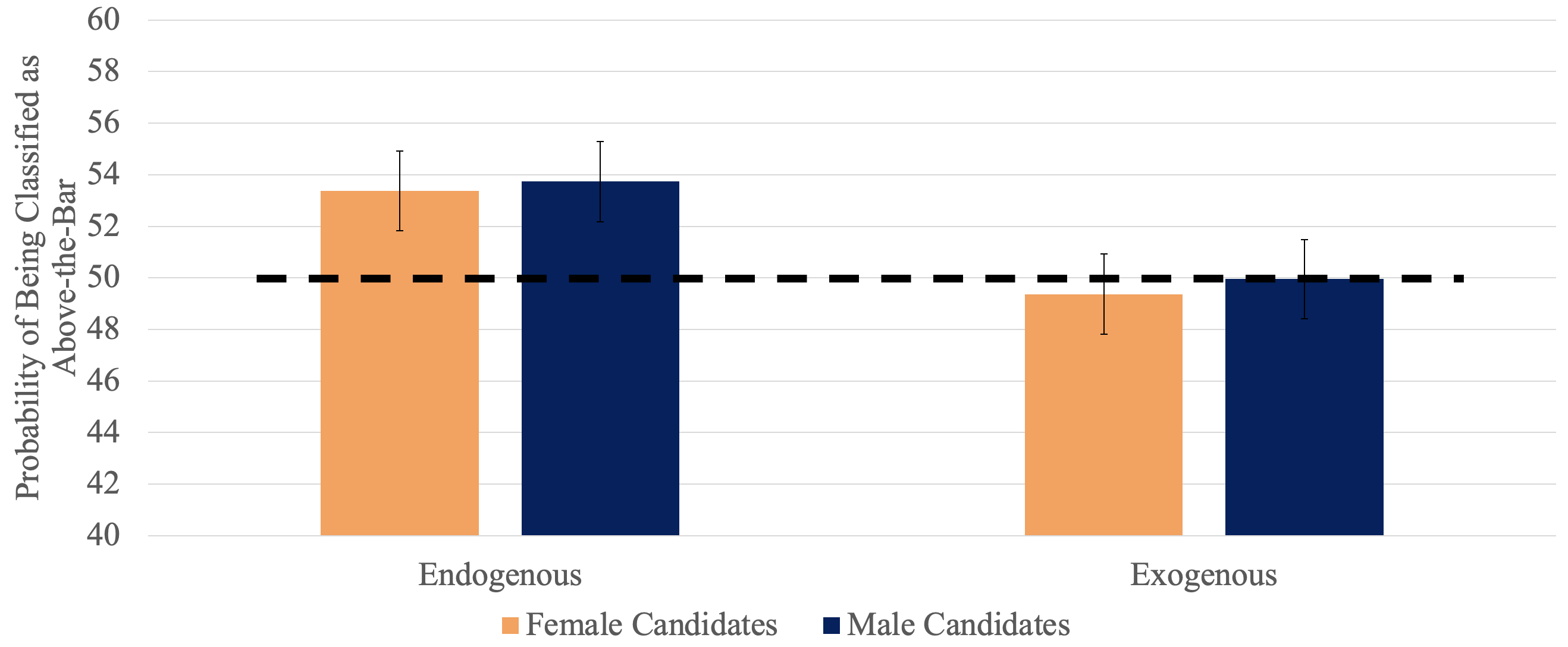}
    \end{center}
  \hfill
	\medskip %
\begin{minipage}{0.99\textwidth} %
{\footnotesize Note: This figure presents the average assessment of male and female candidates, for both treatment groups.  Panel~A shows initial assessments, while Panel~B shows final assessments. The dashed lines show the true percentage of candidates in that group in the pool that are Above-the-Bar. A 95\% confidence interval is shown at the top of each bar.}
\end{minipage}
\end{figure}

Figure \ref{fig:initial_final_by_gender} presents the average assessment of male and female candidates. Panel A presents initial assessments, while Panel B shows the share of male and female candidates who are classified as Above-the-Bar in final evaluations. We find no evidence that evaluators discriminate on the basis of gender. Evaluators provide very similar assessments of male and female candidates---initially and after acquiring information. This is true independent of whether the information was acquired endogenously or exogenously.

Initially, evaluators believe that both male and female candidates have approximately a 54\% chance of being Above-the-Bar on average. In the endogenous treatment, 54\% of male candidates and 53\% of female candidates are ultimately classified as Above-the-Bar in final evaluations, tracking initial assessments; this gender difference is not significant. We see that evaluators in the exogenous treatment make significantly less optimistic final evaluations; more information seems to bring evaluations closer to the truth, on average. There continues to be no significant gender difference, with approximately 50\% of both male and female candidates classified as Above-the-Bar. 

\begin{figure}[h] %
  \hfill
    \begin{center}  
     \caption{Initial Beliefs and Final Evaluations by Relative Advantage}
     \medskip
     
      \label{fig:initial_final_by_reladv}
       \centering	
       Panel A: Initial Assessments\\
       \includegraphics[scale=0.22]{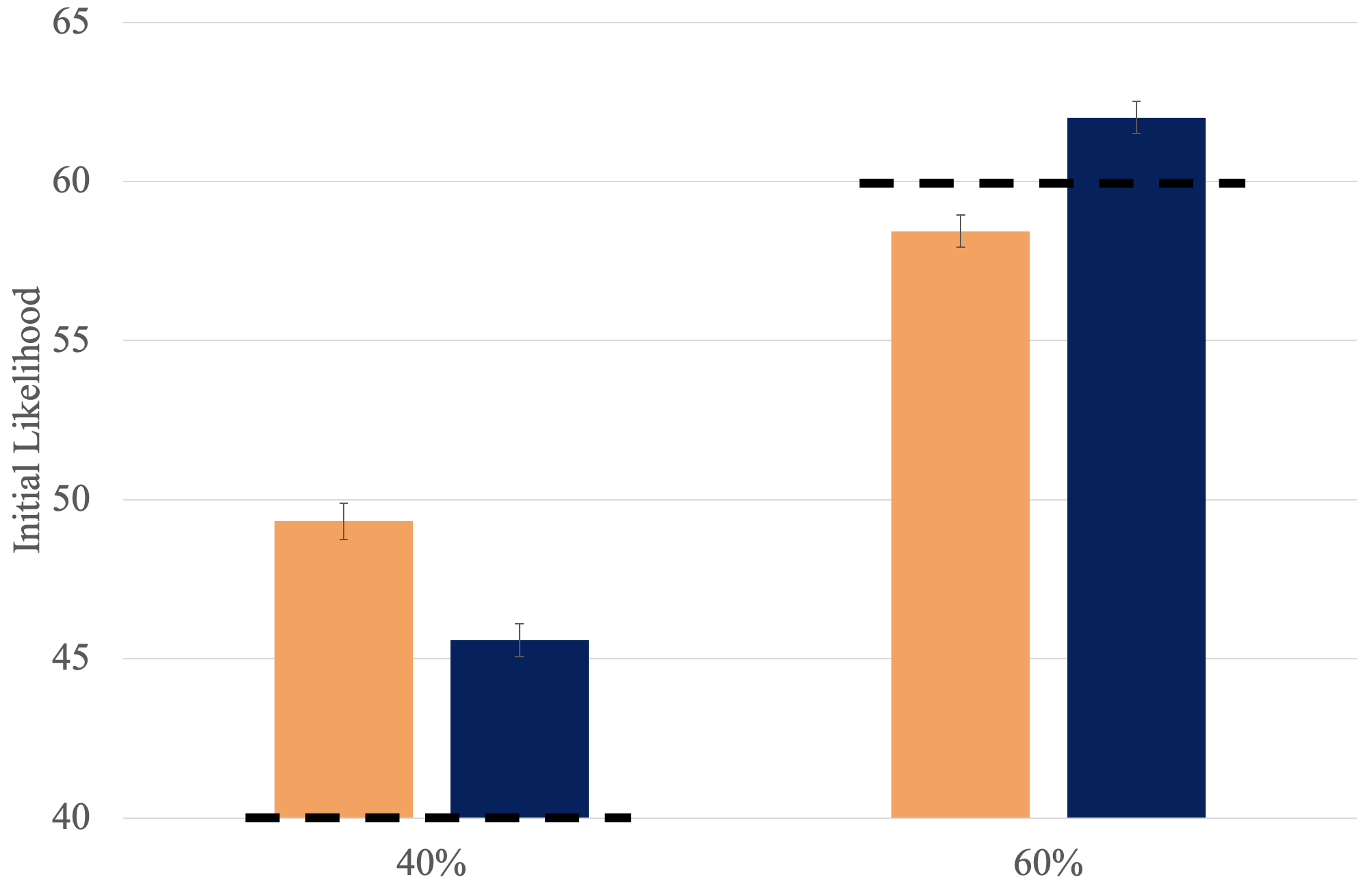}\\ %
       	\medskip
       Panel B: Final Evaluations by Treatment\\
       \includegraphics[scale=0.22]{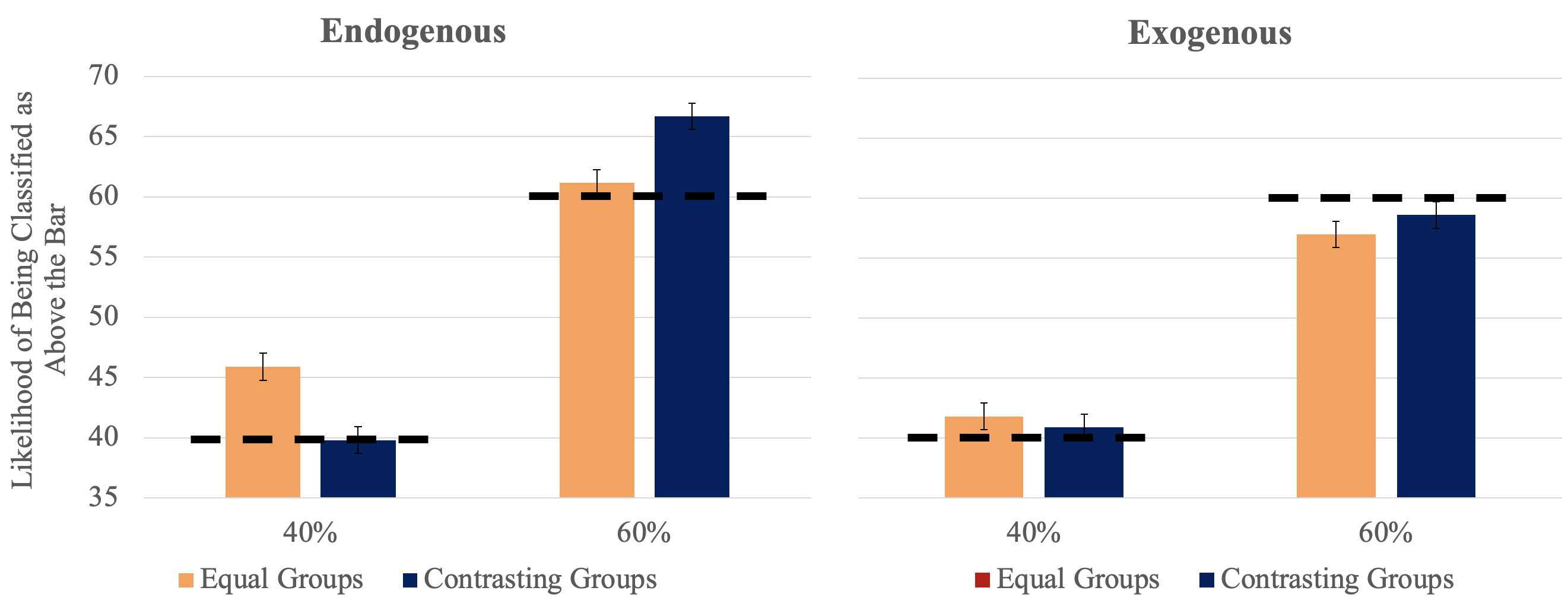}\\ %
    \end{center}
  \hfill
	\medskip %
\begin{minipage}{0.99\textwidth} %
{\footnotesize Note: This figure presents the average assessments broken down by prior likelihood and equal/contrasting grouping, for both treatment groups. Panel A shows initial assessments, while Panel B shows final evaluations. The dashed lines show the true percentage of candidates in that group in the pool that are Above-the-Bar. A 95\% confidence interval is shown at the top of each bar.}
\end{minipage}
\end{figure}

Figure \ref{fig:initial_final_by_reladv} presents the same data, but now comparing assessments on the basis of group differences. Panel A presents initial assessments. We show initial assessments for low-prior groups, comparing candidates from the $40-40$ pool, in which members of both groups have an equal chance of being Above-the-Bar, to low-prior candidates in the $60-40$ or $40-60$ pools, who are at a performance disadvantage relative to the other group. We observe that low-prior candidates are assessed significantly less favorably when their group is relatively disadvantaged. Candidates from high-prior groups are assessed significantly more favorably if their group holds a relative performance advantage compared to when both groups are equal. 

Panel B examines these same comparisons, now focusing on final evaluations by treatment. Within the endogenous treatment, relative performance advantage continues to play a significant role in final evaluations. Nearly 46\% of low-prior candidates are classified as Above-the-Bar when the groups are equal, compared to only 40\% of low-prior candidates from disadvantaged groups. Relative advantage benefits high-prior candidates in the endogenous treatment: 67\% of high-prior candidates are viewed as Above-the-Bar when their group has the performance advantage. Only 61\% are viewed as Above-the-Bar when the groups are equal. 

Final evaluations depend on relative advantage in the endogenous treatment. But, as Figure \ref{fig:initial_final_by_reladv} reveals, this does not hold true in the exogenous treatment. When evaluators must acquire all five signals about each candidate, final evaluations are more accurate on average and there is less of a role for group comparisons in shaping perceptions of candidates. 

Table \ref{t:first_final_slider_and_eval} formalizes these patterns. We predict an evaluator's assessment of a candidate---either initial likelihood estimate, final likelihood estimate, or final binary classification---from our two key variables: whether the candidate is male and the candidate's relative advantage. We control for candidate prior ($40$ or $60$) as well as the candidate's relative advantage in terms of the other profile characteristics. We include round fixed effects and cluster standard errors at the evaluator level. In Columns~\num{1}, \num{3}, and \num{5}, we simply include an indicator for the exogenous treatment. In Columns~\num{2}, \num{4}, and \num{6}, we interact the exogenous treatment with male and relative advantage, asking if their impact is smaller when information is required. 

	\begin{table}[h]
		\begin{center}
			\caption{\label{t:first_final_slider_and_eval} The Impact of Relative Advantage on Assessments}\medskip
                \medskip
                OLS Predicting Initial Beliefs, Final Beliefs, and Final Evaluations \medskip
                
			\small %
			\makebox[\textwidth][c]{\input{Results_Peri/Tables/first_final_slider_and_eval}}
		\end{center}
		\footnotesize
Notes: Standard errors are reported in parentheses and clustered at the evaluator level. *~p$<$0.10, **~p$<$0.05, ***~p$<$0.01. In the first two columns, the dependent variable is the likelihood of the candidate being Above-the-Bar chosen by the evaluator in the first slider. Columns \num{3} and \num{4} report the regressions on the likelihood of the candidate being Above-the-Bar chosen by the evaluator in the last slider before making their final decision. The last two columns report the regressions of the indicator variable taking value 1, if the final evaluation of the candidate was Above-the-Bar, and 0 otherwise. All regressions include round fixed effects and candidate advantage by filler characteristics.
	\end{table}

The results are consistent with our takeaways from Figures \ref{fig:initial_final_by_gender} and \ref{fig:initial_final_by_reladv}. Evaluators do not assess male candidates significantly more positively than female candidates, but relative advantage matters.\footnote{We caution the reader throughout when comparing the coefficients on male and relative advantage. Male is a binary indicator, while relative advantage ranges from $[-20,20]$. To compare the treatment effect of being ``male" to the treatment effect of moving from an equal to an advantaged group, one would need to multiply the coefficient on relative advantage by 20.} We estimate that, initially, one point of relative advantage increases the believed likelihood of that candidate being Above-the-Bar by 0.16 percentage points (p<0.01, Column \num{1}). Acquiring information on average reduces this bias, shrinking the impact of a point of relative advantage by roughly half (to 0.08pp) in final likelihood assessments (p<0.01, Column \num{3}). In terms of final binary evaluations, we estimate that one point of relative advantage increases a candidate's likelihood of being classified as Above-the-Bar by 0.15pp (p<0.01, Column \num{5}). 

Note that male candidates benefit no more from relative advantage than female candidates. In Table \appref{t:first_final_slider_and_eval_gender_int} in the Online Appendix, we expand Table \ref{t:first_final_slider_and_eval} to include the interaction of candidate gender and relative advantage, generally finding no significant relationship.\footnote{One exception is that we find that in initial likelihood assessments men benefit from relative advantage somewhat less than women (by roughly a third, $p<0.01$).} This further reinforces our finding that relative advantage between groups, rather than candidate gender, underlies evaluator decision-making in our setting. 

When we turn to treatment differences, we see that the exogenous provision of information reduces the bias induced by group differences. One point of relative advantage increases an evaluator's final likelihood slider for the candidate by 0.11pp in the endogenous treatment (p<0.01, Column \num{4}), but by only 0.05pp in the exogenous treatment (not statistically significantly different than 0, but we also cannot reject that the impact is the same across the two treatments). The differences are more stark in terms of final binary evaluations. Relative advantage has a significant and sizable impact in the endogenous treatment: we estimate that a candidate from the advantaged group is approximately 5pp more likely to be deemed Above-the-Bar relative to the same candidate when both groups are equal (p<0.01, Column \num{6}, computed by multiplying the coefficient on relative advantage, 0.267, by the amount of relative advantage, 20). But, we estimate no significant impact of relative advantage on final evaluations in the exogenous treatment.

Our analysis reveals that evaluators' initial assessments depend, in part, upon comparisons across groups. When information is provided exogenously, relative differences across groups no longer predict final evaluations. But, when evaluators have the choice of whether and how much information to acquire, final evaluations remain biased by comparisons across groups. Evaluators treat candidates more favorably when their group is advantaged relative to the other, and less favorably when their group is disadvantaged relative to the other, even after the chance to acquire individualizing information about the candidate.

In the following sections, we consider two reasons why access to information may fail to de-bias evaluators. First, we consider under-acquisition of information. Then, we consider biased processing of information.

\subsection{Information Acquisition}\label{sec.infoacq}

In this section, we analyze the information acquisition decisions made by evaluators in the endogenous treatment. Figure \appref{fig:signal_distr_pool} in the Appendix presents the distribution of number of signals acquired by pool, revealing no significant differences in the number of signals acquired across pool type. Importantly, this suggests that the mere existence of differences across groups did not alter the amount of information sought by evaluators on average, perhaps reducing concerns that effort, focus, or attention varied across pools. On average, evaluators acquire 2.3 signals about a candidate. In 25\% of evaluator-candidate observations, evaluators choose to receive no signals about the candidate. The median number of signals acquired is 2.\footnote{Figure \appref{fig:signal_distr_order} in the Appendix presents the average total number of signals acquired by candidate order. We do find that participants acquire fewer signals for the first candidate they assess. We speculate that participants may under-estimate the time-cost or annoyance of the filler task for skipped signals initially. After directly experiencing the filler task following the first candidate, they may revise downwards the net cost of a signal, leading them to acquire more signals for candidates 2 - 5. We show that our results are robust to excluding decisions for the first candidate in Figure \appref{fig:classification_timing_robust} and Tables \appref{t:first_final_slider_and_eval_robust}, \appref{t:info_acquisition_robust}, and \appref{t:whentoacquire_robust} in the Appendix.} 

This represents substantial under-acquisition of information relative to a Bayesian benchmark. Our Bayesian holds a prior belief equal to the candidate prior (so, 40 for a low-prior candidate and 60 for a high-prior candidate) and uses Bayes' rule to update their beliefs.\footnote{For simplicity, we assume a risk-neutral Bayesian decision-maker. However, a Bayesian decision-maker with risk aversion calibrated to the levels typically observed would not meaningfully change our conclusions. For instance, consider the decision to acquire the first signal. The evaluator is choosing between a lottery that pays \$7 with probability $P$ and \$0.50 with probability $(1-P)$, and a lottery that pays \$6.95 with probability $P+Y$ and \$0.45 with probability $(1-P-Y)$. It is not clear that a risk averse decision-maker who properly values the informational content of the signal (the gain in probability $Y$) would prefer the second lottery to the first.} When deciding whether to acquire another signal, they weigh the gain in final evaluation accuracy that a signal (and the option to buy additional signals) would provide against the cost of the signal (5~cents) and choose to buy a signal if it would increase their earnings in expectation. We can compare the behavior of our participants to how this Bayesian would behave. Appendix Figure \appref{fig:underacquisition_vs_bayesian} presents the results, showing the total number of signals per candidate acquired by our evaluators compared to the total number of signals acquired by the Bayesian in these same cases. The Bayesian always buys at least 3 signals; our evaluators acquire fewer than 3 signals in 54\% of cases, suggesting significant under-acquisition on average. We can also use a Bayesian benchmark to compute the expected dollar value of purchasing another signal at the moment a decision-maker chooses to stop acquiring signals. In approximately half of cases, acquiring another signal had an expected value of more than 50 cents (given the evaluator's reported likelihood at the moment they stopped acquiring signals). In a third of cases, the expected value was more than \$1. 

Evaluators in our endogenous treatment are leaving money on the table, both relative to this Bayesian benchmark and relative to evaluators in the exogenous treatment who were forced to acquire information. Even if we factor in a 5 cent signal cost in both treatments, evaluators in the exogenous treatment earn significantly more money from their final evaluations. And, because of the filler task for skipped signals, evaluators in the endogenous treatment are also \textit{not} saving time in the study; in fact, evaluators in the endogenous treatment spend significantly longer on the study on average (roughly 3 minutes). It is somewhat striking that we see significant under-acquisition of information even when signals are relatively cheap in terms of both money and time.\footnote{We cannot precisely quantify the cognitive costs of processing additional signals, so we cannot fully rule out that evaluators in the endogenous treatment are behaving optimally, despite earning less and taking more time on average.}

\subsubsection{When and Why Do Evaluators Stop Acquiring Information?}

What predicts the decision to seek more information about a candidate? First, we show that more biased evaluators---in terms of their propensity to rely on differences across groups in their initial assessments---are less likely to seek additional information. Drawing on the specification of Table~\ref{t:first_final_slider_and_eval}, Column \num{1}, Table \ref{t:info_acquisition} predicts an evaluator's \textbf{initial} assessment of a candidate from the candidate's gender and their group's relative advantage. In Column \num{1}, we consider the data for evaluators who chose to make a final evaluation of the candidate immediately, without acquiring any signals for that candidate. In Column \num{2}, we consider the data from all other evaluators in the endogenous treatment, those that went on to acquire anywhere between one and five signals for the candidate. In both columns, we are focused on the initial assessments of evaluators. We are asking whether, prior to receiving any additional information about the candidate, evaluators who choose to receive more information look different than evaluators who make immediate decisions. Indeed, that seems to be the case. We estimate that evaluators who choose to make their final evaluation immediately are significantly more reliant on group differences in their initial assessments than evaluators who go on to seek more information. One point of relative advantage increases a candidate's initial assessment by 0.14pp for evaluators who choose more information, compared to increasing it by 0.28pp for evaluators who make their final evaluations without any more information. We present an interacted model in Column \num{3}, finding that the extent to which group differences impact initial assessments varies significantly with the evaluator's subsequent information choice (p<0.05).

	\begin{table}[ht]
		\begin{center}
			\caption{\label{t:info_acquisition} Selection Into Acquiring Information} \medskip
   
                OLS Predicting Initial Likelihoods and Final Evaluations by Subsample\medskip
                
			\small %
			\makebox[\textwidth][c]{\input{Results_Peri/Tables/info_acquisition.tex}}
		\end{center}
		\footnotesize
Notes: Standard errors are reported in parentheses and clustered at the evaluator level. *~p$<$0.10, **~p$<$0.05, ***~p$<$0.01. The dependent variable in all columns is the likelihood of the candidate being Above-the-Bar chosen by the evaluator in the first slider before receiving any signals. The first column includes the sample of observations where the evaluation was made without acquiring any information. The second column includes the observations where evaluation was made after receiving at least one signal. The third and fourth columns pool all observations and estimate the effect of receiving any signals on the initial likelihood assessment and final evaluation, respectively. All regressions include controls for round fixed effects and candidate advantage by filler characteristics. All three columns include endogenous treatment only. 
	\end{table}

Column \num{4} shows that evaluators who choose not to receive more information also produce more biased final evaluations. For evaluators who do not seek out information, we estimate that every point of relative advantage increases the chance of a candidate being classified as Above-the-Bar by 0.8pp (p<0.01). For a sense of magnitudes, this translates to a candidate from a high-prior group being 16pp more likely to be classified as Above-the-Bar if they are in a pool with some low-prior candidates rather than in a pool with all high-prior candidates. Among evaluators who seek out more information before making a final decision, relative advantage has no significant impact on final evaluations (p<0.01 for difference across evaluator types). 

Next, we focus on evaluators who did choose to receive more information and explore what predicts the decision to receive another signal about a candidate.  Recall that we define a confirming signal as a signal that is consistent with the candidate's prior: a positive signal for a candidate from a high-prior group, or a negative signal for a candidate from a low-prior group (see Section~\ref{Hyp} for more details). 

Intuitively, we expect that receiving a confirming signal should reduce the likelihood that an evaluator chooses to receive another signal. For example, suppose an evaluator is assessing a candidate from a high-prior group. The evaluator chooses to receive a signal, and the signal is positive. Now, the evaluator knows the candidate both comes from the high-prior group and has generated a positive signal, both consistent with the candidate being Above-the-Bar. This should make the evaluator feel more ready to reach a final evaluation, classifying the candidate as Above-the-Bar, relative to a situation where that same high-prior candidate has generated a negative signal. This logic leads us to expect that, in general, receiving a confirming signal should reduce the likelihood of acquiring another signal. 

Table \ref{t:whentoacquire} tests this hypothesis. We predict an evaluator's decision to acquire another signal from an indicator for whether the evaluator has just received a confirming signal. In Panel A, Columns~\num{1}--\num{4}, we analyze decisions after the first, second, third, and fourth signals, respectively. Note that there is selection into each additional signal, potentially in terms of evaluator characteristics and signal histories, and a correspondingly smaller sample size as we consider later signal realizations. With that note of caution, we investigate whether confirming signals reduce the likelihood of receiving more information. For early signals, we see evidence consistent with the hypothesis. We estimate that receiving a confirming first signal reduces the likelihood of pursuing another signal by 3pp (p<0.01), while receiving a confirming signal as the second signal (independent of whether the first signal was confirming or not), reduces the likelihood of pursuing a third signal by 8pp (p<0.01). We do not find this effect for the third and fourth signals. The final column pools over all signals, continuing to predict the decision to receive more information from whether the directly preceding signal was confirming. We estimate that, on average, a confirming signal reduces the likelihood of acquiring another signal by 3pp (p<0.01).

	\begin{table}[ht!]
		\begin{center}
			\caption{\label{t:whentoacquire} Endogenous Information Acquisition }\bigskip
            OLS Predicting the Decision to Get More Info After... \medskip
            
			\small %
		Panel A: Overall 
                \medskip
			\makebox[\textwidth][c]{\input{Results_Peri/Tables/when_to_acquire_info}}
            \medskip
  
            Panel B: Comparing Pools with Equal Versus Contrasting Pools  
                \medskip			
			\makebox[\textwidth][c]{\input{Results_Peri/Tables/when_to_acquire_info_equal_groups.tex}}
		\end{center}
		\footnotesize
Notes: Standard errors are reported in parentheses and clustered at the evaluator level. *~p$<$0.10, **~p$<$0.05, ***~p$<$0.01. All regressions include controls for relative advantage, male candidate indicator, group prior, initial likelihood assessment, round fixed effects, and candidate advantage by filler characteristics. All three columns include endogenous treatment only. The dependent variable is an indicator variable for deciding to receive a signal after the corresponding signal in each column for the corresponding subsample. The last column pools all observations in each round in the long format.
	\end{table}

We interpret this result as illustrating that confirming signals make it ``easier'' for evaluators to reach a decision about a candidate. This motivates the important question of how contrast between groups interacts with this effect. Consider our previous example, with an evaluator assessing a candidate from a high-prior group who has just received a confirming (positive) signal. Table \ref{t:whentoacquire} revealed that this evaluator will be less likely to seek out additional information (relative to having observed a negative signal). In our next analysis, we ask whether this effect is amplified when there are differences across groups. Consider a high-prior candidate for whom an evaluator has seen a positive signal. We ask whether the impact of that confirming signal will be larger when the candidate's group is advantaged ($60-40$ pool) relative to when the two groups are equal ($60-60$ pool). That is, does the candidate being in the advantaged group make it even more likely that the evaluator feels convinced by the confirming signal, forgoes additional information, and classifies the candidate as Above-the-Bar? 

Considering the case of the low-prior candidate helps to crystallize the hypothesis and motivate our empirical approach. Suppose an evaluator is assessing a low-prior candidate and has just seen a negative signal. Our hypothesis is that it is even more likely that the evaluator feels convinced by this confirming signal when the candidate's group is disadvantaged relative to the other group ($40-60$) compared to when both groups are equal ($40-40$ pool). Pulling the two examples together, we are hypothesizing that confirming signals have a larger impact---in terms of reducing the likelihood of pursuing more information---when the two groups are contrasting relative to when the two groups are equal. 

Panel B of Table \ref{t:whentoacquire} explores this question, augmenting the specifications of Panel A. We include an indicator for whether the candidate pool featured equal groups ($40-40$ or $60-60$), instead of contrasting groups ($60-40$, $40-60$), and we interact this indicator with having just observed a confirming signal. We find that contrast between groups does amplify the impact of confirming signals. Evaluators are 5pp less likely to get more information after seeing a confirming first signal when the groups are contrasting (Column \num{1}, p<0.01). This effect is eliminated when the two groups are equal (p<0.05 for difference across pool types). Responses after the second through fourth signals are directionally consistent with this same pattern, though more noisily estimated. Column \num{5} considers all of the data. We estimate that receiving a confirming signal (relative to a disconfirming signal) reduces the likelihood of seeking more information by 4pp when groups are contrasting (p$<$0.01), but by only 1pp when groups are equal (p<0.01 for difference across pool types). 

This finding suggests that group differences and confirming signals reinforce each other. A positive signal for a high-prior candidate is particularly convincing when their group is also advantaged relative to the other group; a negative signal for a low-prior candidate is particularly convincing when their group is also disadvantaged relative to the other group. 

Our results highlight two ways in which differences across groups shape information acquisition decisions. First, evaluators who choose not to pursue more information are more influenced by group differences on average, both in terms of their initial assessments and in terms of their final evaluations. Second, receiving a confirming signal reduces the likelihood that an evaluator seeks out an additional signal, significantly more so when there are differences across the groups. Both of these findings illustrate how a reliance on group differences in assessments can be reinforced when evaluators must actively seek out additional information. Of course, the other way in which group differences may be reinforced is through biased processing of signals. We explore this in the next section. 

\subsection{Biased Belief Updating}

Evaluators provide beliefs after every signal they acquire, allowing us to investigate how individuals update their beliefs in response to new information. Our focus is understanding whether candidate gender and/or relative advantage predicts updating. While this analysis could be done on the full sample, in the main text we analyze only evaluators in the exogenous treatment. This rules out selection into additional information, isolating the role for biased processing. In the Appendix, we replicate these results with the full sample reported in Table \appref{t:sliders_two_signals_by_prior_full}; while the biases we document are stronger in the exogenous treatment, absent selection, we observe similar patterns when considering the full sample.

To provide intuition and motivate a more unified analysis, we begin by analyzing the data separately for low and high-prior candidates. Table \ref{t:sliders_two_signals_by_prior_exog}, Panel A considers assessments of low-prior candidates. We predict the believed likelihood of the candidate being Above-the-Bar after different signal histories, controlling for relative advantage, an indicator for a male candidate, and our standard controls. In addition, we control for the individual evaluator's initial assessment; any impact we observe of relative advantage on subsequent beliefs is on top of this initial bias.\footnote{Prior to receiving any signals, evaluators believe low-prior candidates have a 2.2pp lower chance of being Above-the-Bar when they are in the disadvantaged group, relative to the equal groups case (p<0.01).} Columns \num{1} and \num{2} consider beliefs after a positive and negative signal, respectively. We estimate that relative advantage has a significant impact on belief updating, but only when the signal is positive. When a low-prior candidate receives a negative signal, it does not matter whether the groups were equal or their group was disadvantaged (Column \num{2}). But, if a low-prior candidate receives a positive signal, the candidate gets a significantly smaller boost from the positive signal when they are disadvantaged (p<0.01, Column \num{1}). When we consider updating after two signals, both positive, both negative, or one of each, we find a directionally similar pattern, though no significant differences. It seems to be the case that a low-prior candidate benefits less from positive news when they are disadvantaged. 

Panel B turns attention to high-prior candidates.\footnote{In terms of initial assessments, prior to receiving signals, evaluators believe high-prior candidates have a 3.2pp greater chance of being Above-the-Bar when they are in the advantaged group relative to when both groups are equal (p<0.01). We control for these initial assessments in the analysis presented.} Columns \num{1} and \num{2} consider beliefs after the first signal, conditional on initial assessment. The results are  flipped relative to low-prior candidates. For high-prior candidates, relative advantage has a significant impact on belief updating after a negative signal, but not after a positive signal. After a negative signal, evaluators maintain significantly more optimistic beliefs for a high-prior candidate when they are advantaged relative to when the two groups are equal (Column \num{2}, p<0.05). We see a similar protective impact of relative advantage after two signals. Conditional on initial assessments, beliefs are 3.6pp more favorable for high-prior candidates with two negative signals when they are advantaged relative to equal (Column \num{5}, p<0.01). The effect of relative advantage is smaller when the signals are mixed (at 1.5pp, Column \num{4}, p<0.05), and non-existent when both signals are positive.

	\begin{table}[htp]
		\begin{center}
			\caption{\label{t:sliders_two_signals_by_prior_exog} Biased Belief Updating}\medskip
            OLS Predicting Beliefs After Given Signal History \medskip
            
			\small
		Panel A: Only Low-Prior Sample (Exogenous):
			\makebox[\textwidth][c]{\input{Results_Peri/Tables/sliders_two_signals_low_prior_exog.tex}}
		Panel B: Only High-Prior Sample (Exogenous):			
			\makebox[\textwidth][c]{\input{Results_Peri/Tables/sliders_two_signals_high_prior_exog.tex}}
		\end{center}
		\footnotesize
Notes: Standard errors are reported in parentheses and clustered at the evaluator level. *~p$<$0.10, **~p$<$0.05, ***~p$<$0.01. This table reports the OLS results predicting beliefs about the likelihood of being Above-the-Bar for only the evaluators in the exogenous treatment. We split the sample by candidate prior and run the analysis on the subsamples including only low-prior candidates and only high-prior candidates in Panel A and B, respectively. The dependent variable in the first two columns is the evaluator's believed likelihood of the candidate being Above-the-Bar after receiving the first signal. The following columns report the regressions on the believed likelihood of the candidate being Above-the-Bar after receiving two signals. All regressions include round fixed effects and candidate advantage by filler characteristics.
	\end{table}

The analysis of Table \ref{t:sliders_two_signals_by_prior_exog} controls for initial likelihoods, attempting to isolate the impact of relative advantage on belief updating. However, it could be that average differences in initial likelihoods mean that indeed even a Bayesian updater (with biased priors) might update differently in response to the same signal depending upon relative advantage. To address this concern, in Online Appendix Table \appref{t:sliders_two_signals_by_prior_exog_bayes} we augment the specification of Table \ref{t:sliders_two_signals_by_prior_exog} by controlling for the Bayesian posterior an individual would hold given their own (biased) belief prior to seeing the signal. The results are largely unchanged. Relative advantage shapes how evaluators update their beliefs after confirming signals, on top of any Bayesian channel. 

Together, the results for low and high-prior candidates seem to tell a clear story. When evaluators observe early confirming signals, relative advantage does not play a significant role in shaping how beliefs are updated. But, when early signals are disconfirming, relative advantage shapes how evaluators respond to the mixed picture. Low-prior candidates seem to get less of a boost out of a positive signal when they are disadvantaged, relative to when the groups are equal; high-prior candidates seem to get more of the benefit of the doubt after a bad signal when they are advantaged, relative to when the groups are equal. These patterns are inconsistent with any Bayesian story for behavior, as candidates with the same prior probability of being Above-the-Bar and the same signal history are assessed differently on the basis of an irrelevant comparison group. 

In Table \ref{t:sliders_beyond_two_signals}, we use the insights from responses to early signals to build and analyze a more general model of behavior. We want to understand the impact of relative advantage depending upon whether the evidence has been mostly confirming (in terms of consistency with the candidate prior). To do so, we predict an evaluator's belief about a candidate after signal $t$ from the share of the $t$ signals that have been confirming. For a low-prior candidate, this is the share of negative signals seen so far; for a high-prior candidate, this is the share of positive signals seen so far. To explore whether biases are larger when the evidence has been more mixed, we interact this share of confirming signals variable with relative advantage. A negative interaction term suggests that relative advantage benefits a candidate relatively less when the evidence has been more consistent with the candidate's prior.

	\begin{table}[ht]
		\begin{center}
			\caption{\label{t:sliders_beyond_two_signals} Biased Belief Updating Beyond Two Signals  (Exogenous Treatment Only)}\medskip
                OLS Predicting Beliefs After Given Signal History  \medskip
                
			\footnotesize
			\makebox[\textwidth][c]{\input{Results_Peri/Tables/sliders_beyond_two_signals.tex}}
		\end{center}
		\footnotesize
Notes: Standard errors are reported in parentheses and clustered at the evaluator level. *~p$<$0.10, **~p$<$0.05, ***~p$<$0.01. The dependent variable is the perceived likelihood of the candidate being Above-the-Bar after receiving $t$ signals in Columns \num{1} and \num{3} for low and high-prior candidates, respectively. Columns \num{2} and \num{4} include only the last round, the final evaluation round, where the dependent variable is whether the candidate is classified as Above-the-Bar for low and high-prior candidates, respectively. All columns include observations from the exogenous information acquisition treatment only. All regressions include round fixed effects and candidate advantage by filler characteristics as well as initial likelihood chosen by the participant before any signals.
	\end{table}

Columns \num{1} and \num{2} consider low-prior candidates. The first column predicts the evaluator's belief of a candidate's likelihood of being Above-the-Bar after signal $t$. As we did in Table \ref{t:sliders_two_signals_by_prior_exog}, we control for the evaluator's initial assessment of the candidate, so that any estimated impact of relative advantage on later beliefs can be interpreted as an impact on updating behavior. The negative coefficient on the share of confirming signals tells us that, as expected, as the share of negative signals increases, evaluator beliefs are more pessimistic. We do not observe a significant interaction of relative advantage with the share of confirming signals, suggesting that bias against disadvantaged groups is, on average, no greater when the evidence has been more mixed (p=0.107).\footnote{To give a sense of effect sizes, we estimate that moving from all positive signals to all negative signals reduces the believed likelihood of being Above-the-Bar by 47pp for a candidate in the $40-40$ pool (coefficient on confirming share in Column~\num{1}). The (insignificant) negative interaction term suggests that this reduction is approximately 49pp for a low-prior candidate in the $40-60$ pool.} In Column~\num{2}, we bring this same empirical approach to analyzing final classifications, asking whether relative advantage plays a larger role in final evaluations when signals are less in line with the candidate prior. Indeed this seems to be the case. As the share of confirming signals increases (negative signals for low-prior candidates), relative advantage plays a significantly smaller role in how that candidate is classified.

Columns \num{3} and \num{4} present the same analysis for high-prior candidates. In these specifications, a larger share of confirming signals indicates more positive signals for the candidate. We see that, as expected, this translates into a greater believed likelihood of the candidate being Above-the-Bar after signal~$t$ and a greater chance of the candidate ultimately being classified as Above-the-Bar. Relative advantage also strongly predicts positive assessments, both in terms of beliefs after signal~$t$ and in final classifications. However, we see a significant and sizable negative interaction. High-prior candidates benefit less from being relatively advantaged when most signals have been positive compared to when signals have been more mixed. 

Our data shows that initial assessments are biased by group differences. These same group differences predict both the decision to get more information and reactions to new information. In particular, group comparisons predict how evaluators react to mixed evidence. When the picture seems clear, i.e., information is more consistent with the candidate prior, evaluators update their beliefs similarly independent of the candidate's relative advantage. On the other hand, group comparisons play a significant role in the case of exceptions, i.e., when information is less consistent with the candidate prior. Being from an advantaged group seems to mitigate the impact of a negative signal. And, positive signals seem to carry less weight for candidates from disadvantaged groups. These forces have implications for the accuracy of evaluations, particularly for candidates who are not typical of their pool.     

\subsection{Accuracy of Final Evaluations}

\begin{figure}[h] %
  \hfill
    \begin{center}  
     \caption{Speed of Classifications}
      \label{fig:classification_timing} \medskip
      
       \centering	
       Panel A: High-Prior Candidates\\
       \includegraphics[scale=0.14]{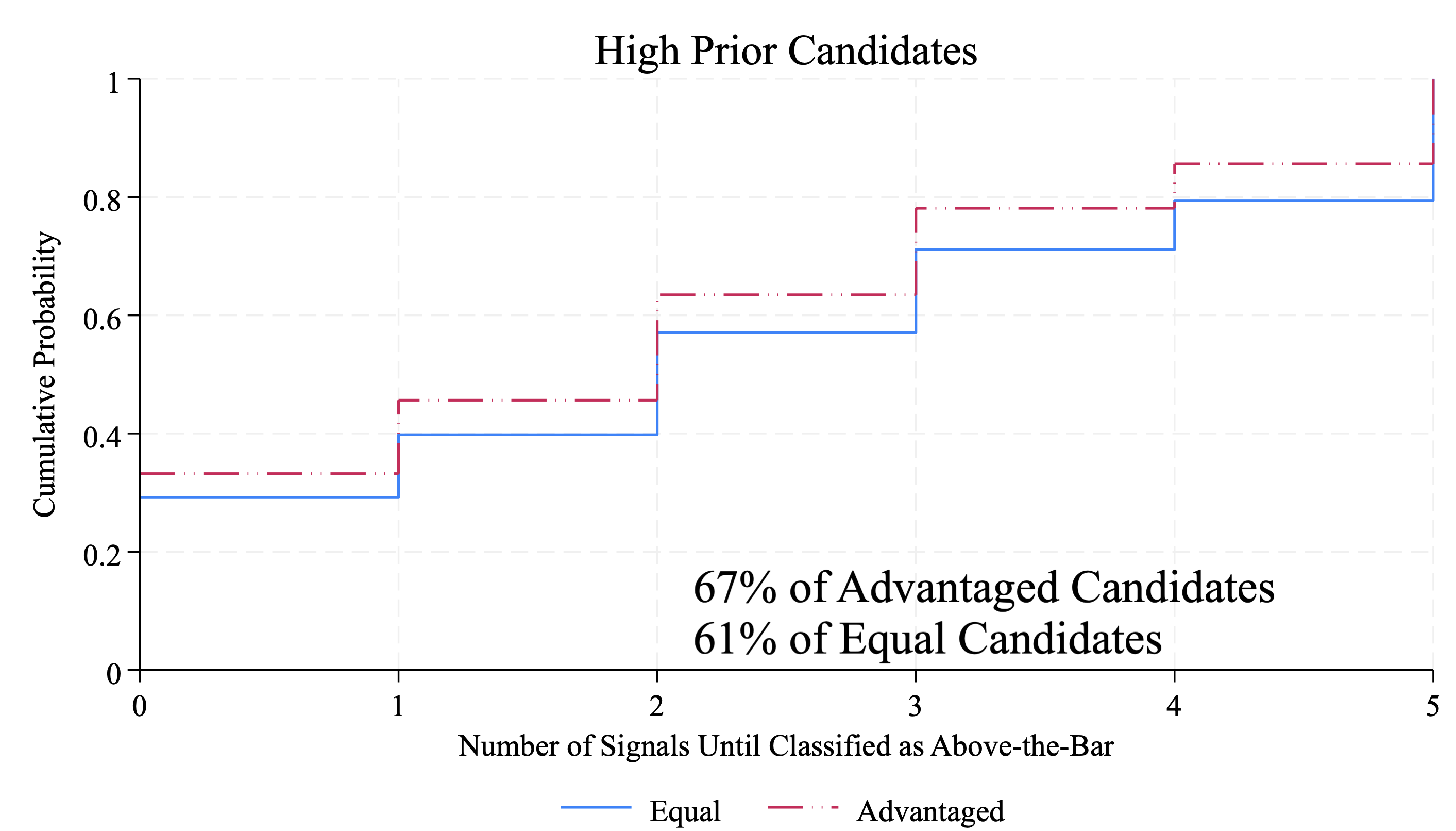}
       \includegraphics[scale=0.14]{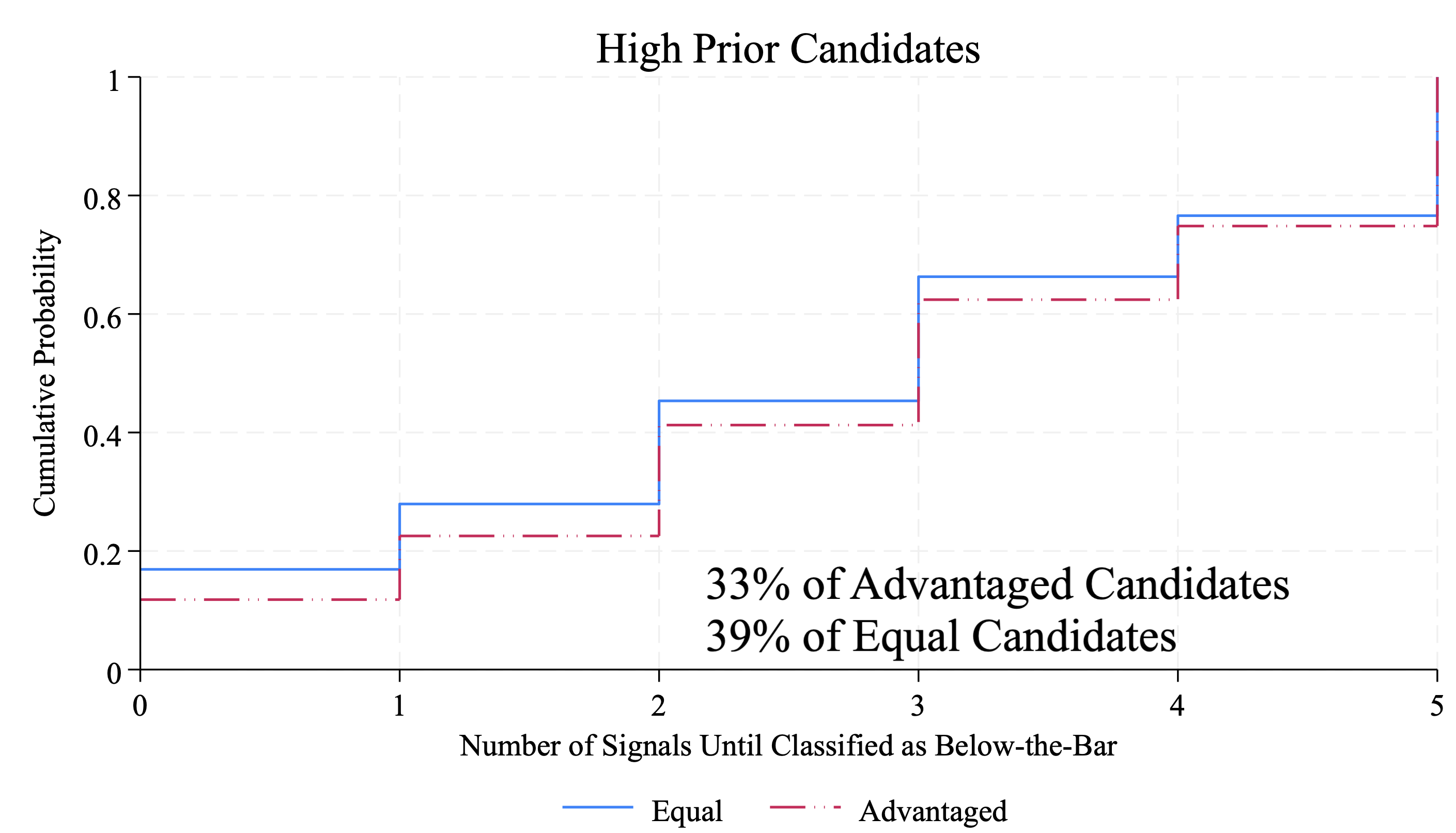}\\
       	\bigskip
            \bigskip
        
       Panel B: Low-Prior Candidates\\
       \includegraphics[scale=0.14]{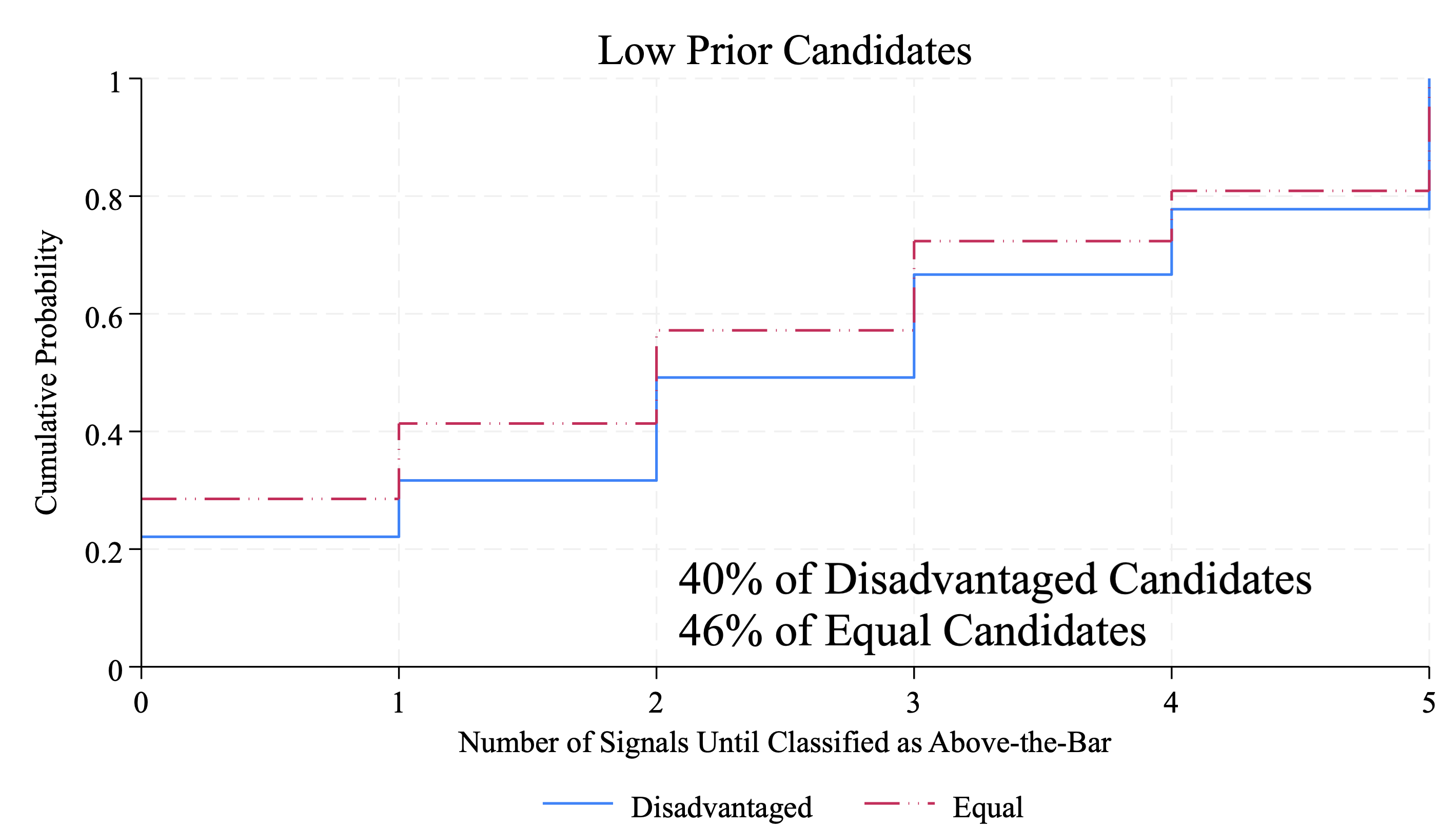}       
       \includegraphics[scale=0.14]{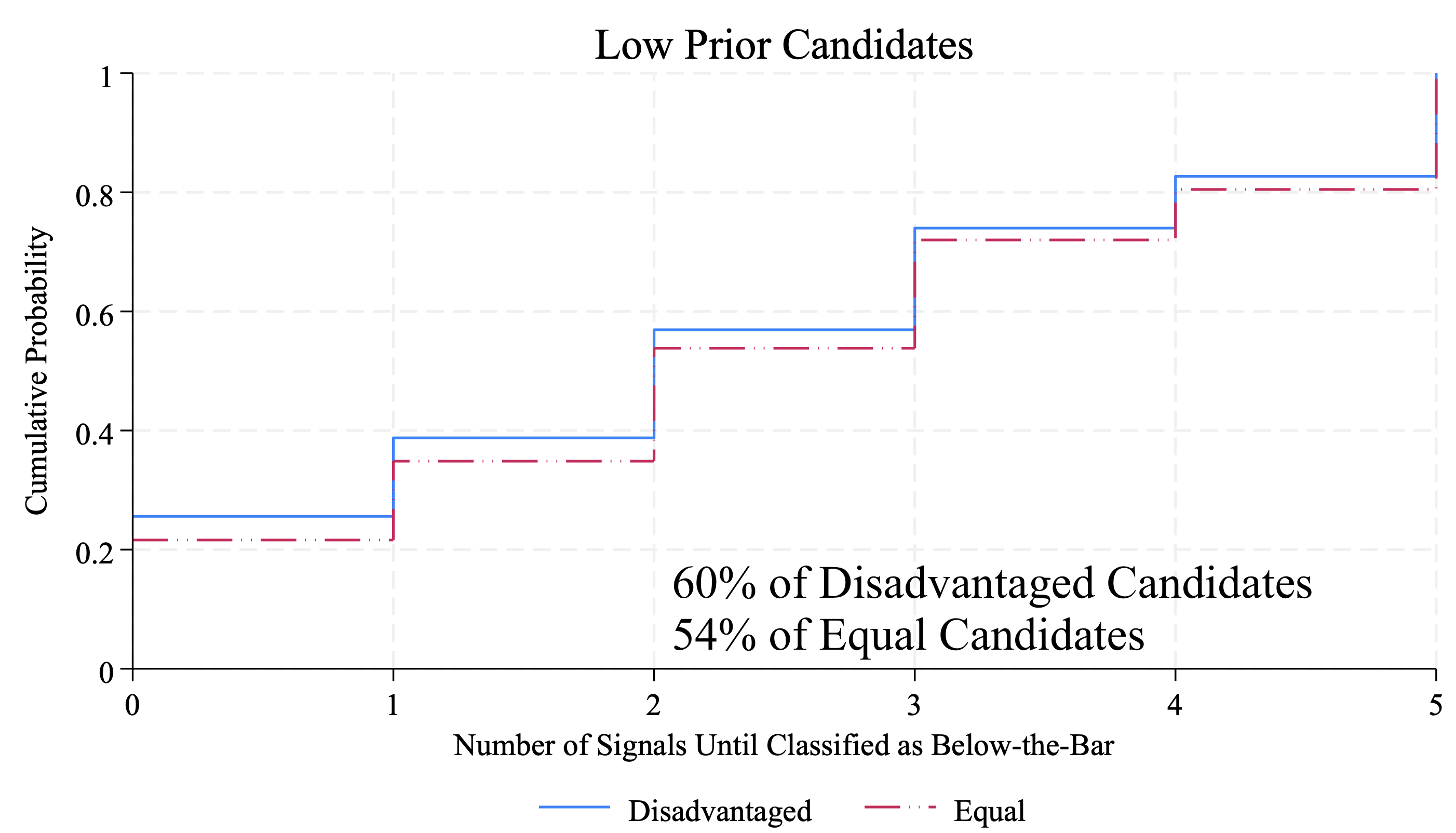}\\
    \end{center}
  \hfill
	\medskip %
\begin{minipage}{0.99\textwidth} %
{\footnotesize Note: Panel A presents the distribution of the number of total signals viewed before a high-prior candidate is classified as Above-the-Bar (left) or Below-the-Bar (right). The percentages reported within the figure are the percentage of high-prior candidates of each type that are classified as Above-the-Bar (left) or Below-the-Bar (right). Panel B presents the same analysis for low-prior candidates. \par}
\end{minipage}
\end{figure}

We have shown that group differences impact both the decision to acquire more information and how that information is incorporated into beliefs. What is the cumulative impact of these decisions on final evaluations? To answer this question, we must focus on the endogenous treatment where both of these factors are at work. Figure \ref{fig:classification_timing} presents CDFs of the number of signals acquired before an evaluator makes their final decision, splitting the data by candidate advantage. Panel A considers only high-prior candidates, showing the speed with which they are classified as Above-the-Bar (left) or Below-the-Bar (right).\footnote{Note that by speed here we mean the number of signals before a final evaluation is reached.} We observe that evaluators are quicker to classify advantaged candidates as Above-the-Bar, relative to candidates from the $60-60$ pool (Panel A -- left). Thus, not only are advantaged candidates more likely to be classified as Above-the-Bar (67\% of high-prior candidates in the $60-40$ pool compared to 61\% of candidates in the $60-60$ pool), they are also classified this way with less evidence on average. Similarly, evaluators are slower (i.e. require more information) to classify advantaged candidates as Below-the-Bar (Panel A -- right).  

Panel B presents this same analysis for low-prior candidates, revealing similar patterns. Evaluators require more signals before classifying a disadvantaged candidate as Above-the-Bar compared to a low-prior candidate from the $40-40$ pool (left); they require fewer signals before classifying that disadvantaged candidate as Below-the-Bar (right). Disadvantaged candidates are more likely to be classified as Below-the-Bar relative to other low-prior candidates (60\% versus 54\%), and they are classified this way more quickly. Across candidate types, evaluators are quicker to classify candidates in line with group priors, particularly so when the groups are contrasting rather than equal. 

Overall, evaluators are quite accurate in their final evaluations. Within the exogenous treatment, 84\% of final evaluations are correct. Accuracy is significantly lower in the endogenous treatment at 76\% (p<0.01 when comparing across treatments). 

\begin{figure}[ht] %
  \hfill
    \begin{center}  
     \caption{Rate of Correct Classifications of Atypical Candidates}
      \label{fig:accuracy} \medskip

       \centering	
        \includegraphics[scale=0.40]{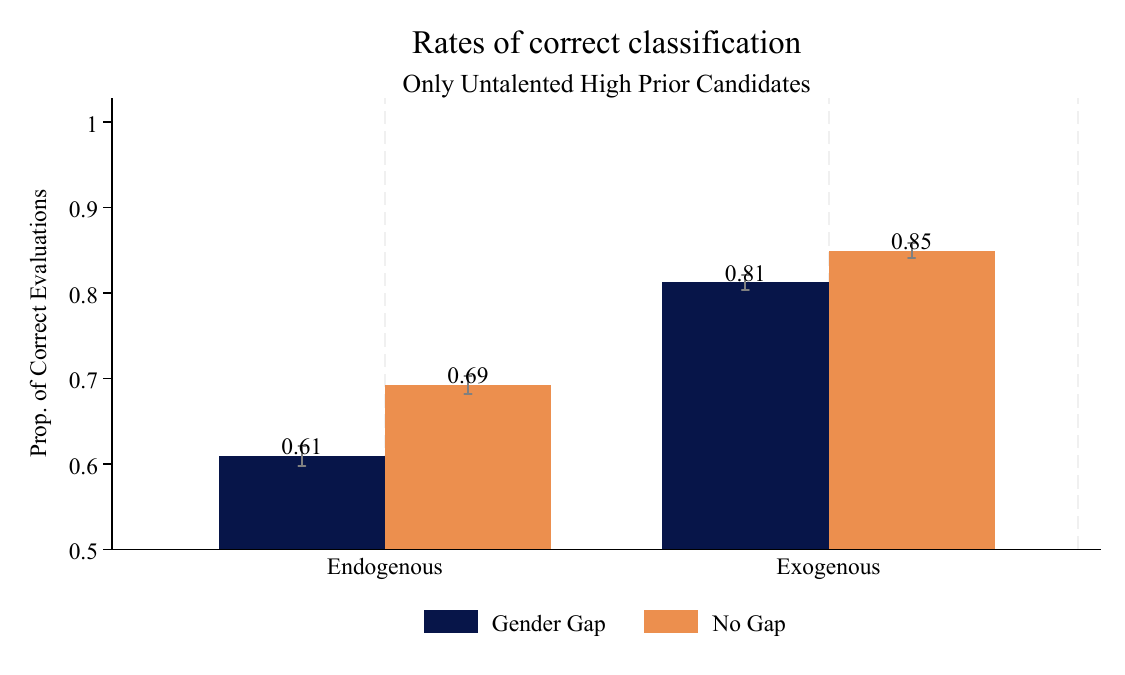}
        \bigskip
        \bigskip
        \includegraphics[scale=0.40]{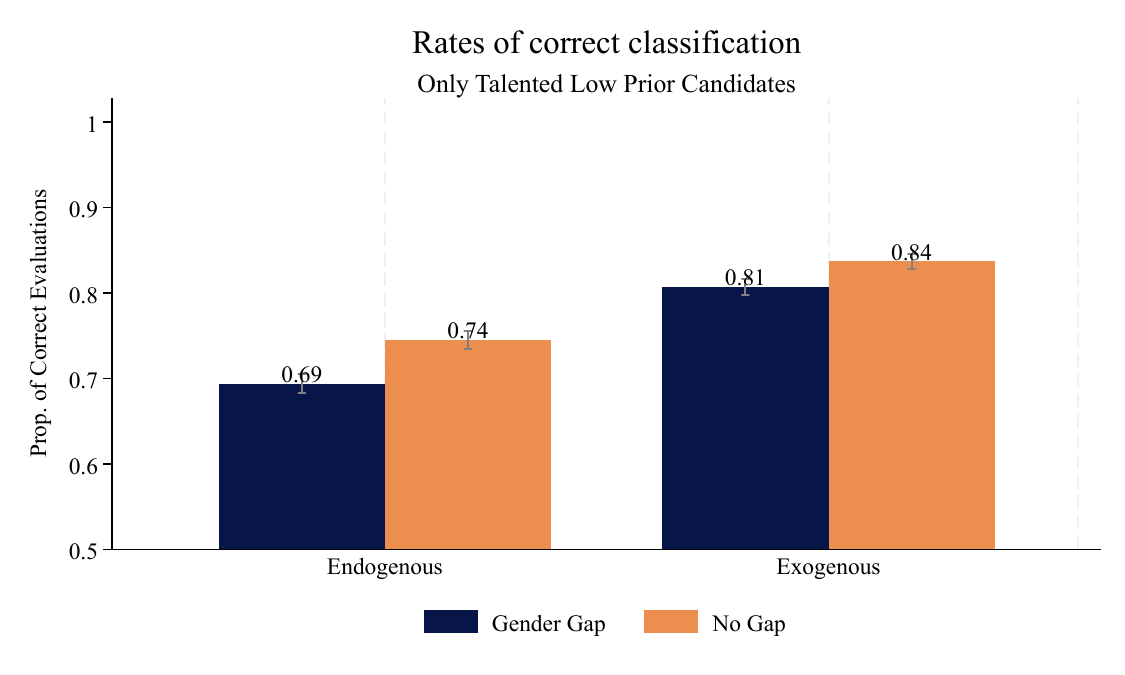}\\
    \end{center}
  \hfill
	\medskip %
\begin{minipage}{0.99\textwidth} %
{\vspace{-1cm}\footnotesize Note: The top graph shows the likelihood of correctly identifying a talented candidate from the low-prior group while the bottom graph shows the likelihood of correctly identifying an untalented candidate from the high-prior group. We show these likelihoods by candidates' relative advantages in their pools both in the exogenous and endogenous information acquisition treatments. Both types of mistakes are more common in the endogenous treatment and in particular when the high-prior candidate has a relative advantage and low-prior candidate has a relative disadvantage.\par}
\end{minipage}
\end{figure}

How do group differences impact the accuracy of final evaluations? On average, they have no significant impact: evaluators classify approximately the same proportion of candidates accurately independent of whether the groups are equal ($40-40$ or $60-60$ pools) or contrasting ($60-40$ or $40-60$ pools).\footnote{Consistent with the treatment differences described in the previous paragraph, this is approximately 84\% of candidates in the exogenous treatment and 76\% of candidates in the endogenous treatment.} But, there is important heterogeneity. Group differences make it more likely that ``typical'' candidates are accurately classified. Above-the-Bar candidates from high-prior groups are more likely to be correctly classified as Above-the-Bar when they are advantaged (relative to the equal groups case); Below-the-Bar candidates from low-prior groups are more likely to be correctly classified as Below-the-Bar when they are disadvantaged (relative to the equal groups case). At the same time, group advantage makes it significantly less likely that an evaluator successfully classifies candidates that are not typical of their group. Figure \ref{fig:accuracy} demonstrates this important pattern. We graph the proportion of correct evaluations for these atypical candidates, splitting the data by treatment assignment and candidate pool.

Not surprisingly, atypical candidates are on average harder to successfully classify, with overall success rates at 76\% compared to 83\% for candidates that are typical of their pool. The more important finding is that these mistakes are significantly more likely when the groups are different. In both treatments, contrast across the groups increases the share of incorrect classifications of atypical candidates. Consistent with the overall pattern, the rate of correct classifications of atypical candidates is lower in the endogenous treatment compared to the exogenous treatment. And, the impact of differences across groups on the likelihood of mistakes is nearly twice as large in the endogenous treatment. Within the exogenous treatment, atypical candidates are classified incorrectly approximately 3pp more often when there are differences across groups; in the endogenous treatment, this gap is 7pp.

The combination of requiring information acquisition and removing contrast between groups has a significant impact on the rate at which atypical candidates are classified correctly. Suppose we wanted to understand how likely it is that an evaluator correctly susses out an untalented Below-the-Bar candidate from a high-prior group. Evaluators in our best case scenario -- where they are required to acquire individual-level information about the candidates and where there are no differences across groups -- correctly classify these candidates 25pp more often than in our worst-case scenario, where evaluators can choose to rely on their priors rather than individual-level signals and where differences across groups bias evaluators toward relatively advantaged candidates. Similarly, Above-the-Bar candidates from low-prior groups are 16pp more likely to be correctly identified as such when evaluators must receive individual-level information and there is no contrast between groups compared to the endogenous treatment with contrast across groups.

\section{Discussion}

We explore how evaluators choose and use information as they assess candidates. Overall, male and female candidates are assessed quite similarly by evaluators; we find no evidence that evaluators, in general, are biased against female candidates, either in terms of their initial assessments, how they update their beliefs, or in what final evaluations they submit. Instead, we find that relative advantage---the extent to which the candidate's group outperforms the other group within the candidate pool---plays a significant role in decision-making. Conditional on a candidate's likelihood of being Above-the-Bar, evaluators provide significantly more favorable absolute assessments of the candidate when that candidate's group has a relative advantage within the pool.

Why do we not observe a general bias against female candidates in our study? A few possibilities are worth considering. First, it is plausible that evaluators recognized that the study had a gender focus, prompting concerns about social desirability. However, we think our data is not entirely consistent with this hypothesis. Consider the data from our 40-60 and 60-40 pools, collected across evaluator. Across these treatments, evaluators treat members of the lower-performing group identically, regardless of whether that group was men or women. While equating treatment of lower-performing men and women within-subject might suggest social desirability bias, the fact that we see similar treatment of low-performing men and women \textit{across-subject} seems less consistent with this account and more consistent with gender being less important than relative advantage in our setting. Perhaps a more plausible explanation for why we do not observe general bias against female candidates is related to our design choice to fix priors. While the candidate performance was on a math and science test, our design likely limits the impact of the the male stereotype of this underlying task. We provide evaluators with precise information about the gender difference in performance on the task within the candidate pool they are hiring from, reducing the extent to which their own potentially biased beliefs may play a role. Future work could consider designs that not only exogenously vary perceived relative advantage but also leave open scope for stereotypical beliefs, such as by providing only noisy signals of group differences in performance. Finally, it is worth explicitly considering that the absence of general bias against female candidates that we observe is revealing something potentially important about the nature of gender bias in other settings as well: in some contexts, what appears to be anti-female bias may actually be relative advantage bias. To the extent that women belong to the perceived lower-performing group, it may be relative comparisons that drive some of the discrimination we observe in other contexts of interest.

In our setting, relative advantage shapes initial assessments and how evaluators update their beliefs in response to new information. As a result, evaluators are quicker to classify high-prior candidates as Above-the-Bar when they are at a relative advantage, requiring fewer signals. They are similarly quicker to dismiss low-prior candidates as Below-the-Bar when they are at a relative disadvantage. 

We find that relative advantage plays a larger role when information is more mixed. Evaluators are more likely to give the benefit of the doubt to high-prior candidates with early negative signals when they are from an advantaged group; low-prior candidates from the disadvantaged group benefit less from early positive signals. 

Evaluators are significantly less likely to make accurate assessments when information must be endogenously acquired. We observe surprisingly low acquisition of individual-level information, despite the fact that information is cheap in our setting in terms of monetary and time costs. This fuels stereotypical judgments: compared to the exogenous treatment, evaluators in the endogenous treatment are significantly less likely to accurately identify an Above-the-Bar candidate from a low-prior group, particularly when their group is disadvantaged. Similarly, evaluators in the endogenous treatment are significantly less likely to accurately identify a Below-the-Bar candidate from a high-prior group, particularly when their group is advantaged.

We observe that evaluators rely on relative advantage to inform their judgments, even when relative advantage is objectively irrelevant. We believe this result has important external relevance. First, while there are of course many real world evaluations that rely on relative comparisons, other important ones do not, such as loan applications or medical diagnoses. This suggests the potential to misuse relative advantage heuristics exists in other, higher-stakes contexts outside of our study. Second, while our participants are not real HR managers making real HR decisions, it is not clear that they are more susceptible to the mis-allocation of attention than others. The vast majority of our participants make internally consistent decisions, update their beliefs in the right direction in response to signals, and spend a reasonable amount of time on the study; our results are unchanged when we exclude participants in the bottom or top 5\% of participation duration. Our participants' tendency to attend to features of the problem that are irrelevant, while failing to acquire or attend to more relevant decision criteria, may speak instead to more general behavioral patterns \citep{gagnon2018channeled}. 

While we can only speculate, it may be the case that evaluators' tendency to rely on relative advantage may reflect the (mis)use of a decision-making heuristic that is helpful in other contexts that do rely on relative advantage, such as when there are scarce positions available in hiring. Future experiments could compare decision-making processes across contexts where positions are or are not scarce, looking at the extent to which decision-makers tend to export relative advantage considerations from one environment to the other. Our results may also have implications for decisions that indeed require the allocation of scarce positions, necessitating relative comparisons. Literature on choice theory (see, e.g., \cite{masatlioglu2012revealed}) suggests that in many cases decision-makers first construct a ``consideration set'' and then from that set exert the cognitive cost of ordering the elements. This cognitive process indicates that it is natural for people to first apply a filter, such as the Above-the-Bar/Below-the-Bar criteria, before making explicit ordering decisions. This suggests that the biases we observe may have implications for how consideration sets are formed. 

Although our experimental design involves only one time period, our
results may have dynamic implications. Consider a context where men in the candidate pool are more likely to be qualified than women in the candidate pool, due to direct or systemic discrimination or other forces. In this world, an evaluator may hold an (accurate) prior like one in our study, that 40\% of women are Above-the-Bar and 60\%
of men are Above-the-Bar. We can use the estimates from our setting to provide a rough approximation of how an evaluation process might unfold. If the evaluator's prior on differences across groups biases decisions as in our experiment, endogenous information acquisition and biased processing could lead to women having approximately a 40\% chance of being hired (classified as Above-the-Bar) and men having approximately a 65\% chance of being hired
(5pp higher than the prior). 

If performance is noisy, it might be difficult to elicit the true type of the candidates even after the candidate is hired. In the next time period, the evaluator might think that the outcomes of the previous time period provide the least biased prior, i.e., that 40\% of women are Above-the-Bar and 65\% of men are Above-the-Bar. This could set into motion a dynamic process through which beliefs of group differences are further exaggerated over time, fueling disparities in evaluations. While of course this is only a speculative back-of-the-envelope calculation, it suggests a channel through which gender gaps could be perpetuated and amplified. Endogenous information acquisition can lead to judgments that reinforce, or even expand, baseline gender gaps.

Our findings suggest that individuals belonging to disadvantaged groups may find themselves facing additional hurdles in order to convey their capabilities or merit effectively. This is particularly true in environments where the process of acquiring information is endogenous.

Making additional information on candidates available has been proposed as a way to reduce discrimination based upon biased beliefs \citep{Bohren_etal_2019}. Our results suggest a potential obstacle for this path, particularly in the case where decision-makers must actively choose to receive information. Evaluators in our study significantly under-acquire information on individuals, despite the signals being readily available, cheap, and informative. Future work should explore the reasons why individuals do not take advantage of this information. Money and time costs are arguably trivial in our setting. Risk preferences also seem to be an unlikely explanation. It could be that evaluators aim to avoid the cognitive cost of incorporating new information into their beliefs and/or that they under-estimate the value of this additional information.  

While we use the exogenous treatment as an important comparison group in our study, it also suggests a potentially promising intervention for remedying discrimination in these contexts. Our results suggest that requiring evaluators to acquire a certain amount of information about all candidates could reduce bias and increase the accuracy of judgments. While the practicality of this intervention might vary across settings of interest, future research should consider how reducing evaluator discretion around information acquisition impacts discrimination in the field.

\printbibliography[heading=bibintoc]

\newpage

\appendix

\counterwithin{figure}{section}
\counterwithin{table}{section}

\section{Supplemental Appendix}

\begin{figure}[hbtp] %
  \hfill
    \begin{center}  
     \caption{Example of Pie Charts Used to Provide Pool Information} \bigskip
      \label{fig:piechart}
      
       \centering	
              \centering	
      
       \includegraphics[scale=0.49]{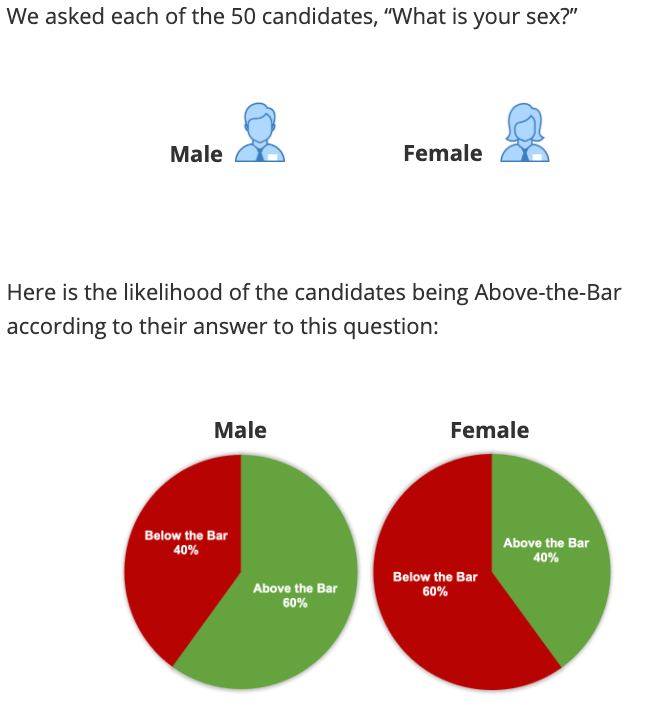}\\
       	\medskip
       
    \end{center}
  \hfill
	\bigskip %
\begin{minipage}{0.99\textwidth} %
{\footnotesize Note: The evaluator is assigned to a pool and then sees full info about the pool through a series of pie charts and a table indicating the proportion of candidates Above-the-Bar by each answer. These figures are illustrating example pie charts for the $60-40$ pool.\par}
\end{minipage}
\end{figure}

\begin{figure}[hbtp] %
  \hfill
    \begin{center}  
     \caption{Example of Summary Table Used to Provide Pool Information} \bigskip
      \label{fig:summarytable}
      
       \centering	
              \centering	
      
       \includegraphics[scale=0.5]{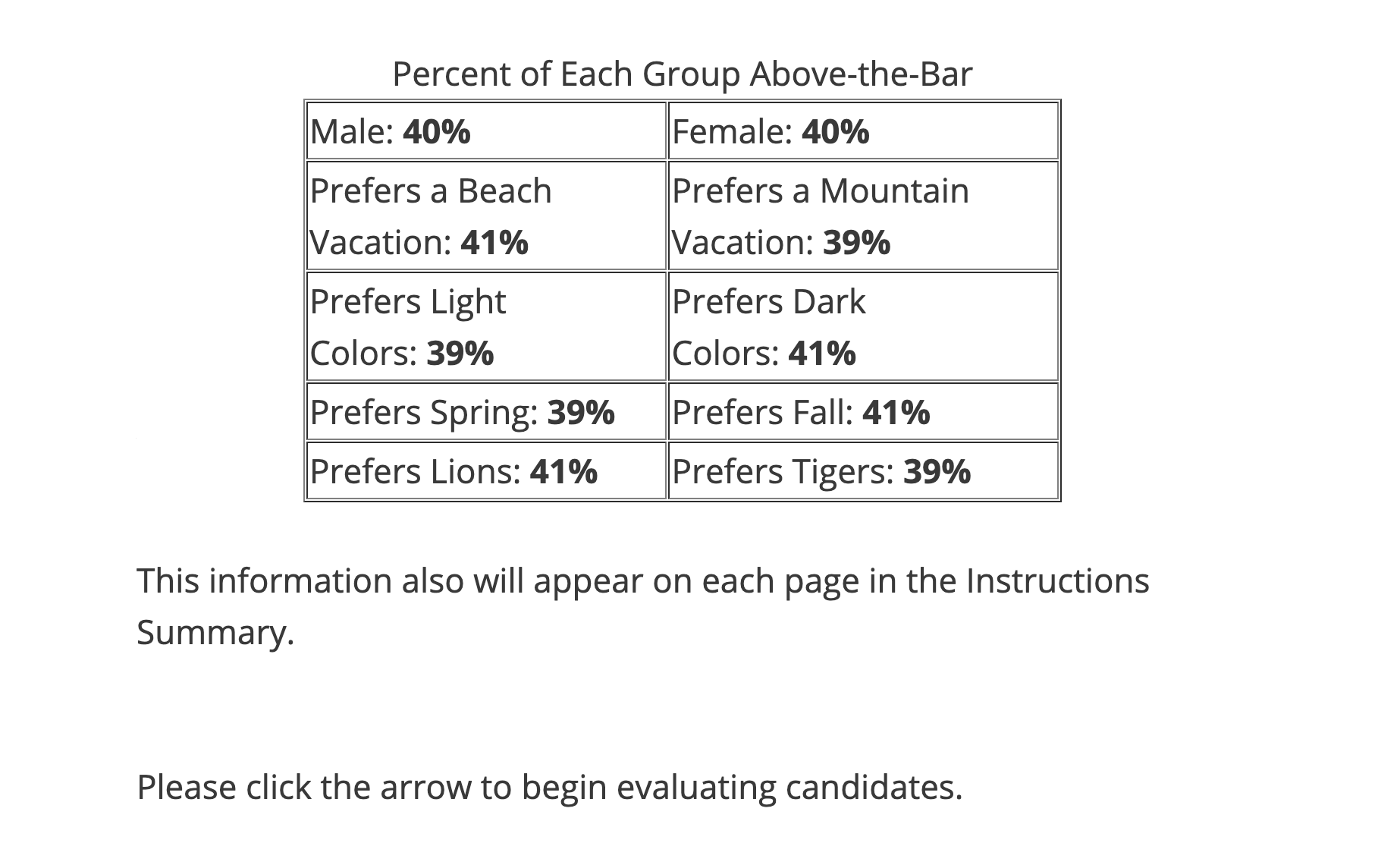}
    \end{center}
  \hfill
	\medskip %
\begin{minipage}{0.99\textwidth} %
{\footnotesize Note: The evaluator is assigned to a pool and then sees full info about the pool with a series of pie charts and a table indicating the proportion of candidates Above-the-Bar by each answer. This figure illustrates an example summary table for the $40-40$ pool.\par}
\end{minipage}
\end{figure}

\begin{figure}[hbtp] %
  \hfill
    \begin{center}  
     \caption{Example Candidate Profile}
      \label{fig:profile}
       \centering	
       \includegraphics[scale=0.5]{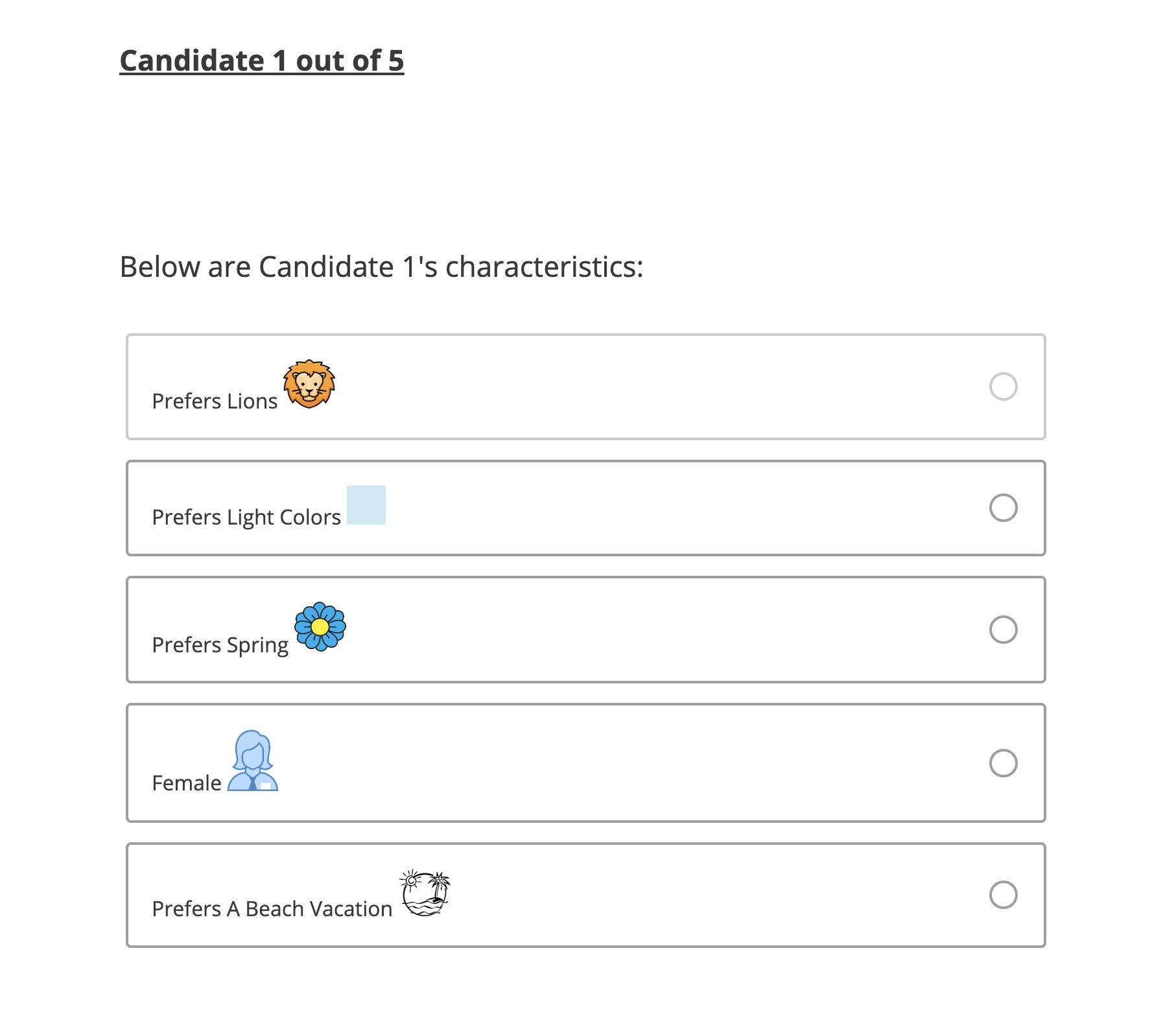}\\
    \end{center}
  \hfill
	\medskip %
\begin{minipage}{0.99\textwidth} %
{\footnotesize Note: This is a screenshot of an example candidate profile.\par}
\end{minipage}
\end{figure}

\begin{figure}[hbtp] %
  \hfill
    \begin{center}  
     \caption{Average number of signals acquired by pool -- Endogenous Treatment Only}
      \label{fig:signal_distr_pool}
       \centering	
              \centering	
       \includegraphics[scale=0.35]{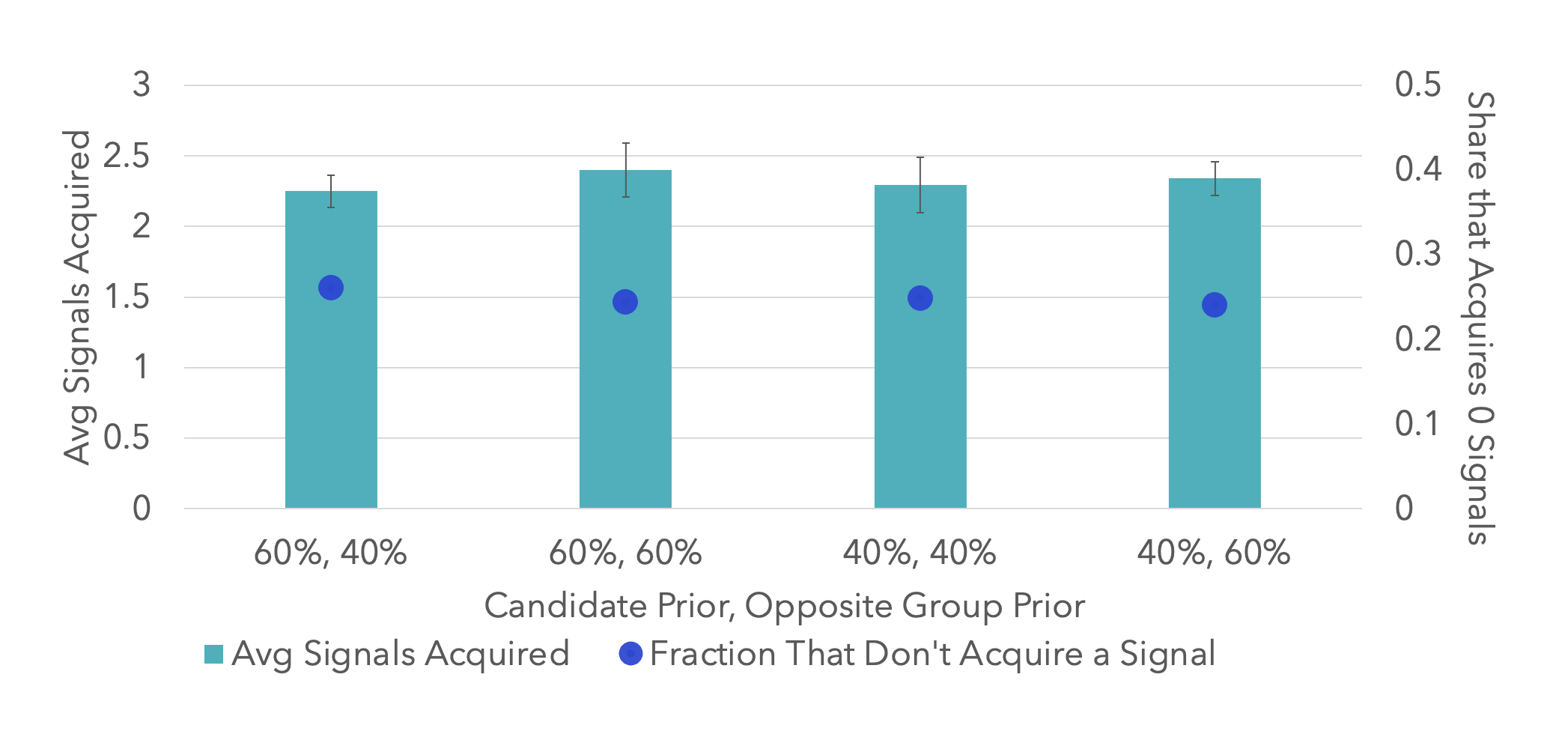}
    \end{center}
  \hfill
	\medskip %
\begin{minipage}{0.99\textwidth} %
{\footnotesize Note: This figure shows the average number of signals acquired by evaluators in the endogenous information acquisition treatment by the candidate pools, as well as the share of candidates for which an evaluator chose to receive 0 signals. \par}
\end{minipage}
\end{figure}

\begin{figure}[hbtp] %
  \hfill
    \begin{center}  
     \caption{Distribution of number of signals acquired by pool -- Endogenous Treatment Only}
      \label{fig:signal_distr}  \bigskip
       \centering	
              \centering	
       \includegraphics[scale=0.5]{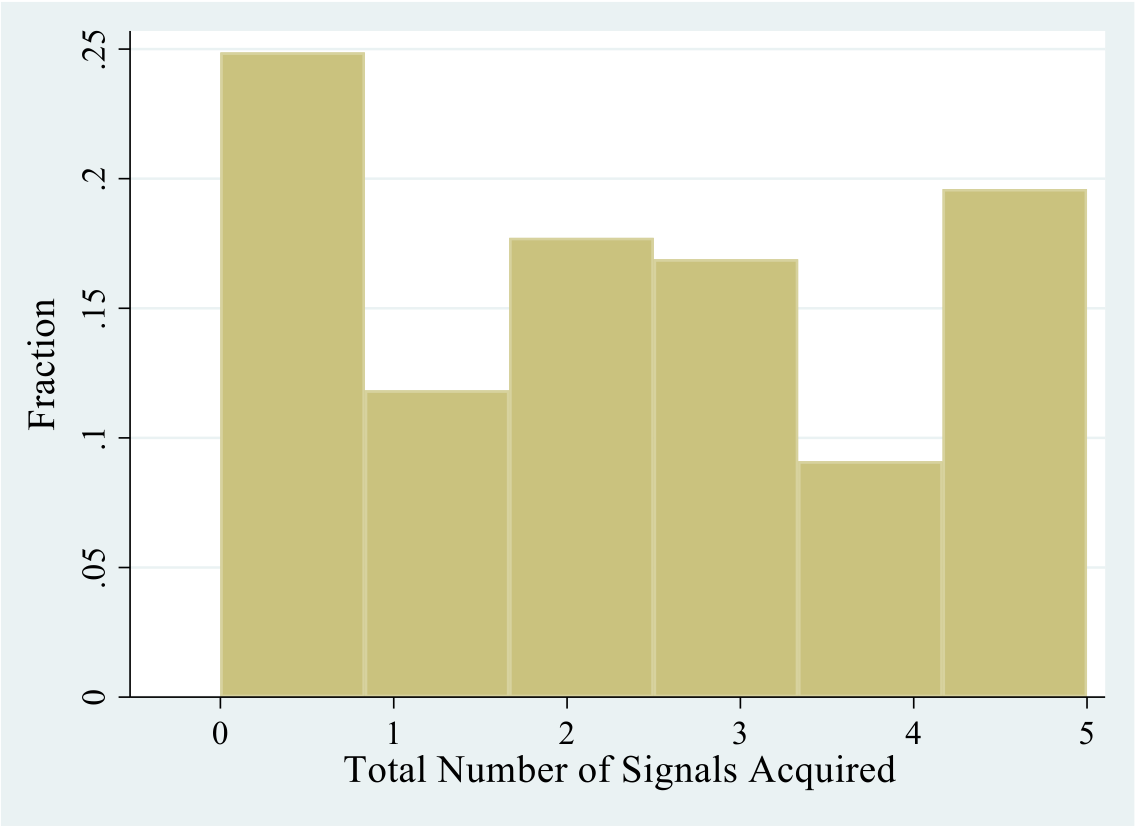}
    \end{center}
  \hfill
	\bigskip %
\begin{minipage}{0.99\textwidth} %
{\footnotesize Note: This is a histogram of the total number of signals acquired for each candidate-evaluator pair in the endogenous treatment.\par}
\end{minipage}
\end{figure}

\begin{figure}[hbtp] %
  \hfill
    \begin{center}  
     \caption{Number of Signals Bought Across Candidates -- Endogenous Treatment Only}
      \label{fig:signal_distr_order}  \bigskip 
       \centering	
              \centering	
       \includegraphics[scale=0.5]{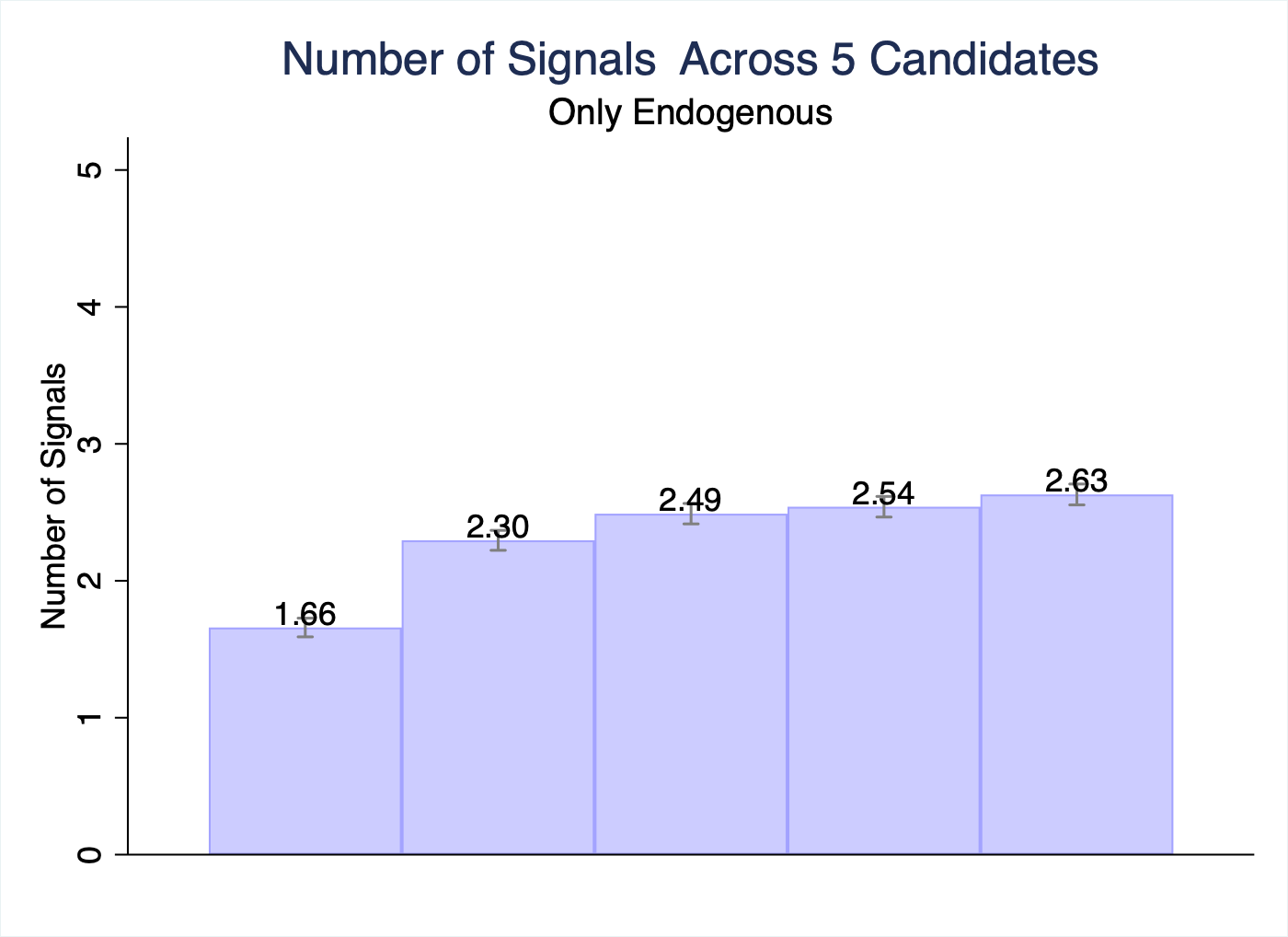}
    \end{center}
  \hfill
	\medskip %
\begin{minipage}{0.99\textwidth} %
{\footnotesize Note: This figure shows the average number of signals bought by the evaluators for each of the 5 candidates they evaluate by candidate order.\par}
\end{minipage}
\end{figure}

\begin{figure}[hp] %
  \hfill
    \begin{center}  
     \caption{Speed of Classifications: Robustness Excluding the First Round}
      \label{fig:classification_timing_robust} \bigskip
      
       \centering	
       Panel A: High-Prior Candidates\\
       \includegraphics[scale=0.30]{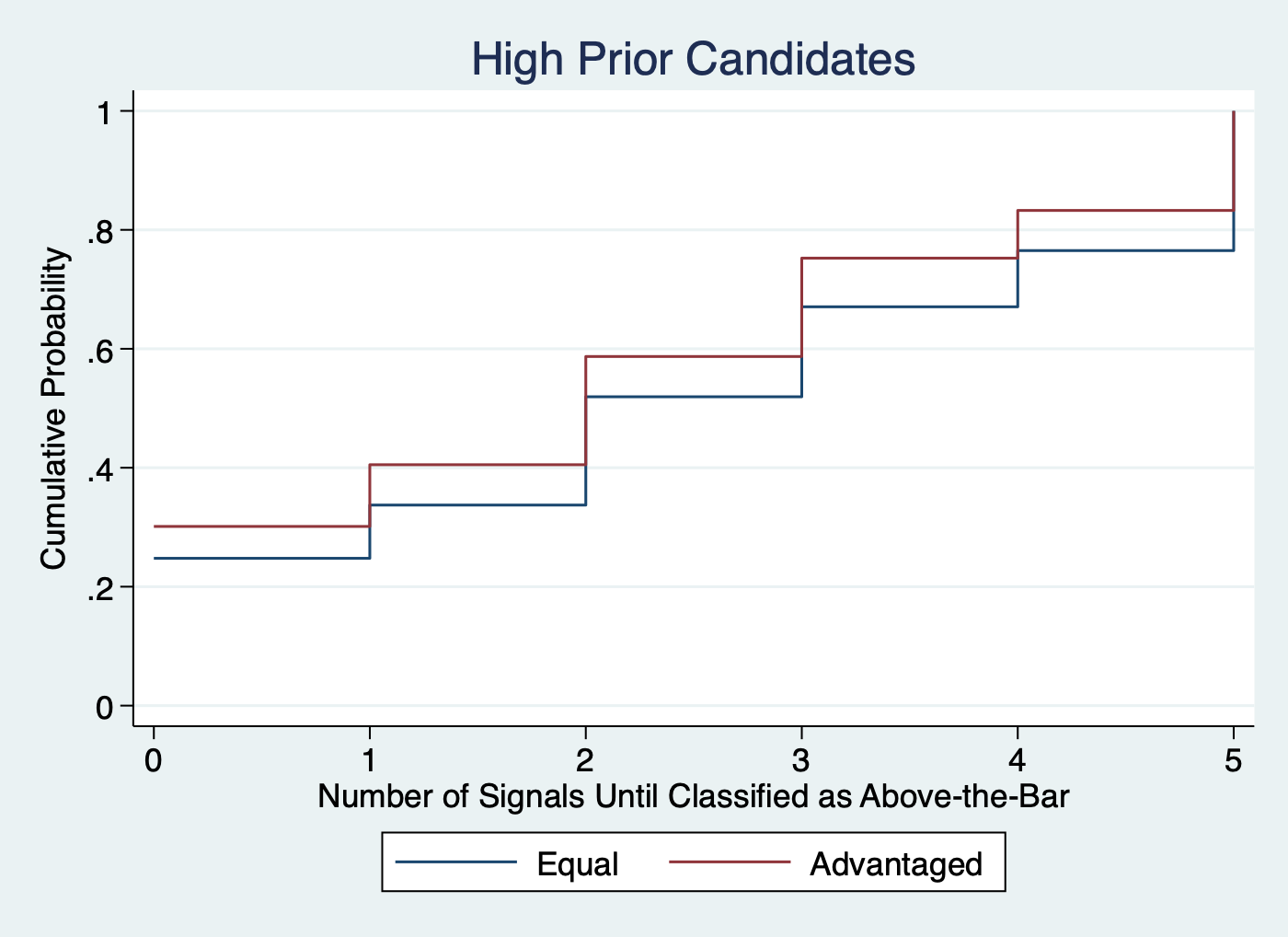}
       \includegraphics[scale=0.30]{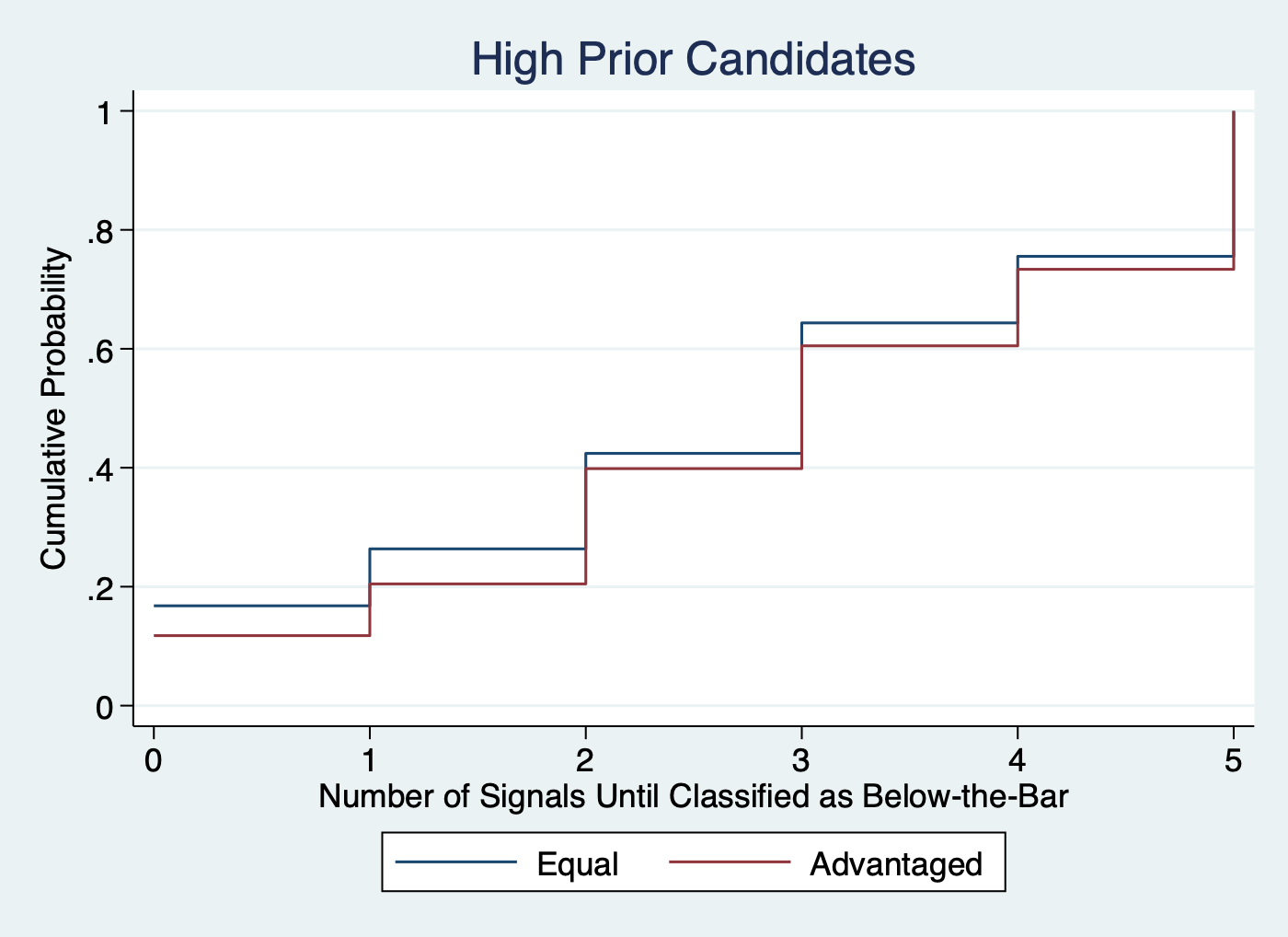}\\
       	\bigskip
            \bigskip
       Panel B: Low-Prior Candidates\\
       \includegraphics[scale=0.30]{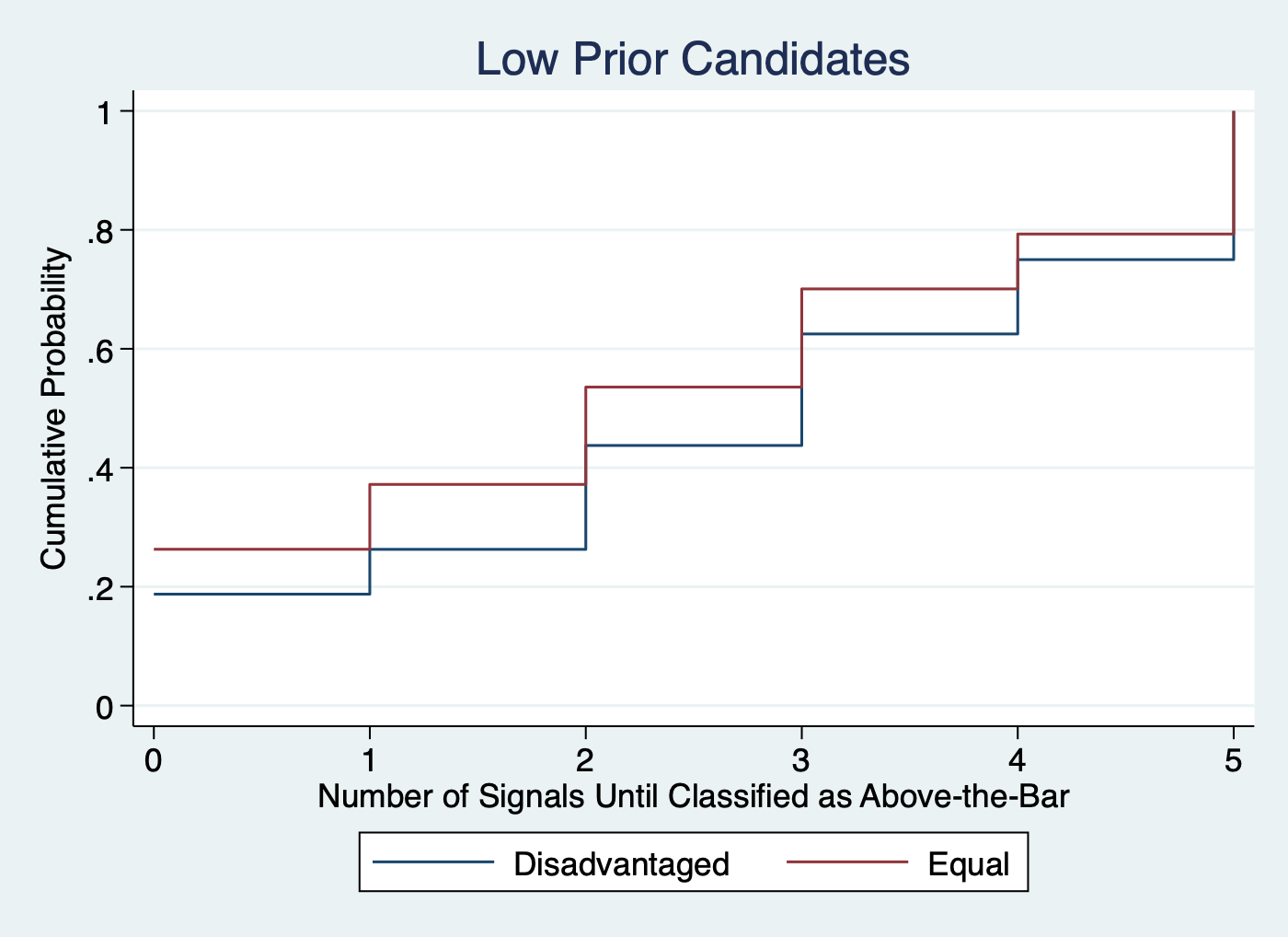}       \includegraphics[scale=0.30]{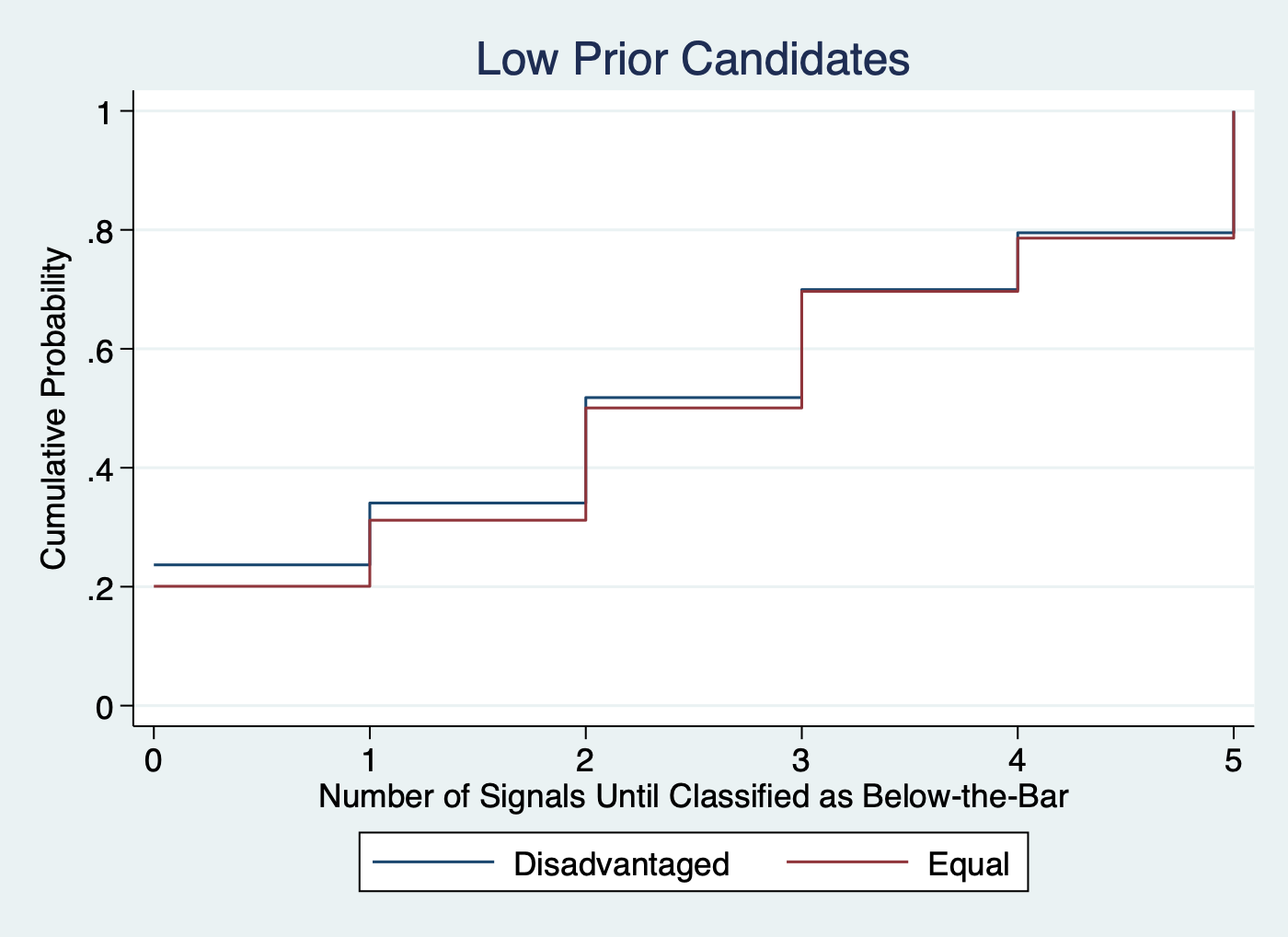}\\
    \end{center}
  \hfill
	\medskip %
\begin{minipage}{0.99\textwidth} %
{\footnotesize Note: This figure replicates Figure \refFromApp{fig:classification_timing} dropping observations from the first candidate for each evaluator. Panel~A presents the distribution of the number of total signals viewed before a high-prior candidate is classified as Above-the-Bar (left) or Below-the-Bar (right). Panel B presents the same analysis for low-prior candidates. When we include round 1 in Figure \refFromApp{fig:classification_timing}, 61\% of equal and 67\% of  advantaged candidates were classified as above the bar among the high prior candidates. When we drop round 1 these rates are 60\% and 65\% respectively. For low prior candidates, including round 1, 46\% equal and 40\% disadvantaged candidates are classified as Above-the-Bar. When we drop round 1, these rates remain mostly unchanged. \par}
\end{minipage}
\end{figure}

\begin{figure}[hbtp] %
  \hfill
    \begin{center}  
     \caption{Under-Acquisition of Information Relative to a Bayesian Benchmark}
      \label{fig:underacquisition_vs_bayesian}  \bigskip
       \centering	
              \centering	
       \includegraphics[scale=0.7]{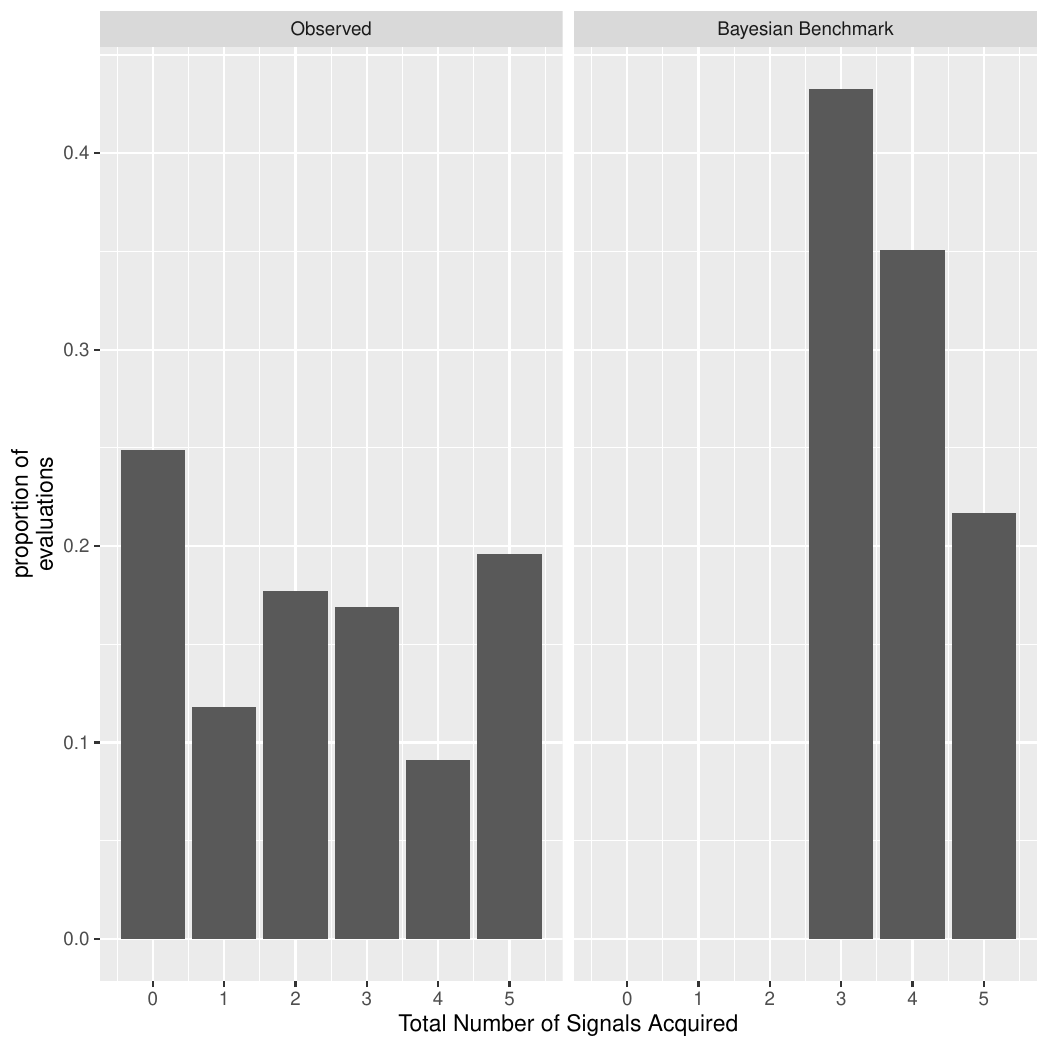}
    \end{center}
  \hfill
	\bigskip %
\begin{minipage}{0.99\textwidth} %
{\footnotesize Note: The first panel displays the frequency of acquiring 0 - 5 signals in our endogenous treatment. The second panel displays the frequency with which a Bayesian decision-maker would acquire 0 - 5 signals in those same instances.\par}
\end{minipage}
\end{figure}

	\begin{table}[ht]
		\begin{center}
			\caption{\label{t:sliders_two_signals_by_prior_full}Updating After Signals by Splitting by Group Prior -- OLS Predicting Beliefs After...}\medskip
			\small
		Panel A: Only Low-Prior Sample:
			\makebox[\textwidth][c]{\input{Results_Peri/Tables/sliders_two_signals_low_prior.tex}}
		Panel B: Only High-Prior Sample:			
			\makebox[\textwidth][c]{\input{Results_Peri/Tables/sliders_two_signals_high_prior.tex}}
		\end{center}
		\footnotesize
Notes: Standard errors are reported in parentheses and clustered at the evaluator level. *~p$<$0.10, **~p$<$0.05, ***~p$<$0.01. This table repeats the analysis in the main text Table \refFromApp{t:sliders_two_signals_by_prior_exog} by including both exogenous and endogenous information acquisition groups. The dependent variable in the first column is the likelihood of the candidate being Above-the-Bar chosen by the evaluator in the first slider before receiving any signals. The following columns report the regressions on the likelihoods of being Above-the-Bar chosen on the sliders after receiving the corresponding first two signals in each column for the corresponding subsample. All regressions include round fixed effects and candidate advantage by filler characteristics.
	\end{table}

	\begin{table}[t]
		\begin{center}
			\caption{\label{t:first_final_slider_and_eval_robust} The Impact of Relative Advantage on Assessments -- Robustness Excluding the First Round}\medskip
                \medskip
                OLS Predicting Initial Beliefs, Final Beliefs, and Final Evaluations \medskip
                
			\small
			\makebox[\textwidth][c]{\input{Results_Peri/Tables/first_final_slider_and_eval_robust}}
		\end{center}
		\footnotesize
Notes: This table tests the robustness of Table \refFromApp{t:first_final_slider_and_eval} to the exclusion of the first round. Standard errors are reported in parentheses and clustered at the evaluator level. *~p$<$0.10, **~p$<$0.05, ***~p$<$0.01. In the first two columns, the dependent variable is the likelihood of the candidate being Above-the-Bar chosen by the evaluator in the first slider. Columns \num{3} and \num{4} report the regressions on the likelihood of the candidate being Above-the-Bar chosen by the evaluator in the last slider before making their final decision. The last two columns report the regressions of the indicator variable taking value 1, if the final evaluation of the candidate was Above-the-Bar, and 0 otherwise. All regressions include round fixed effects and candidate advantage by filler characteristics.
	\end{table}

	\begin{table}[ht]
		\begin{center}
			\caption{\label{t:info_acquisition_robust} Selection Into Acquiring Information -- Robustness Excluding the First Round} \medskip
   
                OLS Predicting Initial Likelihoods and Final Evaluations by Subsample\medskip
                
			\small
			\makebox[\textwidth][c]{\input{Results_Peri/Tables/info_acquisition_robust.tex}}
		\end{center}
		\footnotesize
Notes: This table tests the robustness of Table \refFromApp{t:info_acquisition} to the exclusion of the first round. Standard errors are reported in parentheses and clustered at the evaluator level. *~p$<$0.10, **~p$<$0.05, ***~p$<$0.01. The dependent variable in all columns is the likelihood of the candidate being Above-the-Bar chosen by the evaluator in the first slider before receiving any signals. The first column includes the sample of observations where the evaluation was made without acquiring any information. The second column includes the observations where evaluation was made after receiving at least one signal. The third and fourth columns pool all observations and estimate the effect of receiving any signals on the initial likelihood assessment and final evaluation, respectively. All regressions include controls for round fixed effects and candidate advantage by filler characteristics. All three columns include endogenous treatment only. 
	\end{table}

\begin{table}[htp]
		\begin{center}
			\caption{\label{t:whentoacquire_robust} Endogenous Information Acquisition -- Robustness Excluding the First Round}\bigskip
            OLS Predicting the Decision to Get More Info After... \medskip
            
			\small
		Panel A: Overall 
                \medskip
			\makebox[\textwidth][c]{\input{Results_Peri/Tables/when_to_acquire_info_robust}}
            \bigskip
            \bigskip
  
            Panel B: Comparing Pools with Equal Versus Contrasting Pools  
                \medskip			
			\makebox[\textwidth][c]{\input{Results_Peri/Tables/when_to_acquire_info_equal_groups_robust.tex}}
		\end{center}
		\footnotesize
Notes: This table tests the robustness of the Table \refFromApp{t:whentoacquire} to the exclusion of the first round. Standard errors are reported in parentheses and clustered at the evaluator level. *~p$<$0.10, **~p$<$0.05, ***~p$<$0.01. All regressions include controls for relative advantage, male candidate indicator, group prior, initial likelihood assessment, round fixed effects, and candidate advantage by filler characteristics. All columns include endogenous treatment only. The dependent variable is an indicator variable for deciding to receive a signal after the corresponding signal in each column for the corresponding subsample. The last column pools all observations in each round in the long format.
	\end{table}

	\begin{table}[t]
		\begin{center}
			\caption{\label{t:first_final_slider_and_eval_gender_int} The Impact of Relative Advantage on Assessments: Gender Interaction}\medskip
                \medskip
                OLS Predicting Initial Beliefs, Final Beliefs, and Final Evaluations \medskip
                
			\small
			\makebox[\textwidth][c]{\input{Results_Peri/Tables/first_final_slider_and_eval_gender_int.tex}}
		\end{center}
		\footnotesize
Notes: This table expands the Table \refFromApp{t:first_final_slider_and_eval} adding the interaction of candidate gender and relative advantage. Standard errors are reported in parentheses and clustered at the evaluator level. *~p$<$0.10, **~p$<$0.05, ***~p$<$0.01. In the first three columns, the dependent variable is the likelihood of the candidate being Above-the-Bar chosen by the evaluator in the first slider. Columns \num{4} and \num{5} and \num{6} report the regressions on the likelihood of the candidate being Above-the-Bar chosen by the evaluator in the last slider before making their final decision. The last three columns report the regressions of the indicator variable taking value 1, if the final evaluation of the candidate was Above-the-Bar, and 0 otherwise. All regressions include round fixed effects and candidate advantage by filler characteristics.
	\end{table}

	\begin{table}[htp]
		\begin{center}
			\caption{\label{t:sliders_two_signals_by_prior_exog_bayes} Biased Belief Updating - Controlling for the Bayesian Posterior}\medskip
            OLS Predicting Beliefs After Given Signal History \medskip
            
			\small
		Panel A: Only Low-Prior Sample (Exogenous):
			\makebox[\textwidth][c]{\input{Results_Peri/Tables/sliders_two_signals_low_prior_exog_bayes.tex}}
		Panel B: Only High-Prior Sample (Exogenous):			
			\makebox[\textwidth][c]{\input{Results_Peri/Tables/sliders_two_signals_high_prior_exog_bayes.tex}}
		\end{center}
		\footnotesize  
Notes: This table augments the specification of Table \refFromApp{t:sliders_two_signals_by_prior_exog} by controlling for the Bayesian posterior an individual would hold after seeing the signal given their own (biased) belief prior to seeing the corresponding signal. This would correspond to controlling for the Bayes update after seeing signal 1 in the first two columns and the Bayes update after seeing the signal 2 in the remaining columns. Standard errors are reported in parentheses and clustered at the evaluator level. *~p$<$0.10, **~p$<$0.05, ***~p$<$0.01. This table reports the OLS results predicting beliefs about the likelihood of being Above-the-Bar for only the evaluators in the exogenous treatment. We split the sample by candidate prior and run the analysis on the subsamples including only low-prior candidates and only high-prior candidates in Panel A and B, respectively. The dependent variable in the first two columns is the evaluator's believed likelihood of the candidate being Above-the-Bar after receiving the first signal. The following columns report the regressions on the believed likelihood of the candidate being Above-the-Bar after receiving two signals. All regressions include round fixed effects and candidate advantage by filler characteristics.
	\end{table}

	\begin{table}[t]
		\begin{center}
			\caption{\label{t:first_final_slider_and_eval_rob_dur} The Impact of Relative Advantage on Assessments - Excluding the Evaluators Taking too Short or too Long}\medskip
                \medskip
                OLS Predicting Initial Beliefs, Final Beliefs, and Final Evaluations \medskip
                
			\small
			\makebox[\textwidth][c]{\input{Results_Peri/Tables/first_final_slider_and_eval_dur_rob.tex}}
		\end{center}
		\footnotesize
Notes: This table tests the robustness of Table \refFromApp{t:first_final_slider_and_eval} on the subsample of evaluators where we excluded the evaluators who took too short (bottom 5\%) or too long (top 5\%) to complete the study. Standard errors are reported in parentheses and clustered at the evaluator level. *~p$<$0.10, **~p$<$0.05, ***~p$<$0.01. In the first two columns, the dependent variable is the likelihood of the candidate being Above-the-Bar chosen by the evaluator in the first slider. Columns \num{3} and \num{4} report the regressions on the likelihood of the candidate being Above-the-Bar chosen by the evaluator in the last slider before making their final decision. The last two columns report the regressions of the indicator variable taking value 1, if the final evaluation of the candidate was Above-the-Bar, and 0 otherwise. All regressions include round fixed effects and candidate advantage by filler characteristics.
	\end{table}

	\begin{table}[ht]
		\begin{center}
			\caption{\label{t:info_acquisition_dur_rob} Selection Into Acquiring Information - Excluding the Evaluators Taking too Short or too Long} \medskip
   
                OLS Predicting Initial Likelihoods and Final Evaluations by Subsample\medskip
                
			\small
			\makebox[\textwidth][c]{\input{Results_Peri/Tables/info_acquisition_dur_rob.tex}}
		\end{center}
		\footnotesize
Notes: This table tests the robustness of Table \refFromApp{t:info_acquisition} on the subsample of evaluators where we excluded the evaluators who took too short (bottom 5\%) or too long (top 5\%) to complete the study. Standard errors are reported in parentheses and clustered at the evaluator level. *~p$<$0.10, **~p$<$0.05, ***~p$<$0.01. The dependent variable in all columns is the likelihood of the candidate being Above-the-Bar chosen by the evaluator in the first slider before receiving any signals. The first column includes the sample of observations where the evaluation was made without acquiring any information. The second column includes the observations where evaluation was made after receiving at least one signal. The third and fourth columns pool all observations and estimate the effect of receiving any signals on the initial likelihood assessment and final evaluation, respectively. All regressions include controls for round fixed effects and candidate advantage by filler characteristics. All three columns include endogenous treatment only. 
	\end{table}

	\begin{table}[htp]
		\begin{center}
			\caption{\label{t:whentoacquire_dur_rob} Endogenous Information Acquisition - Excluding the Evaluators Taking too Short or too Long}\bigskip
            OLS Predicting the Decision to Get More Info After... \medskip
            
			\small
		Panel A: Overall 
                \medskip
			\makebox[\textwidth][c]{\input{Results_Peri/Tables/when_to_acquire_info_dur_rob.tex}}
            \bigskip
            \bigskip
  
            Panel B: Comparing Pools with Equal Versus Contrasting Pools  
                \medskip			
			\makebox[\textwidth][c]{\input{Results_Peri/Tables/when_to_acquire_info_equal_groups_dur_rob.tex}}
		\end{center}
		\footnotesize
Notes: This table tests the robustness of Table \refFromApp{t:whentoacquire} on the subsample of evaluators where we excluded the evaluators who took too short (bottom 5\%) or too long (top 5\%) to complete the study. Standard errors are reported in parentheses and clustered at the evaluator level. *~p$<$0.10, **~p$<$0.05, ***~p$<$0.01. All regressions include controls for relative advantage, male candidate indicator, group prior, initial likelihood assessment, round fixed effects, and candidate advantage by filler characteristics. All three columns include endogenous treatment only. The dependent variable is an indicator variable for deciding to receive a signal after the corresponding signal in each column for the corresponding subsample. The last column pools all observations in each round in the long format.
	\end{table}

	\begin{table}[htp]
		\begin{center}
			\caption{\label{t:sliders_two_signals_by_prior_exog_dur_rob} Biased Belief Updating - Excluding the Evaluators Taking too Short or too Long}\medskip
            OLS Predicting Beliefs After Given Signal History \medskip
            
			\small
		Panel A: Only Low-Prior Sample (Exogenous):
			\makebox[\textwidth][c]{\input{Results_Peri/Tables/sliders_two_signals_low_prior_exog_dur_rob.tex}}
		Panel B: Only High-Prior Sample (Exogenous):			
			\makebox[\textwidth][c]{\input{Results_Peri/Tables/sliders_two_signals_high_prior_exog_dur_rob.tex}}
		\end{center}
		\footnotesize
Notes: This table tests the robustness of Table \refFromApp{t:sliders_two_signals_by_prior_exog} on the subsample of evaluators where we excluded the evaluators who took too short (bottom 5\%) or too long (top 5\%) to complete the study. Standard errors are reported in parentheses and clustered at the evaluator level. *~p$<$0.10, **~p$<$0.05, ***~p$<$0.01. This table reports the OLS results predicting beliefs about the likelihood of being Above-the-Bar for only the evaluators in the exogenous treatment. We split the sample by candidate prior and run the analysis on the subsamples including only low-prior candidates and only high-prior candidates in Panel A and B, respectively. The dependent variable in the first two columns is the evaluator's believed likelihood of the candidate being Above-the-Bar after receiving the first signal. The following columns report the regressions on the believed likelihood of the candidate being Above-the-Bar after receiving two signals. All regressions include round fixed effects and candidate advantage by filler characteristics.
	\end{table}

	\begin{table}[htp]
		\begin{center}
			\caption{\label{t:sliders_beyond_two_signals_dur_rob} Biased Belief Updating Beyond Two Signals  (Exogenous Treatment Only) - Excluding the Evaluators Taking too Short or too Long}\medskip
                OLS Predicting Beliefs After Given Signal History  \medskip
                
			\small
			\makebox[\textwidth][c]{\input{Results_Peri/Tables/sliders_beyond_two_signals_dur_rob.tex}}
		\end{center}
		\footnotesize
Notes: This table tests the robustness of Table \refFromApp{t:sliders_beyond_two_signals} on the subsample of evaluators where we excluded the evaluators who took too short (bottom 5\%) or too long (top 5\%) to complete the study. Standard errors are reported in parentheses and clustered at the evaluator level. *~p$<$0.10, **~p$<$0.05, ***~p$<$0.01. The dependent variable is the perceived likelihood of the candidate being Above-the-Bar after receiving $t$ signals in Columns \num{1} and \num{3} for low and high-prior candidates, respectively. Columns \num{2} and \num{4} include only the last round, the final evaluation round, where the dependent variable is whether the candidate is classified as Above-the-Bar for low and high-prior candidates, respectively. All columns include observations from the exogenous information acquisition treatment only. All regressions include round fixed effects and candidate advantage by filler characteristics as well as initial likelihood chosen by the participant before any signals.
	\end{table}

\end{document}

%% file: Results_Peri/Tables/first_final_slider_and_eval.tex
{
\def\sym#1{\ifmmode^{#1}\else\(^{#1}\)\fi}
\begin{tabular}{l*{6}{c}}
\toprule
                    &\multicolumn{2}{c}{Initial Slider}&\multicolumn{2}{c}{Final Slider}&\multicolumn{2}{c}{Final Eval} \\\cmidrule(lr){2-3}\cmidrule(lr){4-5}\cmidrule(lr){6-7}
                    &\multicolumn{1}{c}{(1)}   &\multicolumn{1}{c}{(2)}   &\multicolumn{1}{c}{(3)}   &\multicolumn{1}{c}{(4)}   &\multicolumn{1}{c}{(5)}   &\multicolumn{1}{c}{(6)}   \\
\midrule
Relative Advantage  &       0.157***&       0.164***&      0.0779***&       0.105***&       0.148***&       0.267***\\
                    &    (0.0177)   &    (0.0208)   &    (0.0272)   &    (0.0323)   &    (0.0398)   &    (0.0499)   \\
\addlinespace
Male Candidate      &      0.0105   &       0.128   &       0.145   &       0.440   &       0.267   &       0.342   \\
                    &     (0.292)   &     (0.419)   &     (0.531)   &     (0.705)   &     (0.789)   &     (1.141)   \\
\addlinespace
Group Prior         &       0.456***&       0.456***&       0.550***&       0.550***&       0.778***&       0.778***\\
                    &    (0.0282)   &    (0.0282)   &    (0.0393)   &    (0.0393)   &    (0.0558)   &    (0.0558)   \\
\addlinespace
Exogenous           &       0.242   &       0.357   &      -2.726***&      -2.438***&      -3.854***&      -3.808***\\
                    &     (0.368)   &     (0.464)   &     (0.531)   &     (0.758)   &     (0.757)   &     (1.083)   \\
\addlinespace
Rel Adv x Exog      &               &     -0.0145   &               &     -0.0524   &               &      -0.235***\\
                    &               &    (0.0217)   &               &    (0.0377)   &               &    (0.0565)   \\
\addlinespace
Male x Exog         &               &      -0.230   &               &      -0.578   &               &      -0.103   \\
                    &               &     (0.585)   &               &     (1.060)   &               &     (1.571)   \\
\midrule
Round FE            &         Yes   &         Yes   &         Yes   &         Yes   &         Yes   &         Yes   \\
Advantage by Fillers&         Yes   &         Yes   &         Yes   &         Yes   &         Yes   &         Yes   \\
No. Observations    &       16000   &       16000   &       16000   &       16000   &       16000   &       16000   \\
No. Clusters        &        3200   &        3200   &        3200   &        3200   &        3200   &        3200   \\
Adj. R2             &       0.168   &       0.168   &      0.0411   &      0.0411   &      0.0409   &      0.0419   \\
\bottomrule
\end{tabular}
}

%% file: Results_Peri/Tables/info_acquisition.tex
{
\def\sym#1{\ifmmode^{#1}\else\(^{#1}\)\fi}
\begin{tabular}{l*{4}{c}}
\toprule
                    &\multicolumn{1}{c}{Immediate Evaluation}&\multicolumn{1}{c}{Got More Info}&\multicolumn{2}{c}{Pooled}     \\\cmidrule(lr){2-2}\cmidrule(lr){3-3}\cmidrule(lr){4-5}
                    &\multicolumn{1}{c}{Initial Slider}&\multicolumn{1}{c}{Initial Slider}&\multicolumn{1}{c}{Initial Slider}&\multicolumn{1}{c}{Final Eval}\\
\midrule
Relative Advantage  &        0.28***&        0.14***&        0.24***&        0.78***\\
                    &     (0.056)   &     (0.025)   &     (0.043)   &     (0.085)   \\
\addlinespace
Male Candidate      &       -0.77   &        0.41   &       -0.78   &       -3.26   \\
                    &      (1.04)   &      (0.42)   &      (1.05)   &      (2.19)   \\
\addlinespace
Group Prior         &        0.34***&        0.46***&        0.43***&        0.78***\\
                    &     (0.083)   &     (0.040)   &     (0.038)   &     (0.079)   \\
\addlinespace
Got More Info       &               &               &       -4.89***&       -13.2***\\
                    &               &               &      (0.84)   &      (1.75)   \\
\addlinespace
Got More Info x Rel Adv&               &               &      -0.093** &       -0.70***\\
                    &               &               &     (0.041)   &     (0.088)   \\
\addlinespace
Got More Info x Male&               &               &        1.16   &        4.43*  \\
                    &               &               &      (1.12)   &      (2.53)   \\
\midrule
Round FE            &         Yes   &         Yes   &         Yes   &         Yes   \\
Advantage by Fillers&         Yes   &         Yes   &         Yes   &         Yes   \\
No. Observations    &        1982   &        5988   &        7970   &        7970   \\
No. Clusters        &         717   &        1424   &        1594   &        1594   \\
Adj. R2             &        0.16   &        0.18   &        0.18   &       0.070   \\
\bottomrule
\end{tabular}
}

%% file: Results_Peri/Tables/when_to_acquire_info.tex
{
\def\sym#1{\ifmmode^{#1}\else\(^{#1}\)\fi}
\begin{tabular}{l*{5}{c}}
\toprule
                    &\multicolumn{1}{c}{1st Signal}&\multicolumn{1}{c}{2nd Signal}&\multicolumn{1}{c}{3rd Signal}&\multicolumn{1}{c}{4th Signal}&\multicolumn{1}{c}{Pooled-Last Signal}\\
\midrule
Confirming Signal   &      -0.031***&      -0.079***&      -0.016   &     -0.0068   &      -0.027***\\
                    &    (0.0092)   &     (0.013)   &     (0.016)   &     (0.020)   &    (0.0053)   \\
\midrule
Round FE            &         Yes   &         Yes   &         Yes   &         Yes   &         Yes   \\
Advantage by Fillers&         Yes   &         Yes   &         Yes   &         Yes   &         Yes   \\
No. Observations    &        5988   &        5045   &        3633   &        2287   &       18515   \\
No. Clusters        &        1424   &        1287   &        1138   &         826   &        1424   \\
Adj. R2             &       0.043   &       0.018   &      0.0029   &     0.00012   &      0.0050   \\
\bottomrule
\end{tabular}
}

%% file: Results_Peri/Tables/when_to_acquire_info_equal_groups.tex
{
\def\sym#1{\ifmmode^{#1}\else\(^{#1}\)\fi}
\begin{tabular}{l*{5}{c}}
\toprule
                    &\multicolumn{1}{c}{1st Signal}&\multicolumn{1}{c}{2nd Signal}&\multicolumn{1}{c}{ 3rd Signal}&\multicolumn{1}{c}{4th Signal}&\multicolumn{1}{c}{Pooled-Last Signal}\\
\midrule
Confirming Signal   &      -0.054***&       -0.10***&      -0.025   &      -0.031   &      -0.042***\\
                    &     (0.013)   &     (0.019)   &     (0.024)   &     (0.030)   &    (0.0078)   \\
\addlinespace
Equal Groups        &      -0.026   &      -0.014   &      0.0081   &       0.013   &     -0.0095   \\
                    &     (0.016)   &     (0.021)   &     (0.028)   &     (0.033)   &    (0.0082)   \\
\addlinespace
Equal Groups x Conf. Signal&       0.046** &       0.047*  &       0.019   &       0.044   &       0.029***\\
                    &     (0.018)   &     (0.025)   &     (0.032)   &     (0.040)   &     (0.011)   \\
\midrule
Round FE            &         Yes   &         Yes   &         Yes   &         Yes   &         Yes   \\
Advantage by Fillers&         Yes   &         Yes   &         Yes   &         Yes   &         Yes   \\
No. Observations    &        5988   &        5045   &        3633   &        2287   &       18515   \\
No. Clusters        &        1424   &        1287   &        1138   &         826   &        1424   \\
Adj. R2             &       0.043   &       0.018   &      0.0029   &      0.0017   &      0.0052   \\
\bottomrule
\end{tabular}
}

%% file: Results_Peri/Tables/sliders_two_signals_low_prior_exog.tex
{
\def\sym#1{\ifmmode^{#1}\else\(^{#1}\)\fi}
\begin{tabular}{l*{5}{c}}
\toprule
                    &\multicolumn{1}{c}{1 Positive}&\multicolumn{1}{c}{1 Negative}&\multicolumn{1}{c}{2 Positives}&\multicolumn{1}{c}{2 Mixed}&\multicolumn{1}{c}{2 Negatives}\\
\midrule
Relative Advantage  &       0.111***&    -0.00154   &      0.0522   &      0.0308   &     -0.0410   \\
                    &    (0.0351)   &    (0.0356)   &    (0.0644)   &    (0.0347)   &    (0.0494)   \\
\addlinespace
Male Candidate      &       0.766   &      -0.259   &       0.417   &       0.380   &      -0.909   \\
                    &     (0.601)   &     (0.625)   &     (1.001)   &     (0.634)   &     (0.879)   \\
\addlinespace
Initial Likelihood  &       0.699***&       0.614***&       0.481***&       0.634***&       0.501***\\
                    &    (0.0250)   &    (0.0406)   &    (0.0383)   &    (0.0322)   &    (0.0434)   \\
\midrule
Round FE            &         Yes   &         Yes   &         Yes   &         Yes   &         Yes   \\
Advantage by Fillers&         Yes   &         Yes   &         Yes   &         Yes   &         Yes   \\
No. Observations    &        1809   &        2166   &        1043   &        1543   &        1389   \\
No. Clusters        &         959   &        1042   &         719   &         894   &         864   \\
Adj. R2             &       0.540   &       0.402   &       0.263   &       0.465   &       0.244   \\
\bottomrule
\end{tabular}
}

%% file: Results_Peri/Tables/sliders_two_signals_high_prior_exog.tex
{
\def\sym#1{\ifmmode^{#1}\else\(^{#1}\)\fi}
\begin{tabular}{l*{5}{c}}
\toprule
                    &\multicolumn{1}{c}{1 Positive}&\multicolumn{1}{c}{1 Negative}&\multicolumn{1}{c}{2 Positives}&\multicolumn{1}{c}{2 Mixed}&\multicolumn{1}{c}{2 Negatives}\\
\midrule
Relative Advantage  &    -0.00456   &      0.0936** &      0.0207   &      0.0748** &       0.175***\\
                    &    (0.0251)   &    (0.0404)   &    (0.0418)   &    (0.0326)   &    (0.0644)   \\
\addlinespace
Male Candidate      &      -0.241   &       0.409   &      -0.182   &      0.0725   &       1.271   \\
                    &     (0.455)   &     (0.705)   &     (0.724)   &     (0.586)   &     (1.130)   \\
\addlinespace
Initial Likelihood  &       0.722***&       0.665***&       0.544***&       0.701***&       0.443***\\
                    &    (0.0254)   &    (0.0406)   &    (0.0360)   &    (0.0289)   &    (0.0486)   \\
\midrule
Round FE            &         Yes   &         Yes   &         Yes   &         Yes   &         Yes   \\
Advantage by Fillers&         Yes   &         Yes   &         Yes   &         Yes   &         Yes   \\
No. Observations    &        2184   &        1871   &        1457   &        1515   &        1083   \\
No. Clusters        &        1050   &         978   &         863   &         895   &         728   \\
Adj. R2             &       0.628   &       0.383   &       0.391   &       0.522   &       0.146   \\
\bottomrule
\end{tabular}
}

%% file: Results_Peri/Tables/sliders_beyond_two_signals.tex
{
\def\sym#1{\ifmmode^{#1}\else\(^{#1}\)\fi}
\begin{tabular}{l*{4}{c}}
\toprule
                    &\multicolumn{2}{c}{Low Prior}  &\multicolumn{2}{c}{High Prior} \\\cmidrule(lr){2-3}\cmidrule(lr){4-5}
                    &\multicolumn{1}{c}{Belief after Signal t}&\multicolumn{1}{c}{Above-the-Bar?}&\multicolumn{1}{c}{Belief after Signal t}&\multicolumn{1}{c}{Above-the-Bar?}\\
\midrule
Relative Advantage  &       0.074*  &        0.22** &        0.16***&        0.27***\\
                    &     (0.045)   &     (0.099)   &     (0.048)   &      (0.10)   \\
\addlinespace
Male Candidate      &        1.04   &        1.46   &        1.43*  &        2.45   \\
                    &      (0.75)   &      (1.83)   &      (0.80)   &      (1.85)   \\
\addlinespace
Share of Confirming Signals&       -0.47***&       -1.25***&        0.50***&        1.29***\\
                    &     (0.013)   &     (0.025)   &     (0.011)   &     (0.020)   \\
\addlinespace
Sh. of Conf. Signals x Rel. Adv.&     -0.0012   &     -0.0037** &     -0.0019***&     -0.0029** \\
                    &   (0.00076)   &    (0.0015)   &   (0.00068)   &    (0.0014)   \\
\addlinespace
Sh. of Conf. Signals x Male&      -0.022*  &      -0.017   &      -0.016   &      -0.038   \\
                    &     (0.012)   &     (0.026)   &     (0.011)   &     (0.024)   \\
\midrule
Round FE            &         Yes   &         Yes   &         Yes   &         Yes   \\
Advantage by Fillers&         Yes   &         Yes   &         Yes   &         Yes   \\
No. Observations    &       19875   &        3975   &       20275   &        4055   \\
No. Clusters        &        1180   &        1180   &        1198   &        1198   \\
Adj. R2             &        0.51   &        0.61   &        0.51   &        0.62   \\
\bottomrule
\end{tabular}
}

%% file: Results_Peri/Tables/sliders_two_signals_low_prior.tex
{
\def\sym#1{\ifmmode^{#1}\else\(^{#1}\)\fi}
\begin{tabular}{l*{6}{c}}
\toprule
                    &\multicolumn{1}{c}{No Signals}&\multicolumn{1}{c}{1 Positive}&\multicolumn{1}{c}{1 Negative}&\multicolumn{1}{c}{2 Positives}&\multicolumn{1}{c}{2 Mixed}&\multicolumn{1}{c}{2 Negatives}\\
\midrule
Relative Advantage  &       0.150***&       0.107***&     -0.0229   &      0.0880*  &    -0.00160   &     -0.0835** \\
                    &    (0.0274)   &    (0.0266)   &    (0.0266)   &    (0.0487)   &    (0.0252)   &    (0.0399)   \\
\addlinespace
Male Candidate      &       0.679   &       0.231   &      -0.220   &      -0.523   &       0.247   &      -0.345   \\
                    &     (0.448)   &     (0.457)   &     (0.472)   &     (0.802)   &     (0.458)   &     (0.708)   \\
\addlinespace
Exogenous           &      0.0637   &      -2.160***&       1.930***&      -6.547***&       0.754   &       5.157***\\
                    &     (0.544)   &     (0.519)   &     (0.520)   &     (0.914)   &     (0.467)   &     (0.815)   \\
\addlinespace
Initial Likelihood  &               &       0.692***&       0.599***&       0.419***&       0.643***&       0.457***\\
                    &               &    (0.0200)   &    (0.0284)   &    (0.0319)   &    (0.0246)   &    (0.0332)   \\
\midrule
Round FE            &         Yes   &         Yes   &         Yes   &         Yes   &         Yes   &         Yes   \\
Advantage by Fillers&         Yes   &         Yes   &         Yes   &         Yes   &         Yes   &         Yes   \\
No. Clusters        &        7882   &        3131   &        3794   &        1705   &        2479   &        2257   \\
No. Observations    &        2326   &        1706   &        1871   &        1192   &        1497   &        1439   \\
Adj. R2             &      0.0483   &       0.507   &       0.385   &       0.211   &       0.483   &       0.214   \\
\bottomrule
\end{tabular}
}

%% file: Results_Peri/Tables/sliders_two_signals_high_prior.tex
{
\def\sym#1{\ifmmode^{#1}\else\(^{#1}\)\fi}
\begin{tabular}{l*{6}{c}}
\toprule
                    &\multicolumn{1}{c}{No Signals}&\multicolumn{1}{c}{1 Positive}&\multicolumn{1}{c}{1 Negative}&\multicolumn{1}{c}{2 Positives}&\multicolumn{1}{c}{2 Mixed}&\multicolumn{1}{c}{2 Negatives}\\
\midrule
Relative Advantage  &       0.162***&      0.0181   &      0.0756** &      0.0393   &      0.0801***&       0.123** \\
                    &    (0.0237)   &    (0.0194)   &    (0.0296)   &    (0.0320)   &    (0.0246)   &    (0.0498)   \\
\addlinespace
Male Candidate      &      -0.648   &     -0.0633   &       0.113   &     -0.0135   &    -0.00698   &       0.743   \\
                    &     (0.401)   &     (0.339)   &     (0.517)   &     (0.562)   &     (0.448)   &     (0.873)   \\
\addlinespace
Exogenous           &       0.406   &      -2.722***&       2.443***&      -6.252***&      -0.313   &       6.937***\\
                    &     (0.476)   &     (0.395)   &     (0.571)   &     (0.632)   &     (0.487)   &     (0.994)   \\
\addlinespace
Initial Likelihood  &               &       0.695***&       0.654***&       0.517***&       0.670***&       0.443***\\
                    &               &    (0.0202)   &    (0.0282)   &    (0.0294)   &    (0.0240)   &    (0.0383)   \\
\midrule
Round FE            &         Yes   &         Yes   &         Yes   &         Yes   &         Yes   &         Yes   \\
Advantage by Fillers&         Yes   &         Yes   &         Yes   &         Yes   &         Yes   &         Yes   \\
No. Observations    &        8118   &        3905   &        3188   &        2408   &        2476   &        1750   \\
No. Clusters        &        2381   &        1907   &        1745   &        1455   &        1537   &        1213   \\
Adj. R2             &      0.0641   &       0.569   &       0.381   &       0.350   &       0.480   &       0.165   \\
\bottomrule
\end{tabular}
}

%% file: Results_Peri/Tables/first_final_slider_and_eval_robust.tex
{
\def\sym#1{\ifmmode^{#1}\else\(^{#1}\)\fi}
\begin{tabular}{l*{6}{c}}
\toprule
                    &\multicolumn{2}{c}{Initial Slider}&\multicolumn{2}{c}{Final Slider}&\multicolumn{2}{c}{Final Eval} \\\cmidrule(lr){2-3}\cmidrule(lr){4-5}\cmidrule(lr){6-7}
                    &\multicolumn{1}{c}{(1)}   &\multicolumn{1}{c}{(2)}   &\multicolumn{1}{c}{(3)}   &\multicolumn{1}{c}{(4)}   &\multicolumn{1}{c}{(5)}   &\multicolumn{1}{c}{(6)}   \\
\midrule
Relative Advantage  &       0.145***&       0.170***&      0.0775** &      0.0912** &       0.160***&       0.255***\\
                    &    (0.0183)   &    (0.0215)   &    (0.0302)   &    (0.0355)   &    (0.0446)   &    (0.0550)   \\
\addlinespace
Male Candidate      &      -0.116   &       0.191   &       0.409   &       0.359   &       0.504   &       0.307   \\
                    &     (0.313)   &     (0.447)   &     (0.590)   &     (0.786)   &     (0.879)   &     (1.263)   \\
\addlinespace
Group Prior         &       0.426***&       0.426***&       0.533***&       0.533***&       0.712***&       0.712***\\
                    &    (0.0289)   &    (0.0289)   &    (0.0439)   &    (0.0439)   &    (0.0633)   &    (0.0633)   \\
\addlinespace
Exogenous           &       0.265   &       0.566   &      -2.338***&      -2.390***&      -3.006***&      -3.220***\\
                    &     (0.377)   &     (0.484)   &     (0.598)   &     (0.846)   &     (0.864)   &     (1.221)   \\
\addlinespace
Rel Adv x Exog      &               &     -0.0478** &               &     -0.0271   &               &      -0.186***\\
                    &               &    (0.0228)   &               &    (0.0411)   &               &    (0.0625)   \\
\addlinespace
Male x Exog         &               &      -0.603   &               &       0.105   &               &       0.428   \\
                    &               &     (0.626)   &               &     (1.177)   &               &     (1.751)   \\
\midrule
Round FE            &         Yes   &         Yes   &         Yes   &         Yes   &         Yes   &         Yes   \\
Advantage by Fillers&         Yes   &         Yes   &         Yes   &         Yes   &         Yes   &         Yes   \\
No. Observations    &       12800   &       12800   &       12800   &       12800   &       12800   &       12800   \\
No. Clusters        &        3200   &        3200   &        3200   &        3200   &        3200   &        3200   \\
Adj. R2             &       0.152   &       0.152   &      0.0387   &      0.0386   &      0.0361   &      0.0367   \\
\bottomrule
\end{tabular}
}

%% file: Results_Peri/Tables/info_acquisition_robust.tex
{
\def\sym#1{\ifmmode^{#1}\else\(^{#1}\)\fi}
\begin{tabular}{l*{4}{c}}
\toprule
                    &\multicolumn{1}{c}{Immediate Evaluation}&\multicolumn{1}{c}{Got More Info}&\multicolumn{2}{c}{Pooled}     \\\cmidrule(lr){2-2}\cmidrule(lr){3-3}\cmidrule(lr){4-5}
                    &\multicolumn{1}{c}{Initial Slider}&\multicolumn{1}{c}{Initial Slider}&\multicolumn{1}{c}{Initial Slider}&\multicolumn{1}{c}{Final Eval}\\
\midrule
Relative Advantage  &        0.29***&        0.14***&        0.25***&        0.81***\\
                    &     (0.064)   &     (0.026)   &     (0.049)   &     (0.098)   \\
\addlinespace
Male Candidate      &        0.26   &        0.13   &        0.30   &       -2.71   \\
                    &      (1.22)   &      (0.45)   &      (1.21)   &      (2.54)   \\
\addlinespace
Group Prior         &        0.31***&        0.43***&        0.40***&        0.71***\\
                    &     (0.093)   &     (0.042)   &     (0.039)   &     (0.089)   \\
\addlinespace
Got More Info       &               &               &       -3.58***&       -10.7***\\
                    &               &               &      (0.95)   &      (2.01)   \\
\addlinespace
Got More Info x Rel Adv&               &               &      -0.090*  &       -0.72***\\
                    &               &               &     (0.047)   &      (0.10)   \\
\addlinespace
Got More Info x Male&               &               &       -0.18   &        3.59   \\
                    &               &               &      (1.29)   &      (2.93)   \\
\midrule
Round FE            &         Yes   &         Yes   &         Yes   &         Yes   \\
Advantage by Fillers&         Yes   &         Yes   &         Yes   &         Yes   \\
No. Observations    &        1432   &        4944   &        6376   &        6376   \\
No. Clusters        &         562   &        1405   &        1594   &        1594   \\
Adj. R2             &        0.14   &        0.17   &        0.16   &       0.058   \\
\bottomrule
\end{tabular}
}

%% file: Results_Peri/Tables/when_to_acquire_info_robust.tex
{
\def\sym#1{\ifmmode^{#1}\else\(^{#1}\)\fi}
\begin{tabular}{l*{5}{c}}
\toprule
                    &\multicolumn{1}{c}{1st Signal}&\multicolumn{1}{c}{2nd Signal}&\multicolumn{1}{c}{3rd Signal}&\multicolumn{1}{c}{4th Signal}&\multicolumn{1}{c}{Pooled-Last Signal}\\
\midrule
Confirming Signal   &      -0.025***&      -0.076***&     -0.0099   &      0.0033   &      -0.022***\\
                    &    (0.0095)   &     (0.013)   &     (0.017)   &     (0.020)   &    (0.0057)   \\
\midrule
Round FE            &         Yes   &         Yes   &         Yes   &         Yes   &         Yes   \\
Advantage by Fillers&         Yes   &         Yes   &         Yes   &         Yes   &         Yes   \\
No. Observations    &        4944   &        4319   &        3183   &        2029   &       15871   \\
No. Clusters        &        1405   &        1280   &        1119   &         809   &        1405   \\
Adj. R2             &       0.010   &      0.0097   &      0.0016   &     -0.0022   &      0.0010   \\
\bottomrule
\end{tabular}
}

%% file: Results_Peri/Tables/when_to_acquire_info_equal_groups_robust.tex
{
\def\sym#1{\ifmmode^{#1}\else\(^{#1}\)\fi}
\begin{tabular}{l*{5}{c}}
\toprule
                    &\multicolumn{1}{c}{1st Signal}&\multicolumn{1}{c}{2nd Signal}&\multicolumn{1}{c}{ 3rd Signal}&\multicolumn{1}{c}{4th Signal}&\multicolumn{1}{c}{Pooled-Last Signal}\\
\midrule
Confirming Signal   &      -0.048***&      -0.090***&      -0.024   &      -0.021   &      -0.034***\\
                    &     (0.013)   &     (0.020)   &     (0.025)   &     (0.030)   &    (0.0082)   \\
\addlinespace
Equal Groups        &      -0.032** &     -0.0062   &     -0.0044   &     0.00045   &      -0.010   \\
                    &     (0.016)   &     (0.021)   &     (0.029)   &     (0.034)   &    (0.0082)   \\
\addlinespace
Equal Groups x Conf. Signal&       0.045** &       0.026   &       0.028   &       0.045   &       0.023** \\
                    &     (0.019)   &     (0.027)   &     (0.034)   &     (0.041)   &     (0.011)   \\
\midrule
Round FE            &         Yes   &         Yes   &         Yes   &         Yes   &         Yes   \\
Advantage by Fillers&         Yes   &         Yes   &         Yes   &         Yes   &         Yes   \\
No. Observations    &        4944   &        4319   &        3183   &        2029   &       15871   \\
No. Clusters        &        1405   &        1280   &        1119   &         809   &        1405   \\
Adj. R2             &       0.011   &      0.0090   &      0.0012   &     -0.0014   &     0.00095   \\
\bottomrule
\end{tabular}
}

%% file: Results_Peri/Tables/first_final_slider_and_eval_gender_int.tex
{
\def\sym#1{\ifmmode^{#1}\else\(^{#1}\)\fi}
\begin{tabular}{l*{9}{c}}
\toprule
                    &\multicolumn{3}{c}{Initial Slider}             &\multicolumn{3}{c}{Final Slider}               &\multicolumn{3}{c}{Final Eval}                 \\\cmidrule(lr){2-4}\cmidrule(lr){5-7}\cmidrule(lr){8-10}
                    &\multicolumn{1}{c}{(1)}   &\multicolumn{1}{c}{(2)}   &\multicolumn{1}{c}{(3)}   &\multicolumn{1}{c}{(4)}   &\multicolumn{1}{c}{(5)}   &\multicolumn{1}{c}{(6)}   &\multicolumn{1}{c}{(7)}   &\multicolumn{1}{c}{(8)}   &\multicolumn{1}{c}{(9)}   \\
\midrule
Relative Advantage  &       0.157***&       0.164***&       0.189***&      0.0779***&       0.105***&      0.0834** &       0.148***&       0.267***&       0.131***\\
                    &    (0.0177)   &    (0.0208)   &    (0.0210)   &    (0.0272)   &    (0.0323)   &    (0.0333)   &    (0.0398)   &    (0.0499)   &    (0.0477)   \\
\addlinespace
Male Candidate      &      0.0105   &       0.128   &      0.0119   &       0.145   &       0.440   &       0.145   &       0.267   &       0.342   &       0.267   \\
                    &     (0.292)   &     (0.419)   &     (0.292)   &     (0.531)   &     (0.705)   &     (0.531)   &     (0.789)   &     (1.141)   &     (0.790)   \\
\addlinespace
Group Prior         &       0.456***&       0.456***&       0.457***&       0.550***&       0.550***&       0.550***&       0.778***&       0.778***&       0.778***\\
                    &    (0.0282)   &    (0.0282)   &    (0.0282)   &    (0.0393)   &    (0.0393)   &    (0.0393)   &    (0.0558)   &    (0.0558)   &    (0.0558)   \\
\addlinespace
Exogenous           &       0.242   &       0.357   &       0.248   &      -2.726***&      -2.438***&      -2.725***&      -3.854***&      -3.808***&      -3.858***\\
                    &     (0.368)   &     (0.464)   &     (0.368)   &     (0.531)   &     (0.758)   &     (0.531)   &     (0.757)   &     (1.083)   &     (0.757)   \\
\addlinespace
Rel Adv x Exog      &               &     -0.0145   &               &               &     -0.0524   &               &               &      -0.235***&               \\
                    &               &    (0.0217)   &               &               &    (0.0377)   &               &               &    (0.0565)   &               \\
\addlinespace
Male x Exog         &               &      -0.230   &               &               &      -0.578   &               &               &      -0.103   &               \\
                    &               &     (0.585)   &               &               &     (1.060)   &               &               &     (1.571)   &               \\
\addlinespace
Rel Adv x Male      &               &               &     -0.0643***&               &               &     -0.0111   &               &               &      0.0338   \\
                    &               &               &    (0.0239)   &               &               &    (0.0369)   &               &               &    (0.0526)   \\
\midrule
Round FE            &         Yes   &         Yes   &         Yes   &         Yes   &         Yes   &         Yes   &         Yes   &         Yes   &         Yes   \\
Advantage by Fillers&         Yes   &         Yes   &         Yes   &         Yes   &         Yes   &         Yes   &         Yes   &         Yes   &         Yes   \\
No. Observations    &       16000   &       16000   &       16000   &       16000   &       16000   &       16000   &       16000   &       16000   &       16000   \\
No. Clusters        &        3200   &        3200   &        3200   &        3200   &        3200   &        3200   &        3200   &        3200   &        3200   \\
Adj. R2             &       0.168   &       0.168   &       0.169   &      0.0411   &      0.0411   &      0.0411   &      0.0409   &      0.0419   &      0.0409   \\
\bottomrule
\end{tabular}
}

%% file: Results_Peri/Tables/sliders_two_signals_low_prior_exog_bayes.tex
{
\def\sym#1{\ifmmode^{#1}\else\(^{#1}\)\fi}
\begin{tabular}{l*{5}{c}}
\toprule
                    &\multicolumn{1}{c}{1 Positive}&\multicolumn{1}{c}{1 Negative}&\multicolumn{1}{c}{2 Positives}&\multicolumn{1}{c}{2 Mixed}&\multicolumn{1}{c}{2 Negatives}\\
\midrule
Relative Advantage  &       0.131***&     0.00228   &      0.0292   &      0.0297   &     -0.0471   \\
                    &    (0.0370)   &    (0.0367)   &    (0.0479)   &    (0.0415)   &    (0.0412)   \\
\addlinespace
Male Candidate      &       0.830   &      -0.330   &      -0.166   &      -0.510   &      -0.658   \\
                    &     (0.636)   &     (0.653)   &     (0.751)   &     (0.781)   &     (0.709)   \\
\midrule
Round FE            &         Yes   &         Yes   &         Yes   &         Yes   &         Yes   \\
Advantage by Fillers&         Yes   &         Yes   &         Yes   &         Yes   &         Yes   \\
No. Observations    &        1809   &        2166   &        1043   &        1543   &        1389   \\
No. Clusters        &         959   &        1042   &         719   &         894   &         864   \\
Adj. R2             &       0.488   &       0.357   &       0.566   &       0.282   &       0.452   \\
\bottomrule
\end{tabular}
}

%% file: Results_Peri/Tables/sliders_two_signals_high_prior_exog_bayes.tex
{
\def\sym#1{\ifmmode^{#1}\else\(^{#1}\)\fi}
\begin{tabular}{l*{5}{c}}
\toprule
                    &\multicolumn{1}{c}{1 Positive}&\multicolumn{1}{c}{1 Negative}&\multicolumn{1}{c}{2 Positives}&\multicolumn{1}{c}{2 Mixed}&\multicolumn{1}{c}{2 Negatives}\\
\midrule
Relative Advantage  &      0.0158   &       0.104** &      0.0323   &       0.120***&       0.125** \\
                    &    (0.0293)   &    (0.0429)   &    (0.0346)   &    (0.0392)   &    (0.0541)   \\
\addlinespace
Male Candidate      &      -0.260   &       0.483   &     -0.0868   &     -0.0994   &       1.243   \\
                    &     (0.511)   &     (0.748)   &     (0.703)   &     (0.723)   &     (0.957)   \\
\midrule
Round FE            &         Yes   &         Yes   &         Yes   &         Yes   &         Yes   \\
Advantage by Fillers&         Yes   &         Yes   &         Yes   &         Yes   &         Yes   \\
No. Observations    &        2184   &        1871   &        1457   &        1515   &        1083   \\
No. Clusters        &        1050   &         978   &         863   &         895   &         728   \\
Adj. R2             &       0.547   &       0.321   &       0.497   &       0.294   &       0.340   \\
\bottomrule
\end{tabular}
}

%% file: Results_Peri/Tables/first_final_slider_and_eval_dur_rob.tex
{
\def\sym#1{\ifmmode^{#1}\else\(^{#1}\)\fi}
\begin{tabular}{l*{6}{c}}
\toprule
                    &\multicolumn{2}{c}{Initial Slider}&\multicolumn{2}{c}{Final Slider}&\multicolumn{2}{c}{Final Eval} \\\cmidrule(lr){2-3}\cmidrule(lr){4-5}\cmidrule(lr){6-7}
                    &\multicolumn{1}{c}{(1)}   &\multicolumn{1}{c}{(2)}   &\multicolumn{1}{c}{(3)}   &\multicolumn{1}{c}{(4)}   &\multicolumn{1}{c}{(5)}   &\multicolumn{1}{c}{(6)}   \\
\midrule
Relative Advantage  &       0.159***&       0.165***&      0.0744***&       0.103***&       0.170***&       0.287***\\
                    &    (0.0183)   &    (0.0214)   &    (0.0287)   &    (0.0337)   &    (0.0420)   &    (0.0521)   \\
\addlinespace
Male Candidate      &      -0.102   &      0.0295   &      -0.167   &       0.303   &      -0.258   &     -0.0324   \\
                    &     (0.303)   &     (0.427)   &     (0.559)   &     (0.730)   &     (0.836)   &     (1.188)   \\
\addlinespace
Group Prior         &       0.469***&       0.469***&       0.567***&       0.567***&       0.770***&       0.770***\\
                    &    (0.0294)   &    (0.0294)   &    (0.0414)   &    (0.0414)   &    (0.0588)   &    (0.0588)   \\
\addlinespace
Exogenous           &     -0.0348   &      0.0989   &      -2.800***&      -2.322***&      -3.740***&      -3.518***\\
                    &     (0.389)   &     (0.487)   &     (0.561)   &     (0.800)   &     (0.796)   &     (1.141)   \\
\addlinespace
Rel Adv x Exog      &               &     -0.0123   &               &     -0.0589   &               &      -0.238***\\
                    &               &    (0.0224)   &               &    (0.0400)   &               &    (0.0597)   \\
\addlinespace
Male x Exog         &               &      -0.268   &               &      -0.961   &               &      -0.448   \\
                    &               &     (0.607)   &               &     (1.119)   &               &     (1.659)   \\
\midrule
Round FE            &         Yes   &         Yes   &         Yes   &         Yes   &         Yes   &         Yes   \\
Advantage by Fillers&         Yes   &         Yes   &         Yes   &         Yes   &         Yes   &         Yes   \\
No. Observations    &       14420   &       14420   &       14420   &       14420   &       14420   &       14420   \\
No. Clusters        &        2884   &        2884   &        2884   &        2884   &        2884   &        2884   \\
Adj. R2             &       0.177   &       0.177   &      0.0417   &      0.0417   &      0.0414   &      0.0424   \\
\bottomrule
\end{tabular}
}

%% file: Results_Peri/Tables/info_acquisition_dur_rob.tex
{
\def\sym#1{\ifmmode^{#1}\else\(^{#1}\)\fi}
\begin{tabular}{l*{4}{c}}
\toprule
                    &\multicolumn{1}{c}{Immediate Evaluation}&\multicolumn{1}{c}{Got More Info}&\multicolumn{2}{c}{Pooled}     \\\cmidrule(lr){2-2}\cmidrule(lr){3-3}\cmidrule(lr){4-5}
                    &\multicolumn{1}{c}{Initial Slider}&\multicolumn{1}{c}{Initial Slider}&\multicolumn{1}{c}{Initial Slider}&\multicolumn{1}{c}{Final Eval}\\
\midrule
Relative Advantage  &        0.30***&        0.13***&        0.26***&        0.87***\\
                    &     (0.059)   &     (0.026)   &     (0.045)   &     (0.090)   \\
\addlinespace
Male Candidate      &       -0.97   &        0.37   &       -1.01   &       -4.11*  \\
                    &      (1.10)   &      (0.43)   &      (1.10)   &      (2.33)   \\
\addlinespace
Group Prior         &        0.34***&        0.48***&        0.44***&        0.76***\\
                    &     (0.088)   &     (0.041)   &     (0.039)   &     (0.082)   \\
\addlinespace
Got More Info       &               &               &       -5.02***&       -13.2***\\
                    &               &               &      (0.88)   &      (1.82)   \\
\addlinespace
Got More Info x Rel Adv&               &               &       -0.12***&       -0.78***\\
                    &               &               &     (0.042)   &     (0.092)   \\
\addlinespace
Got More Info x Male&               &               &        1.35   &        5.09*  \\
                    &               &               &      (1.18)   &      (2.67)   \\
\midrule
Round FE            &         Yes   &         Yes   &         Yes   &         Yes   \\
Advantage by Fillers&         Yes   &         Yes   &         Yes   &         Yes   \\
No. Observations    &        1766   &        5614   &        7380   &        7380   \\
No. Clusters        &         654   &        1329   &        1476   &        1476   \\
Adj. R2             &        0.18   &        0.19   &        0.19   &       0.072   \\
\bottomrule
\end{tabular}
}

%% file: Results_Peri/Tables/when_to_acquire_info_dur_rob.tex
{
\def\sym#1{\ifmmode^{#1}\else\(^{#1}\)\fi}
\begin{tabular}{l*{5}{c}}
\toprule
                    &\multicolumn{1}{c}{1st Signal}&\multicolumn{1}{c}{2nd Signal}&\multicolumn{1}{c}{3rd Signal}&\multicolumn{1}{c}{4th Signal}&\multicolumn{1}{c}{Pooled-Last Signal}\\
\midrule
Confirming Signal   &      -0.030***&      -0.082***&      -0.011   &      -0.020   &      -0.027***\\
                    &    (0.0094)   &     (0.013)   &     (0.017)   &     (0.020)   &    (0.0055)   \\
\midrule
Round FE            &         Yes   &         Yes   &         Yes   &         Yes   &         Yes   \\
Advantage by Fillers&         Yes   &         Yes   &         Yes   &         Yes   &         Yes   \\
No. Observations    &        5614   &        4745   &        3406   &        2139   &       17359   \\
No. Clusters        &        1329   &        1209   &        1068   &         778   &        1329   \\
Adj. R2             &       0.043   &       0.018   &      0.0020   &     0.00059   &      0.0049   \\
\bottomrule
\end{tabular}
}

%% file: Results_Peri/Tables/when_to_acquire_info_equal_groups_dur_rob.tex
{
\def\sym#1{\ifmmode^{#1}\else\(^{#1}\)\fi}
\begin{tabular}{l*{5}{c}}
\toprule
                    &\multicolumn{1}{c}{1st Signal}&\multicolumn{1}{c}{2nd Signal}&\multicolumn{1}{c}{ 3rd Signal}&\multicolumn{1}{c}{4th Signal}&\multicolumn{1}{c}{Pooled-Last Signal}\\
\midrule
Confirming Signal   &      -0.054***&       -0.11***&      -0.016   &      -0.046   &      -0.041***\\
                    &     (0.013)   &     (0.020)   &     (0.025)   &     (0.031)   &    (0.0080)   \\
\addlinespace
Equal Groups        &      -0.028*  &      -0.010   &       0.018   &       0.026   &     -0.0074   \\
                    &     (0.016)   &     (0.022)   &     (0.029)   &     (0.034)   &    (0.0085)   \\
\addlinespace
Equal Groups x Conf. Signal&       0.047** &       0.049*  &       0.011   &       0.044   &       0.028** \\
                    &     (0.019)   &     (0.026)   &     (0.033)   &     (0.041)   &     (0.011)   \\
\midrule
Round FE            &         Yes   &         Yes   &         Yes   &         Yes   &         Yes   \\
Advantage by Fillers&         Yes   &         Yes   &         Yes   &         Yes   &         Yes   \\
No. Observations    &        5614   &        4745   &        3406   &        2139   &       17359   \\
No. Clusters        &        1329   &        1209   &        1068   &         778   &        1329   \\
Adj. R2             &       0.043   &       0.018   &      0.0024   &      0.0033   &      0.0050   \\
\bottomrule
\end{tabular}
}

%% file: Results_Peri/Tables/sliders_two_signals_low_prior_exog_dur_rob.tex
{
\def\sym#1{\ifmmode^{#1}\else\(^{#1}\)\fi}
\begin{tabular}{l*{5}{c}}
\toprule
                    &\multicolumn{1}{c}{1 Positive}&\multicolumn{1}{c}{1 Negative}&\multicolumn{1}{c}{2 Positives}&\multicolumn{1}{c}{2 Mixed}&\multicolumn{1}{c}{2 Negatives}\\
\midrule
Relative Advantage  &       0.112***&    -0.00313   &      0.0404   &      0.0222   &     -0.0400   \\
                    &    (0.0370)   &    (0.0368)   &    (0.0688)   &    (0.0342)   &    (0.0535)   \\
\addlinespace
Male Candidate      &       0.757   &      -0.662   &      -0.173   &       0.627   &      -1.174   \\
                    &     (0.636)   &     (0.668)   &     (1.058)   &     (0.695)   &     (0.949)   \\
\addlinespace
Initial Likelihood  &       0.695***&       0.661***&       0.466***&       0.648***&       0.541***\\
                    &    (0.0273)   &    (0.0317)   &    (0.0404)   &    (0.0316)   &    (0.0444)   \\
\midrule
Round FE            &         Yes   &         Yes   &         Yes   &         Yes   &         Yes   \\
Advantage by Fillers&         Yes   &         Yes   &         Yes   &         Yes   &         Yes   \\
No. Observations    &        1586   &        1881   &         916   &        1348   &        1203   \\
No. Clusters        &         840   &         909   &         632   &         777   &         747   \\
Adj. R2             &       0.529   &       0.446   &       0.246   &       0.472   &       0.272   \\
\bottomrule
\end{tabular}
}

%% file: Results_Peri/Tables/sliders_two_signals_high_prior_exog_dur_rob.tex
{
\def\sym#1{\ifmmode^{#1}\else\(^{#1}\)\fi}
\begin{tabular}{l*{5}{c}}
\toprule
                    &\multicolumn{1}{c}{1 Positive}&\multicolumn{1}{c}{1 Negative}&\multicolumn{1}{c}{2 Positives}&\multicolumn{1}{c}{2 Mixed}&\multicolumn{1}{c}{2 Negatives}\\
\midrule
Relative Advantage  &     -0.0236   &      0.0713*  &     -0.0112   &      0.0898** &       0.148** \\
                    &    (0.0257)   &    (0.0410)   &    (0.0385)   &    (0.0350)   &    (0.0692)   \\
\addlinespace
Male Candidate      &     -0.0295   &       0.213   &       0.129   &      -0.256   &       1.212   \\
                    &     (0.467)   &     (0.740)   &     (0.680)   &     (0.598)   &     (1.216)   \\
\addlinespace
Initial Likelihood  &       0.732***&       0.697***&       0.579***&       0.722***&       0.444***\\
                    &    (0.0252)   &    (0.0396)   &    (0.0317)   &    (0.0294)   &    (0.0521)   \\
\midrule
Round FE            &         Yes   &         Yes   &         Yes   &         Yes   &         Yes   \\
Advantage by Fillers&         Yes   &         Yes   &         Yes   &         Yes   &         Yes   \\
No. Observations    &        1914   &        1659   &        1278   &        1321   &         974   \\
No. Clusters        &         921   &         858   &         761   &         776   &         651   \\
Adj. R2             &       0.650   &       0.420   &       0.460   &       0.563   &       0.137   \\
\bottomrule
\end{tabular}
}

%% file: Results_Peri/Tables/sliders_beyond_two_signals_dur_rob.tex
{
\def\sym#1{\ifmmode^{#1}\else\(^{#1}\)\fi}
\begin{tabular}{l*{4}{c}}
\toprule
                    &\multicolumn{2}{c}{Low Prior}  &\multicolumn{2}{c}{High Prior} \\\cmidrule(lr){2-3}\cmidrule(lr){4-5}
                    &\multicolumn{1}{c}{Belief after Signal t}&\multicolumn{1}{c}{Above-the-Bar?}&\multicolumn{1}{c}{Belief after Signal t}&\multicolumn{1}{c}{Above-the-Bar?}\\
\midrule
Relative Advantage  &       0.051   &        0.22** &        0.13***&        0.29***\\
                    &     (0.048)   &      (0.11)   &     (0.051)   &      (0.11)   \\
\addlinespace
Male Candidate      &        0.70   &        0.39   &        1.37   &        2.03   \\
                    &      (0.80)   &      (1.98)   &      (0.86)   &      (1.98)   \\
\addlinespace
Share of Confirming Signals&       -0.48***&       -1.26***&        0.50***&        1.29***\\
                    &     (0.014)   &     (0.024)   &     (0.012)   &     (0.022)   \\
\addlinespace
Sh. of Conf. Signals x Rel. Adv.&     -0.0010   &     -0.0037** &     -0.0017** &     -0.0031** \\
                    &   (0.00081)   &    (0.0015)   &   (0.00073)   &    (0.0015)   \\
\addlinespace
Sh. of Conf. Signals x Male&      -0.022*  &     -0.0052   &      -0.015   &      -0.033   \\
                    &     (0.012)   &     (0.028)   &     (0.012)   &     (0.026)   \\
\midrule
Round FE            &         Yes   &         Yes   &         Yes   &         Yes   \\
Advantage by Fillers&         Yes   &         Yes   &         Yes   &         Yes   \\
No. Observations    &       17335   &        3467   &       17865   &        3573   \\
No. Clusters        &        1030   &        1030   &        1051   &        1051   \\
Adj. R2             &        0.52   &        0.62   &        0.51   &        0.62   \\
\bottomrule
\end{tabular}
}